\documentclass[
  journal=pasa,
  manuscript=research-paper, %% or "review"
  year=2024,
  volume=xx,
]{cup-journal}

\usepackage{graphicx}
\usepackage{xcolor}
\usepackage{microtype,siunitx}
\usepackage{amssymb}
\usepackage{amsmath}
\usepackage{CJKutf8}
\usepackage[backref=page]{hyperref}
\definecolor{dodgerblue}{RGB}{30, 144, 255}
\definecolor{crimson}{RGB}{220, 20, 60}
\definecolor{darkerblue}{RGB}{0, 0, 139}
\hypersetup{colorlinks,citecolor=dodgerblue,linkcolor=blue,urlcolor=darkerblue}
\usepackage{enumitem, xcolor}
\let\svitem\item

\usepackage{booktabs} % table formatting
\usepackage{threeparttable,longtable}  % table formatting
\usepackage{makecell}
\usepackage[skip=0.5ex]{subcaption}
\usepackage{cprotect}
\usepackage{multirow}

\definecolor{dodgerblue}{RGB}{30, 144, 255}

\newcommand{\corrs}[1]{{#1}}

\newcommand{\cracofrb}{FRB\,20231027A}

\title{The CRAFT Coherent (CRACO) upgrade I: System Description and Results of the 110-ms Radio Transient Pilot Survey}

\author{Z.~Wang (\begin{CJK*}{UTF8}{gbsn}王子腾\end{CJK*})}
\affiliation{International Centre for Radio Astronomy Research, Curtin University, Bentley, WA 6102, Australia}

\author{K.~W.~Bannister}
\affiliation{CSIRO, Space And Astronomy, PO Box 76, Epping, NSW 1710, Australia}

\author{V.~Gupta}
\affiliation{CSIRO, Space And Astronomy, PO Box 76, Epping, NSW 1710, Australia}

\author{X.~Deng}
\affiliation{CSIRO, Space And Astronomy, PO Box 76, Epping, NSW 1710, Australia}

\author{M.~Pilawa}
\affiliation{CSIRO, Space And Astronomy, PO Box 76, Epping, NSW 1710, Australia}

\author{J.~Tuthill}
\affiliation{CSIRO, Space And Astronomy, PO Box 76, Epping, NSW 1710, Australia}

\author{J.~D.~Bunton}
\affiliation{CSIRO, Space And Astronomy, PO Box 76, Epping, NSW 1710, Australia}

\author{C.~Flynn}
\affiliation{Center for Astrophysics and Supercomputing, Swinburne University of Technology, Post Office Box 218, Hawthorn, VIC 3122, Australia}
\alsoaffiliation{ARC Centre of Excellence for Gravitational Wave Discovery (OzGrav), Post Office Box 218, Hawthorn, VIC 3122, Australia}

\author{M.~Glowacki}
\affiliation{International Centre for Radio Astronomy Research, Curtin University, Bentley, WA 6102, Australia}
\alsoaffiliation{Institute for Astronomy, University of Edinburgh, Royal Observatory, Edinburgh, EH9 3HJ, United Kingdom}

\author{A.~Jaini}
\affiliation{Centre for Astrophysics and Supercomputing, Swinburne University of Technology, Hawthorn, VIC, 3122, Australia}

\author{Y.~W.~J.~Lee}
\affiliation{Sydney Institute for Astronomy, School of Physics, The University of Sydney, Sydney, 2006, NSW, Australia}
\alsoaffiliation{ARC Centre of Excellence for Gravitational Wave Discovery (OzGrav), Post Office Box 218, Hawthorn, VIC 3122, Australia}
\alsoaffiliation{CSIRO, Space And Astronomy, PO Box 76, Epping, NSW 1710, Australia}

\author{E.~Lenc}
\affiliation{CSIRO, Space And Astronomy, PO Box 76, Epping, NSW 1710, Australia}

\author{J.~Lucero}
\affiliation{Advanced Micro Devices Inc, 2485 Augustine Dr, Santa Clara, CA 95054, USA}

\author{A.~Paek}
\affiliation{Advanced Micro Devices Inc, 2485 Augustine Dr, Santa Clara, CA 95054, USA}

\author{R.~Radhakrishnan}
\affiliation{Advanced Micro Devices Inc, 2485 Augustine Dr, Santa Clara, CA 95054, USA}
\alsoaffiliation{Cadence Design Systems, Inc, 2655 Seely Ave, Ste 5, San Jose, CA 95134, USA}

\author{N.~Thyagarajan}
\affiliation{CSIRO, Space And Astronomy, PO Box 1130, Bently, WA 6102, Australia}

\author{P.~Uttarkar}
\affiliation{Centre for Astrophysics and Supercomputing, Swinburne University of Technology, Hawthorn, VIC, 3122, Australia}

\author{Y.~Wang}
\affiliation{Centre for Astrophysics and Supercomputing, Swinburne University of Technology, Hawthorn, VIC, 3122, Australia}
\alsoaffiliation{ARC Centre of Excellence for Gravitational Wave Discovery (OzGrav), Post Office Box 218, Hawthorn, VIC 3122, Australia}

\author{N.~D.~R.~Bhat}
\affiliation{International Centre for Radio Astronomy Research, Curtin University, Bentley, WA 6102, Australia}

\author{C.~W.~James}
\affiliation{International Centre for Radio Astronomy Research, Curtin University, Bentley, WA 6102, Australia}

\author{V.~A.~Moss}
\affiliation{CSIRO, Space And Astronomy, PO Box 76, Epping, NSW 1710, Australia}
\alsoaffiliation{Sydney Institute for Astronomy, School of Physics, The University of Sydney, Sydney, 2006, NSW, Australia}

\author{Tara~Murphy}
\affiliation{Sydney Institute for Astronomy, School of Physics, The University of Sydney, Sydney, 2006, NSW, Australia}
\alsoaffiliation{ARC Centre of Excellence for Gravitational Wave Discovery (OzGrav), Post Office Box 218, Hawthorn, VIC 3122, Australia}

\author{J.~E.~Reynolds}
\affiliation{CSIRO, Space And Astronomy, PO Box 76, Epping, NSW 1710, Australia}

\author{R.~M.~Shannon}
\affiliation{Centre for Astrophysics and Supercomputing, Swinburne University of Technology, Hawthorn, VIC, 3122, Australia}

\author{L.~G.~Spitler}
\affiliation{Max Planck Institute for Radio Astronomy, Auf dem H\"{u}gel 69, 53121 Bonn, Germany}

\author{A.~Tzioumis}
\affiliation{CSIRO, Space And Astronomy, PO Box 76, Epping, NSW 1710, Australia}

\author{M.~Caleb}
\affiliation{Sydney Institute for Astronomy, School of Physics, The University of Sydney, Sydney, 2006, NSW, Australia}

\author{A.~T.~Deller}
\affiliation{Centre for Astrophysics and Supercomputing, Swinburne University of Technology, Hawthorn, VIC, 3122, Australia}
\alsoaffiliation{ARC Centre of Excellence for Gravitational Wave Discovery (OzGrav), Post Office Box 218, Hawthorn, VIC 3122, Australia}

\author{A.~C.~Gordon}
\affiliation{Center for Interdisciplinary Exploration and Research in Astrophysics (CIERA) and Department of Physics and Astronomy, Northwestern University, Evanston, IL 60208, USA}

\author{L.~Marnoch}
\affiliation{School of Mathematical and Physical Sciences, Macquarie University, NSW 2109, Australia}
\alsoaffiliation{Astrophysics and Space Technologies Research Centre, Macquarie University, Sydney, NSW 2109, Australia}
\alsoaffiliation{CSIRO, Space And Astronomy, PO Box 76, Epping, NSW 1710, Australia}
\alsoaffiliation{ARC Centre of Excellence for All-Sky Astrophysics in 3 Dimensions (ASTRO 3D), Sydney, Australia}

\author{S.~D.~Ryder}
\affiliation{School of Mathematical and Physical Sciences, Macquarie University, NSW 2109, Australia}
\alsoaffiliation{Astrophysics and Space Technologies Research Centre, Macquarie University, Sydney, NSW 2109, Australia}

\author{S.~Simha}
\affiliation{Center for Interdisciplinary Exploration and Research in Astrophysics (CIERA) and Department of Physics and Astronomy, Northwestern University, Evanston, IL 60208, USA}
\alsoaffiliation{Department of Astronomy and Astrophysics, University of Chicago, 5640 South Ellis Avenue, Chicago, IL, 60637, USA}

\author{C.~S.~Anderson}
\affiliation{Research School of Astronomy \& Astrophysics, Australian National University, Canberra ACT, Australia 2610}

\author{L.~Ball}
\affiliation{SKA Observatory, Jodrell Bank, Lower Withington, Macclesfield, Cheshire SK11 9FT, UK}
\alsoaffiliation{CSIRO, Space And Astronomy, PO Box 76, Epping, NSW 1710, Australia}

\author{D.~Brodrick}
\affiliation{Advanced Instrumentation Technology Centre, Research School of Astronomy \& Astrophysics, Australian National University, Canberra, Australia}

\author{F.~R.~Cooray}
\affiliation{CSIRO, Space And Astronomy, PO Box 76, Epping, NSW 1710, Australia}

\author{N.~Gupta}
\affiliation{Inter-University Centre for Astronomy and Astrophysics, Post Bag 4, Ganeshkhind, Pune 411 007, India}

\author{D.~B.~Hayman}
\affiliation{CSIRO, Space And Astronomy, PO Box 76, Epping, NSW 1710, Australia}

\author{A.~Ng}
\affiliation{CSIRO, Space And Astronomy, PO Box 76, Epping, NSW 1710, Australia}

\author{S.~E.~Pearce}
\affiliation{SKA Observatory, Jodrell Bank, Lower Withington, Macclesfield, Cheshire SK11 9FT, UK}
\alsoaffiliation{CSIRO, Space And Astronomy, PO Box 76, Epping, NSW 1710, Australia}

\author{C.~Phillips}
\affiliation{CSIRO, Space And Astronomy, PO Box 76, Epping, NSW 1710, Australia}

\author{M.~A.~Voronkov}
\affiliation{CSIRO, Space And Astronomy, PO Box 76, Epping, NSW 1710, Australia}

\author{T.~Westmeier}
\affiliation{International Centre for Radio Astronomy Research, The University of Western Australia, 35 Stirling Highway, Crawley, WA 6009, Australia}

\email[Ziteng Wang]{ziteng.wang@curtin.edu.au}

\def\code#1{\texttt{#1}}

\doi{10.1017/pasa.20XX.XX}

\received {dd Mmm YYYY}
\revised  {dd Mmm YYYY}
\accepted {dd Mmm YYYY}
\published{22 September 2020}

 %% First letter not capped
\begin{document}

\begin{abstract}
We present the first results from a new backend on the Australian Square Kilometre Array Pathfinder, the Commensal Realtime ASKAP Fast Transient COherent (CRACO) upgrade.
CRACO records millisecond time resolution visibility data, and searches for dispersed fast transient signals including fast radio bursts (FRB), pulsars, and ultra-long period objects (ULPO).
With the visibility data, CRACO can localise the transient events to arcsecond-level precision after the detection.
Here, we describe the CRACO system and report the result from a sky survey carried out by CRACO at 110-ms resolution during its commissioning phase.
During the survey, CRACO detected two FRBs (including one discovered solely with CRACO, \cracofrb), reported more precise localisations for four pulsars, discovered two new RRATs, and detected one known ULPO, GPM~J1839$-$10, through its sub-pulse structure.
% We present the detection of sub-pulse structure in GPM~J1839$-$10, which CRACO is sensitive to, despite the much-longer burst duration of $\sim30-300$\,s
We present a sensitivity calibration of CRACO, finding that it achieves the expected sensitivity of 11.6\,Jy\,ms to bursts of 110\,ms duration or less.
% The results of this pilot survey demonstrate the capability of CRACO to detect FRBs, pulsars, and ULPOs.
CRACO is currently running at a 13.8\,ms time resolution and aims at a 1.7\,ms time resolution before the end of 2024.
The planned CRACO has an expected sensitivity of 1.5\,Jy\,ms to bursts of 1.7\,ms duration or less, and can detect $10\times$ more FRBs than the current CRAFT incoherent sum system (i.e., 0.5$-$2 localised FRBs per day), enabling us to better constrain \corrs{the models for FRBs} and use them as cosmological probes.
% CRACO is also sensitive to the sub-pulse structures from ULPOs, a newly emerging population in radio surveys. 
% More ULPO discoveries will be beneficial in understanding their nature, distribution, and potential connections between FRBs, pulsars and ULPOs. 
\end{abstract}

\section{Introduction}
\label{sec:intro}

% What is Fast Radio Burst
Fast radio bursts (FRBs) are highly energetic astrophysical phenomena characterized by millisecond bursts at radio frequencies \citep[e.g.,][]{2007Sci...318..777L, 2013Sci...341...53T} whose nature is still unclear. 
Most FRBs are known to be extragalactic because their dispersion measures (DMs) are greater than the expected values from the Milky Way. In 2020, \citet{2020Natur.587...54C} and \citet{2020Natur.587...59B} detected a bright millisecond-duration radio burst from a Galactic magnetar (SGR~1935+2154), albeit with a lower luminosity compared to other FRBs, which suggests that magnetars at least are one source of FRBs.
The precise localisation and the host galaxy identification of FRBs started after the discovery of the first repeating FRB \citep[FRB\,121102;][]{2014ApJ...790..101S, 2016Natur.531..202S, 2016ApJ...833..177S}. \citet{2017Natur.541...58C} localised FRB~20121102A to $\sim$100\,milliarcsecond precision with the Karl G. Jansky Very Large Array \citep[VLA;][]{2011ApJ...739L...1P} from interferometric imaging.
With more localised FRBs detected, we are able to gain more insight into the nature of FRBs (e.g., constraining the emission mechanism models, understanding their progenitors, and measuring their surrounding properties), and use them to probe the cosmological parameters \citep[e.g.,][]{2022MNRAS.516.4862J} and extragalactic baryon distribution \citep{2020Natur.581..391M}.

% What is CRAFT
The Australian Square Kilometre Array Pathfinder \citep[ASKAP;][]{2021PASA...38....9H} is a 36$\times$12-m antenna radio interferometer located at {\em  Inyarrimanha Ilgari Bundara}, the CSIRO Murchison Radio-astronomy Observatory, operating at frequencies between 700 and 1800 MHz. The phased array feed receivers equipped on each ASKAP antenna enable a wide field of view, $\sim$30\,deg$^2$.
The Commensal Real-time ASKAP Fast Transients \citep[CRAFT;][]{ShannonICS, 2017ApJ...841L..12B} survey is a fast transients program running on ASKAP since 2016 with FRBs as the primary target. 
In the first two years, CRAFT operated in fly's eye mode where a sub-array of approximately 12~antennas were each pointed towards different directions, and detected 23 FRBs with an arcmin precision localisation \citep{2018Natur.562..386S, 2019MNRAS.486...70B, 2019MNRAS.486..166Q, 2019ApJ...872L..19M}.
To enable interferometric localisation, CRAFT then operated ASKAP in online incoherent sum (ICS) mode, with all antennas pointing to the same direction. The CRAFT backend squared-summed the signals from each antenna at an integration time of $\approx$~1\,ms and frequency resolution of 1\,MHz and searched for dispersed signals within.
% Introduce CELEBI
CRAFT also included a ``voltage triggering'' system, which saved the raw antenna voltages of FRB detections. \citet{2023A&C....4400724S} developed the CRAFT Effortless Localisation and Enhanced Burst Inspection (CELEBI) pipeline to correlate, calibrate, image and beamform the voltage data to improve the sky localisation to that which is possible from the longest baselines in the array ($\sim$6 km). This process allows the FRB positional uncertainty to be reduced to the level of $\sim 10.0 / (2 \times S/N)$ arcsec, where $S/N$ is the signal-to-noise ratio of the source. ASKAP accurately localised over 20 FRBs in the ICS mode using the data from the voltage triggers, enabling host galaxy identification and redshifts to be obtained \citep{2019Sci...365..565B}. This yielded, amongst other things, a first independent measure of the cosmological baryon density \citep{2020Natur.581..391M}; properties of the FRB host galaxy population \citep[e.g.,][]{2020ApJ...903..152H, 2022AJ....163...69B, 2023ApJ...954...80G}; properties of the cosmic web \citep[e.g.,][]{2019MNRAS.485..648P, 2020ApJ...901..134S}; FRB environments within nearby host galaxies \citep{2021ApJ...917...75M}; spectropolarimetry at $\mu$sec timescales of FRBs \citep{2020MNRAS.497.3335D, 2020ApJ...891L..38C}; and FRB followup of repeating FRBs at Parkes/Murriyang \citep{2021MNRAS.500.2525K}. However, the detection rate of FRBs is limited by the sensitivity of the ICS system, which is a factor of $N_{\rm ant}^{1/2}$ (where $N_{\mathrm{ant}}$ is the number of antennas ) less than achievable with coherent processing.

% Introduce CRACO
In this paper, we describe CRACO, the ``CRAFT COherent upgrade'', which brings fully coherent, real-time burst detection. Planning for this mode began in 2018, with the first light taking place in 2023. 
Results of the commissioning and operation of the first coarse-time-resolution mode of CRACO, in which dispersed radio transients are searched for offline in 110-ms image data, are described here. The target temporal resolution for the full system is 1.7 ms. 
CRACO was designed to yield a detection SNR improvement of order $\sqrt{N_{\mathrm{ant}}}=6$ (under the assumption of full phasing accuracy and all dishes in operation) over the existing incoherent system.
In practice, phasing efficiency is typically 0.90--0.95 of the theoretical maximum, and considering that not all dishes are available at any given time, we aimed at a sensitivity improvement of $\approx 5$. For FRBs, this sensitivity improvement is expected to yield a $5^{3/2} \approx 12\times$ improvement in the FRB detection rate assuming a Euclidean distribution \citep[e.g.,][]{2016MNRAS.461..984O}. Based on the detection rates of FRBs with CRAFT in the ICS mode \citep[][]{2019PASA...36....9J,ShannonICS}, CRACO thus has the potential to yield $\approx 0.5$ to 2 localisable FRBs per day.  

CRACO has several science drivers based on this sensitivity upgrade: (1) the substantially increased localised FRB discovery rate discussed above; (2) the discovery of higher-$z$ FRBs than in CRAFT/ICS; (3) an increased discovery rate for slow radio transients including rotating radio transients (RRATs) and intermittent pulsars; (4) the study of the interplanetary scintillation of Active Galactic Nuclei (AGN); and (5) space weather.

The first transient survey, CRACO 110-ms Pilot Survey (CRACO-PS), was performed at 110-ms time resolution and operated from 2023 April to 2023 November. This initially rather coarse time resolution allowed us to operate a search program in an under-explored region of transient space, i.e.\ relatively long-duration, dispersed radio transients, or "not so fast radio bursts". Simultaneously, we used this survey to commission the system and work towards higher time resolution.

In this paper, we present a brief system description including the bandpass calibration, RFI rejection strategies, search pipeline, candidates post-processing pipeline, system equivalent flux density estimations, cross-checks of CRACO's performance with the Parkes/Murriyang radio telescope, and known issues in the current system in Sections~\ref{sec:system} and \ref{sec:system_perform}.
In Section~\ref{sec:survey}, we describe the design of CRACO-PS and report the results of the survey in Section~\ref{sec:result}, which includes the detection of two FRBs (one discovered by CRACO, but missed by the ICS system), the coordinate corrections of four pulsars, and the detection of two new RRATs.
In Section~\ref{sec:dis}, we discuss the inferred FRB rates based on the CRACO detections and their implications. We also discuss the prospects for detecting new FRBs, pulsars, and \corrs{ultra-long period objects (ULPOs)} with CRACO, and the further planned improvement of the system. Conclusions of this work are presented in Section~\ref{sec:conclusion}.
% In this paper, we report the operational results and discoveries of this 110-ms mode of CRACO. We describe RFI rejection strategies, array phasing, bandpass calibration, the system SEFD, its single pulse sensitivity, cross-checks of CRACO's performance with the Parkes/Murriyang radio telescope, re-detections of known pulsars, the DM dedispersion via FPGAs, and SNR losses of the de-dispersion algorithm. Initial results of the 110-ms survey are reported, which includes 2 Fast Radio Burstrs, XX Ultra-long Period variables (ULPs), XX transients etc.  

\begin{figure*}[htb!]
    \centering
    \includegraphics[width=0.7\textwidth]{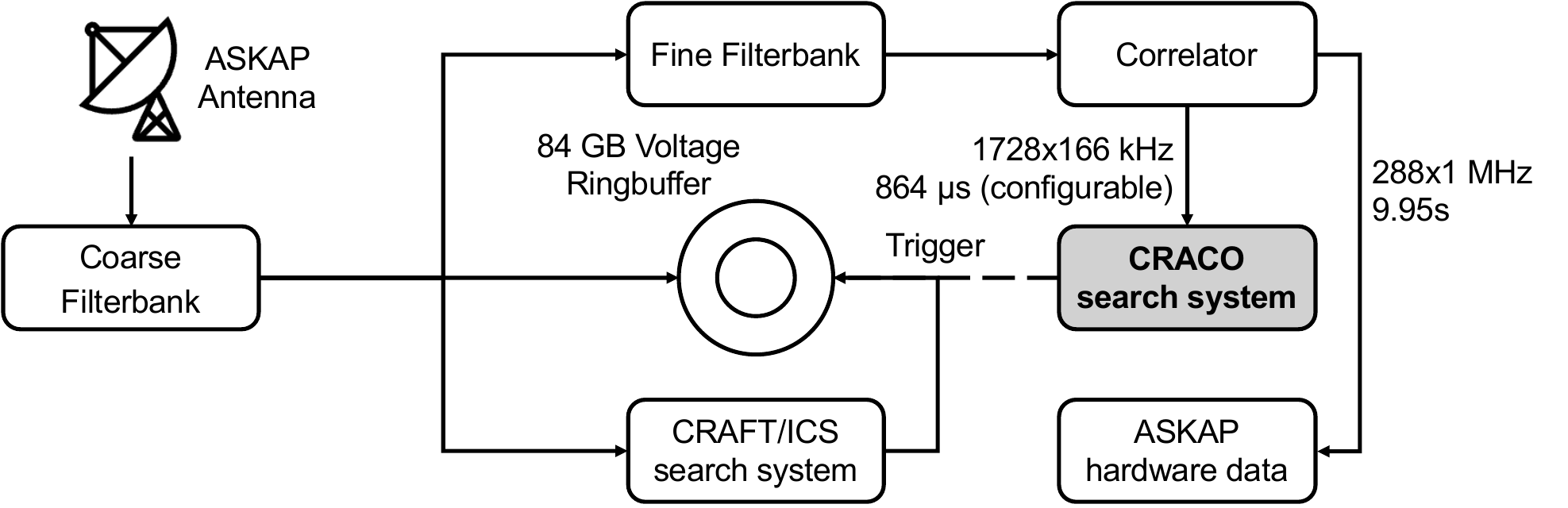}
    \caption{Data stream flowchart for the ASKAP hardware, CRAFT/ICS, and CRACO systems. The dashed line indicates the triggering process from CRACO to the ICS system remains to be implemented.}
    \label{fig:craco_askap_data_flow}
\end{figure*}

\begin{figure*}[htb!]
    \centering
    \includegraphics[width=0.85\textwidth]{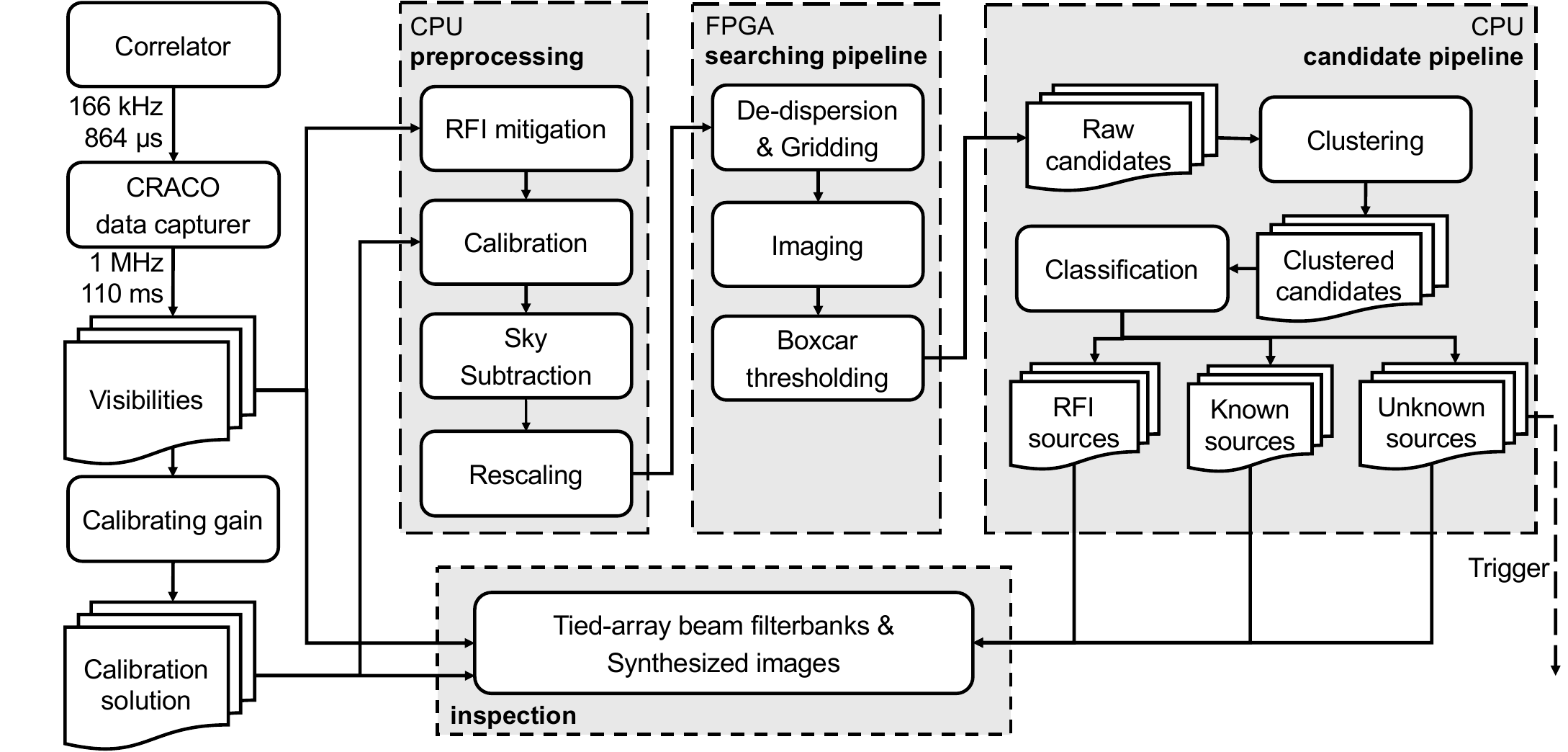}
    \caption{CRACO search system flowchart. The dashed line means the trigger process was not yet implemented when this survey was undertaken.}
    \label{fig:craco_search_flow}
\end{figure*}

\section{CRACO System Design and Operation}
\label{sec:system}

% In this section, we describe briefly the CRACO hardware and software. We then describe how data are recorded, how data are calibrated for a given observation, how calibrated data are processed to get raw candidates, and how raw candidates are filtered to search for real radio transients.
In this section, we briefly describe the design principles, hardware and software system sufficient for the results of the 110-ms science commissioning survey. 
A full description of CRACO will be presented in Bannister et al (in prep.). The ASKAP system is described in detail in \citet{2021PASA...38....9H}.
We show two schematic diagrams in Figures~\ref{fig:craco_askap_data_flow} and \ref{fig:craco_search_flow} describing CRACO operations.
We list the pipeline parameters in Table~\ref{tab:searchpipe_param} and the survey parameters in Table~\ref{tab:survey_param}.

\begin{table*}[htb!]
    \centering
    \begin{tabular}{lll}
        \hline\hline
        Parameter & Description & Value\\
        \hline
        \texttt{NT} & number of samples per block & 256 (hardcoded) \\
        \hline
        Data Preprocessing & & \\
        \hline
        \texttt{FLAG\_RADIUS} & radius parameter used in IQRM & \\
        \texttt{FLAG\_THRESHOLD} & threshold used in IQRM & 5.0 \\
        \texttt{TARGET\_RMS} & the target standard deviation used in the rescaling & 512 \\
        \hline
        Searching & & \\
        \hline
         & shape of $uv$ cells and image pixels & $256\times256$ \\
         \texttt{FOV} & maximum field of view in square degrees & 1.1 \\
         \texttt{OS} & minimum oversampling factor across the synthesised beam & 2.1 \\
         \texttt{NPlanUpdate} & number of blocks when \texttt{CRACOPlan} updates & 12\\
         \texttt{NCIN} & number of channels on FDMT input & 32 \\
         \texttt{NDOUT} & number of DMs the output of FDMT & 186 \\
         \texttt{MAX\_DM} & largest DM to process (in samples) & 1024 (hardcoded) \\
         \texttt{NDM} & number of DM trials to process (in samples) & see Table~\ref{tab:survey_param}\\
         \texttt{NBOX} & number of boxcar width trials (in samples) & 8 \\
         \texttt{THRESHOLD} & SNR threshold for producing candidates & 6 \\
        \hline\hline
    \end{tabular}
    \caption{Parameters used in the search pipeline}
    \label{tab:searchpipe_param}
\end{table*}

\begin{table*}[hbt!]
    \centering
    \caption{CRACO 110-ms Pilot Survey search parameters. $\Omega\Delta$t indicates the search volume in units of deg$^2$\,hr. SNR is the signal-to-noise ratio threshold we used to filter for candidates. \texttt{NDM} is the maximum DM (in units of time samples) used in the search.}
    \label{tab:survey_param}
    \begin{threeparttable}
        \begin{tabular}{ccccccc}
        \hline\hline
        Phase & SBID & Date & bandwidth & $\Omega\Delta$t$^\ddagger$ & SNR & \texttt{NDM} \\
         & & (UTC) & (MHz) & (deg$^2$\,hr) & & \\
        \hline
        1 & 49721--53010 & from 2023-04-27 to 2023-09-12 & 120 & 4688 & 10 & 140 \\
        2 & 53011--54771 & from 2023-09-12 to 2023-11-04 & 240 & 2452 & 8 & 280 \\
        \hline
        \end{tabular}
        \begin{tablenotes}
        \item[$^\ddagger$] We use a field of view of 1.1\,deg in diameter to estimate the search volume. Due to the $uv$ coverage, a small number of schedule blocks had a somewhat smaller (or larger) FoV.
        \end{tablenotes}
    \end{threeparttable}
\end{table*}

CRACO implements millisecond timescale imaging of the field of view (FoV). It adopts a similar approach to \textit{RealFAST} \citep{2018ApJS..236....8L}, the transient detection hardware/software implemented on the Karl G. Jansky Very Large Array \citep[VLA;][]{2011ApJ...739L...1P}. 
Visibility data are received and recorded on disk. In an offline process, the visibilities are read, averaged, calibrated and sky subtracted, then processed with the search pipeline. CRACO forms $256 \times 256$ pixel images with approximately $\sim$15.5\,arcsec resolution (depending on observing frequency and the $uv$-coverage). We typically use all baselines out to antenna 24, which achieves an acceptable compromise between FoV and sensitivity. The pipeline dedisperses the visibilities at $\sim$200 Dispersion Measure (DM) trials (depending on the observing frequency and the recording bandwidth), computes 8 sliding window boxcars, and thresholds the result. The pipeline is implemented on 20 Xilinx Alveo U280 FPGA processing cards. 
\corrs{The number of DM trials represents the maximum dispersive time delay, measured in units of samples, that can be searched. Further detailed discussion is in Section~\ref{subsubsec:boxthre}.}

CRACO aims at a final target 1.728 ms time resolution of the full field of view, yielding 23 Terapixels per second to be searched for transient events. Achieving this is a 3-stage process, in which we aim for time resolutions of 110.592 ms (this paper), 13.824 ms (currently operational) and a final 1.728 ms (slated for late-2024).

\subsection{CRACO hardware and Data Recording}

ASKAP operates in three frequency bands (low, mid, and high) between 700\,MHz and 1800\,MHz. Within each band, observers can choose up to 288\,MHz of bandwidth for the observations\footnote{CRAFT fly's eye and ICS observations, and raw antenna voltages, which do not use the correlator, can access a slightly higher bandwidth of 336 MHz.}. 
We show a schematic diagram describing the data flow for three different systems (ASKAP hardware, CRAFT/ICS, and CRACO) in Figure~\ref{fig:craco_askap_data_flow}.
The signals from each PAF element on each antenna are digitised and channelised to produce 1\,MHz coarse filterbanks.
In typical observations, digital beamformers apply weights derived through observations of the Sun and on-dish calibrator to form 36 dual polarisation beams, where the beam arrangements are called a ``footprint'' (see \citet{2021PASA...38....9H} for typical ASKAP footprints).
The spacing between adjacent beams (also called ``pitch'') varies between 0.75 and 1.05\,deg, which largely depends on the observing frequency and science requirements.
% The CRAFT/ICS search system integrates the data product from the beamformer and generates total power filterbanks with $336\times1$\,MHz channels at 864\,$\mu$sec resolution for each beam. 
An 84 GB ring-buffer is maintained in the beamformer stage, capable of holding 0.9--14.2 seconds, which can be traded for the quantisation number of bits. Triggers from the transient search pipeline yield voltage dumps from this ring-buffer, for later use localising the events, and for high time and frequency resolution analysis. 

Beamformed signals are then split into finer channels with a resolution of 1/54\,MHz and sent to the correlator which computes visibilities for each baseline.
The correlator computes two data products. The low time resolution product used for normal spectral line and continuum processing is 15,552$\times$ 1/54\,kHz channels with a time resolution of 90 $\times$ 110.592\,ms (approximately 9.95\,s) with all four Stokes parameters. The correlator also produces a high time resolution product. This product has a configurable integration time of 0.864, 1.728ms or 3.456\,ms. The correlator applies a fixed factor of 6 frequency averaging to achieve 1/9~MHz channels.  It has two stokes modes, either forming ``pseudo stokes I'' (XX+YY, calibration is not applied), or ``XX and YY''. The ``XX and YY'' mode cannot be used in the 0.864ms mode due to data rate constraints. CRACO-PS captured data in the 3.4\,ms pseudo stokes I mode. Data are captured from the network as RoCE packets and saved to disk. All subsequent processing is performed independently per beam.
% and formed 1~MHz, 110~ms, stokes I products in CPU in the following pipeline processing. %before writing to disk.
We only recorded 120\,MHz bandwidth in the first three months and increased it to 240\,MHz \footnote{We did not go for the full bandwidth, 288\,MHz, during CRACO-PS.} afterwards (see Table~\ref{tab:survey_param}).
We recorded both the real and imaginary parts of visibilities as 32-bit floating point numbers. With one polarisation, 240 channels (240\,MHz bandwidth), a time resolution of 110.592\,ms, and 435 baselines (30 antennas), the data rate is 27.2\,GB\,hour$^{-1}$ for each beam. 

We did not calculate baseline vectors (i.e., $uvw$ values) in real time.
We used ASKAP metadata files (with $uvw$ values for all baselines every 9.95\,s) to interpolate the $uvw$ values in the CRACO data during the following offline search.
The ASKAP system performs fringe rotation of the data to the phase/delay centres corresponding to the nominal pointing position of each beam and this is the same position $uvw$ values are calculated for.

\subsection{Bandpass Calibration}\label{subsec:calib}

% note a how to choose the calibration, and how we did that in real-time
% why we don't choose to use ASKAP normal calibration

For almost all ASKAP observations, the bright Southern hemisphere source PKS~B1934--638 is used as the bandpass calibrator. Observations of this source incur significant operational overhead, taking approximately 2 hours to complete. As CRACO requires solutions in real time, these bandpass observations cannot be used, as they often take place after the target observations.
% directly after configuration changes (e.g., changes of band, footprint, beam weights, etc). 
CRACO uses self-calibration to derive the desired calibration solutions. Almost any ASKAP field is suitable for self-calibration, as there is usually $>1$~Jy of flux in any beam, and an excellent sky model is available in the form of the RACS catalogue \citep{2020PASA...37...48M, 2021PASA...38...58H}.

% This might be helpful in terms of terms - https://research.csiro.au/ratechnologies/wp-content/uploads/sites/295/2022/11/PAFAR2022-Chippendale-PAF_Calibration_Systems.pdf

\begin{figure}[htb!]
    \centering
    \includegraphics[width=0.99\textwidth]{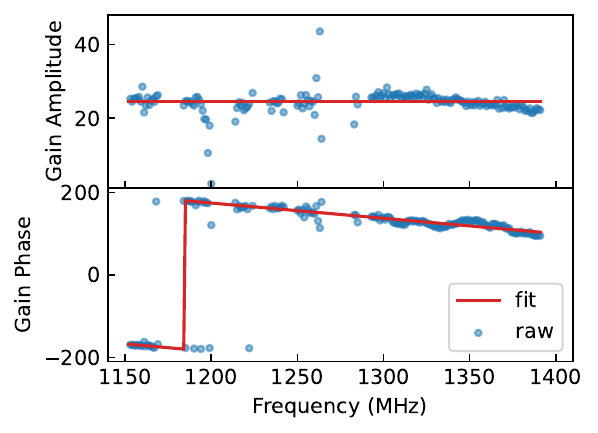}
    \caption{Raw and model-fitted calibration solution of gain and phase for a typical 10-minute observation. Blue circles show the calibration solution per frequency channel derived from the sky model directly, while the red line shows the fitted linear calibration solution. 
    % Outlier rejection for the phase fit is done using XXX. 
    }
    \label{fig:craco_calib_fit}
\end{figure}

The ASKAP bandpass is relatively stable over several days, but small changes in the antenna phases can lower the candidate detection SNR. CRACO automatically generates calibration solutions daily to account for these small changes in antenna phases.
We only use data recorded with an elevation angle $\theta > 20^\circ$ and a declination $\delta < +40^\circ$ to perform the calibration. If there is more than one observation available for calibration, we used the following order to decide which one to use: 1) normal ASKAP bandpass (i.e., PKS~B1934--638); 2) other extragalactic field; 3) galactic field. New calibration solutions are also generated after a change in beamformer weights or observing frequency. 
We use $\sim$10\,minutes of CRACO data from the selected observation to perform the field calibration to obtain the bandpass solution\footnote{Although the calibration procedure we use is more similar to delay calibration, we still refer to it as bandpass calibration since we fit for both amplitude and phase when generating the calibration solution.}.
For a given field, we construct a sky model by extracting all bright (peak flux density $>$5\,mJy/beam) sources within a 2\,degree radius of the field centre from the RACS-Low DR1 catalogue \citep{2020PASA...37...48M,2021PASA...38...58H}\footnote{For observations conducted at a different frequency, we used a spectral index $\alpha=-0.83$ to scale the flux density to the observed frequency.}.
We then calibrate the CRACO visibilities data against this sky model to get the bandpass solutions of gain and phase for each frequency channel using the least squares fit.
The existence of radio frequency interference (RFI) in calibration observations can lead to wrong or missing solutions for the affected channels, which will in turn lower the search sensitivity. 
To mitigate these effects from the RFI, we fit these frequency-dependent solutions with a linear model. This procedure is performed for each antenna, using the median amplitude value of all channels as the model amplitude and by fitting for a linear trend of the unwrapped phase values (see Figure~\ref{fig:craco_calib_fit} for an example). 
We used the fitted model as the final solution and applied it in the following processing procedures.

\begin{figure*}[htb!]
    \centering
    \includegraphics[width=1.0\textwidth]{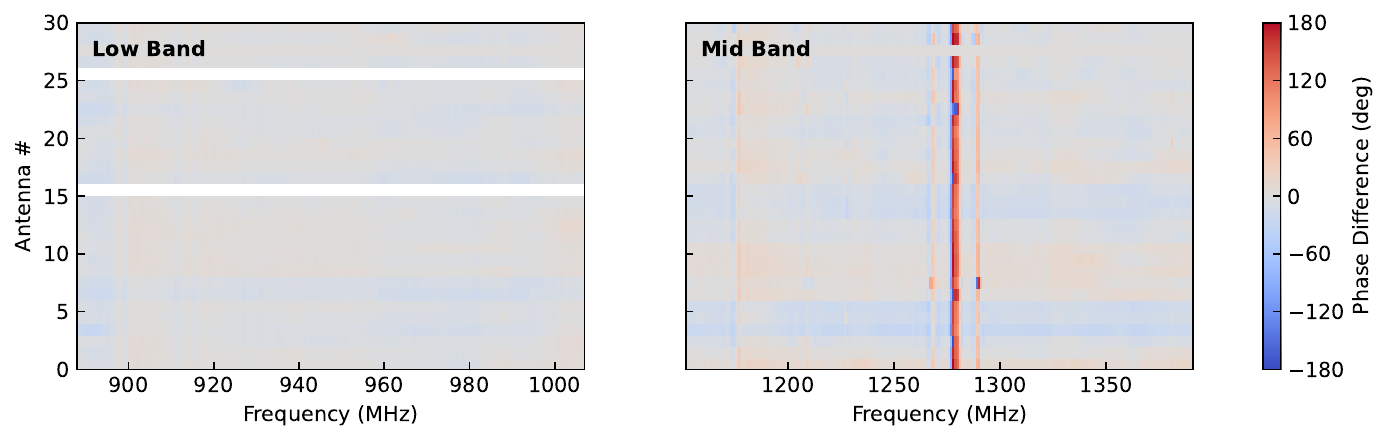}
    \caption{Phase differences between the calibration solutions derived from CRACO field observations and the normal ASKAP PKS~B1934--638 observations.
    The left panel shows the differences between the phase solutions derived from SB51896 (CRACO) and SB51887 (ASKAP Hardware) in low band (centred at 947.5\,MHz), and the right panel shows the differences between the phase solutions derived from SB53094 (CRACO) and SB53091 (ASKAP Hardware) in mid band (centred at 1271.5\,MHz). We show the comparison for Beam 0 (note that the location of this beam within the PAF varies with the configuration footprint). The large differences in phase (at a small set of frequencies) in the mid band data are mainly caused by radio frequency interference (e.g., satellite navigation systems in $\sim$1150--1300\,MHz). These sources of RFI are mitigated against by the linear fitting method illustrated in Fig.~\ref{fig:craco_calib_fit}.
    } \label{fig:craco_hw_calib_compare}
\end{figure*}

To verify the field calibration method, we compared our solutions derived from CRACO observations and ASKAP hardware observations of the bright source PKS~B1934--638. 
In Figure~\ref{fig:craco_hw_calib_compare}, we show the phase differences between the two methods\footnote{We do not consider solution amplitude here because we normalise the amplitude at a later stage in the processing.}.
Except for the frequency ranges affected by the RFI, the phase solutions largely agree with each other (the standard deviations of the differences are $\sim5\,$deg and $\sim10\,$deg for the low band and mid band, respectively).

% verify the field calibration

% \tocheck{Plot comparison of bandpass between different ways}

\subsection{Data Preprocessing}

There are three steps applied to the raw data in the CRACO system: RFI mitigation, calibration, and normalisation. These steps are performed on central processing units (CPUs).
All data were processed block by block, with each block containing 256 samples of the data.

\subsubsection{RFI Mitigation}

There are two steps to remove RFI in the data, namely static flagging and dynamic flagging.

In static flagging, we flag a list of known bad frequency ranges, which are constantly bad and are not well-handled by dynamic flagging.
We have continually updated the list as operations continue, and currently flag the following frequency ranges: 920--960\,MHz, and 1150--1320\,MHz. 
% In the mid-band data,  only half the original band is available for the search.

Given the high volume of data, we cannot use standard flagging tools. Therefore, for the dynamic flagging, we first incoherently sum the visibility amplitudes across all baselines, forming ``cross-amplitude sum" (CAS) filterbanks. 
We use Inter-Quartile Range Mitigation\footnote{\url{https://github.com/v-morello/iqrm}} \citep[IQRM;][]{2022MNRAS.510.1393M} to mask bad channels automatically. 
We set the IQRM radius parameter $r=64$ and $r=180$ for 120\,MHz and 240\,MHz bandwidth observation respectively, and set the threshold to $t=5.0$.
The parameters were tuned empirically. This method effectively removes RFI and leaves the rest of the band clean.

% These frequency ranges are usually occupied for the mobile and satellite communication.

\subsubsection{Calibration}

For a given observation, we selected the gain solution derived from the observation that shared the same beamforming weights and was observed no more than one day apart.
If there was no gain solution available, we would produce a new gain solution based on the method discussed in Section~\ref{subsec:calib}.

\subsubsection{Sky Subtraction and Rescaling}\label{subsubsec:normalize}

Most radio sources in the sky have a relatively constant flux over the timescales we are probing with CRACO. To search for transients, we need to subtract the non-variable sky from the data and rescale it so that it has a zero mean and a target standard deviation (STD). 
For each block, we calculated the sky model as the complex mean values for each baseline and channel across a block, and the rescaling factor as the STD of the raw data absolute value divided by the target STD.
We subtracted the sky model from the raw data and divided it by the rescaling factor.
Since we calculated the sky model as the mean value across a block, the system was only sensitive to pulse-like emission narrower than the block size (conservatively, on a timescale $\lesssim$20\,sec for the block size of 28\,sec using in 110\,ms observations), and all wider transients would be subtracted out.

\subsection{Searching}

The searching process comprises three main parts: de-dispersion and gridding, imaging, and boxcar filtering and thresholding. 
To achieve a better performance, the search was conducted on field-programmable gate arrays (FPGAs).

\subsubsection{De-dispersion and Gridding}\label{subsubsec:dedisp_and_grid}

% Similar to general interferometric imaging, we need to grid the data on the $uv$-plane, but we also need to perform de-dispersion to search for the dispersed signal.
Before making images, we de-disperse and grid the calibrated visibilities into four-dimension data ($u$, $v$, DM, and time).
In the process, we calculate for each baseline and each channel which $uv$ cell the Stokes I correlation amplitudes should be assigned to (we term this assignment \texttt{CRACOPlan}).
The \texttt{CRACOPlan} is calculated every \texttt{NPlanUpdate} (12 by default, see Table~\ref{tab:searchpipe_param}) blocks, which is a result of the balance between positional accuracy and search efficiency.

The $uv$ grid was fixed to $256\times256$ $uv$ cells, and therefore the size of the corresponding synthesized image was $256\times256$ pixels as well.
The image pixel size and the $uv$ cell size depend on two parameters in the search pipeline: the maximum field of view (\texttt{FOV}; 1.1\,deg per beam, by default), and the minimum oversampling factor (\texttt{OS}; 2.1 by default). 
We first calculated the size of the synthesised beam using the maximum length of the $u$, $v$ values in a given block, 
$$
b_l = \cfrac{c}{u_{\rm max}f}; 
b_m = \cfrac{c}{v_{\rm max}f},
$$
where $c$ is the speed of light and $f$ is the centre frequency. 
If the field size when we oversample the synthesised beam at a rate of \texttt{OS} is larger than \texttt{FOV}, we will increase the oversampling factor to match the FOV. Otherwise, we will lower the FOV so that the minimum OS is fulfilled.
Therefore the image pixel sizes are
$$
p_l = \min \left(\texttt{FOV}/256, b_l/\texttt{OS}\right), 
p_m = \min \left(\texttt{FOV}/256, b_m/\texttt{OS}\right)
,$$
and the corresponding $uv$ cell sizes are
$$
c_u = 1 / \min\left(\texttt{FOV}, 256 b_l/\texttt{OS}\right), c_u = 1 / \min\left(\texttt{FOV}, 256 b_m/\texttt{OS}\right)
.$$

To de-disperse the data more efficiently, we divide the whole bandwidth into several chunks of \texttt{NCIN} channels and pad zero values for the chunk of less than \texttt{NCIN} channels.
We first use the Fast Dispersion Measure Transform \citep[FDMT; ][]{2017ApJ...835...11Z} algorithm to de-disperse each \texttt{NCIN} channels chunk prior to gridding. We then assign the partially de-dispersed data onto $uv$ cells based on the \texttt{CRACOPlan} calculated previously, in which we use brute force to de-disperse the results for all chunks, and thus de-dispersed the data across the whole observing band.

\subsubsection{Imaging}

As we have already subtracted the non-variable sky in Section~\ref{subsubsec:normalize}, there will be only a few sources in the data, and dirty images produced with a simple fast Fourier transform are sufficient for searching.

\begin{table*}[hbt!]
    \centering
    \caption{Search pipeline output parameters and descriptions}
    \begin{tabular}{l l}
    \hline\hline
    Column Name & Description \\
    \hline
    \texttt{rawsn} & signal-to-noise ratio of the candidate in hardware units\\
    \texttt{lpix} & candidate coordinate along the $l$-axis in pixel units \\
    \texttt{mpix} & candidate coordinate along the $m$-axis in pixel units \\
    \texttt{dm} & dispersion measure of the candidate in units of samples \\
    \texttt{boxc\_width} & boxcar width of the candidate in units of samples\\
    \texttt{time} & the time elapsed from the start of a block to the candidate detection in units of samples \\ 
    \texttt{iblk} & the index of the block in which the candidate is detected \\
    \texttt{total\_sample} & the time elapsed from the start of the observation to the candidate detection in the unit of sample (equal to 256$\times$\texttt{iblk} + \texttt{time}) \\
    \hline
    \texttt{snr} & signal-to-noise ratio of the candidate in physical units \\
    \texttt{ra\_deg} & right ascension of the candidate in degrees\\
    \texttt{dec\_deg} & declination of the candidate in degrees\\
    \texttt{dm\_pccm3} & dispersion measure of the candidate in units of pc\,cm$^{-3}$ \\
    \texttt{obstime\_sec} & the time elapsed from the start of the observation to the candidate detection in units of seconds \\
    \texttt{mjd} & the modified Julian date of the candidate detection \\
    \hline\hline
    \end{tabular}
    \label{tab:search_coldescr}
\end{table*}

\subsubsection{Boxcar filtering and thresholding}\label{subsubsec:boxthre}

The pipeline can search for transient events of one to eight samples in length, using 8 ``boxcar'' filters. 
The output from the previous imaging step has an equivalent shape of the number of DM trials $\times$ the number of samples in time $\times$ the number of boxcar widths $\times$ the number of $l$-pixels $\times$ the number of $m$-pixels (\texttt{ndm}$\times$256$\times$8$\times$256$\times$256).
The pipeline divided the whole block of data into small chunks with a shape of 1$\times$8$\times$8$\times$1$\times$16, and calculated the maximum signal-to-noise ratio value for each small chunk.
If the maximum signal-to-noise ratio value exceeded the signal-to-noise ratio threshold, the pipeline would write the event to the candidate list.
% At 110ms, this results in \todo{XXX} search trials per second.

We list the candidate properties computed by the pipeline in Table~\ref{tab:search_coldescr}.
The raw DM value (\texttt{dm} in Table~\ref{tab:search_coldescr}) is reported in units of time samples in the pipeline, i.e. it is the dispersive time delay in samples across the observing band. We convert the raw DM to physical units (\texttt{dm\_pccm3} in Table~\ref{tab:search_coldescr}) using:
\begin{equation}\label{eq:dm_convert}
    {\rm DM} = \cfrac{{\rm DM}_{\rm raw}\left(\Delta t / {\rm ms}\right)}{4.15\left[\left(\nu_1/{\rm GHz}\right)^{-2} - \left(\nu_2/{\rm GHz}\right)^{-2}\right]}\,{\rm pc\,cm}^{-3}
\end{equation}
% {\rm DM} = \cfrac{{\rm DM}_{\rm raw}\left(\cfrac{\Delta t}{\rm ms}\right)}{4.15\left[\left(\cfrac{\nu_1}{\rm GHz}\right)^{-2} - \left(\cfrac{\nu_2}{\rm GHz}\right)^{-2}\right]}
where $\Delta t$ is the time resolution (110\,ms), and $\nu_1$ and $\nu_2$ are the edge frequencies of the observing bandwidth.
The error on the raw DM reported in this work was given as $+$/$-$ 1 time sample across the band and was converted to physical units according to Equation~\ref{eq:dm_convert}.
% If a candidate was detected with a \texttt{dm} of $dm$, we used the following value to report the measured DM with uncertainty from the pipeline: $dm_{-1}^{+1}$ (and used the equation above to get the one with physical unit). Similarly, if a candidate was detected with a \texttt{dm} of 0, we quoted \texttt{dm\_pccm3} value corresponding to \texttt{dm} = 1 as an upper limit.

\subsection{Candidate Post-processing}

We reduce the number of candidates to manageable levels using a post-processing pipeline, which consists primarily of two stages: candidate ``Clustering'' and candidate ``Classification''. ``Clustering'' refers to the identification of burst candidates which arise at similar time, boxcar width and dispersion measure, as these are likely to originate from a single celestial event. ``Classification'' refers to the identification of known sky sources (which we can filter out of the final candidate lists), as well as events that are most likely RFI. As such, the aim of the post-processing stage is to obtain burst detections almost exclusively from new transient sources.

\subsubsection{Candidate clustering}
For a given burst, particularly those with high SNR, the detection pipeline can generate several (or several tens of) candidates with a range of parameters (i.e. DMs, boxcar widths, and positions, etc.). We employed the DBSCAN algorithm \citep{1996kddm.conf..226E} to group these these candidates into clusters. We used three data series \texttt{total\_sample}, \texttt{dm}, and \texttt{boxcar\_widths} (all in units of samples), normalising them by dividing a factor of 3, 5, and 4, respectively to perform the clustering step. We set \texttt{eps = 1},  the maximum distance separating events in the same cluster, for \texttt{DBSCAN}. For each cluster, we then calculated the standard deviation for \texttt{lpix} and \texttt{mpix}, (hereinafter referred to as \texttt{lpix\_rms} and \texttt{mpix\_rms}), to be used in further steps. These are termed ``temporal clusters'' in what follows.

\subsubsection{RFI amelioration}
Although we implemented RFI flagging strategies in the search pipeline, and the ASKAP/CRACO RFI environment is rather clean, some unflagged RFI can still leak through, leading to spurious candidates that exceed our detection threshold.
We employed simple heuristics to filter out these RFI candidates.
We evaluated and refined our RFI amelioration heuristics by examining a large number of candidate clusters, including those due to RFI, scintillating continuum sources, and single pulses from pulsars.
RFI is often impulsive, and accordingly generates candidates that are clustered within a relatively narrow time window, but spread widely across the FoV because the RFI is in the near field and exhibits a curved wavefront across the array. For clusters where either \texttt{lpix\_rms} or \texttt{mpix\_rms} $>2$, we conducted additional clustering (hereinafter referred to as ``spatial clustering''), within the cluster itself. 
This clustering used the \texttt{lpix} and \texttt{mpix} series, setting \texttt{eps = 3} as the hyperparameter for \texttt{DBSCAN}. 
For a given temporal cluster, if the number of spatial clusters is $\le2$, we consider each spatial cluster as a unique candidate. However, for clusters with an excessive number of spatial clusters ($>2$), we group them into a single cluster. We classify a candidate as being RFI if the number of candidates within the temporal cluster is $\le 2$ (astronomical sources should be detected as multiple candidates), or the number of the candidates within the spatial cluster is $>2$ (RFI candidates spread widely across the FoV). For candidate clusters not flagged as RFI, we select the candidate with the highest SNR for further analysis.

\subsubsection{Classifying known radio transient sources}
Two known classes of objects, pulsars and scintillating continuum sources (hereafter ``scintillators''), can exhibit rapid millisecond timescale flux variations, which result in candidate detections from the search pipeline. To filter out these known sources, we cross-matched the candidates against two catalogues: the Australia Telescope National Facility Pulsar Catalogue \citep[PSRCAT,][]{2005AJ....129.1993M} for pulsars, and the RACS-low1 catalogue \citep{2021PASA...38...58H} for scintillators.
In both cases, we did not use the entire catalogue. For PSRCAT, we excluded pulsars with positional uncertainties greater than 15\,arcsec. This criterion helps us to 1) avoid excluding real, unknown sources due to coincidental matches, and 2) obtain better localisations for poorly localised pulsars.
For the RACS-low1 catalogue, we selected sources with a flux density $S > 0.3$ Jy, as sources need to be quite bright for scintillation to yield candidates after the pipeline. Taking into account the typical beam size of the 23-antenna data ($\sim$1\,arcmin), we set a threshold of 60\,arcsec for cross-matching to these sources in CRACO-PS, which excludes $\sim$20\,deg$^2$, $\sim$0.1\% of the ASKAP visible sky, from the search.

% {\bf To do: quantification of the previous paragraph required here

% %To do: what about the results of mock FRB injection? did we do that at 110 ms resolution? ANDY mentions this later (VG did do some tests, these will be mentioned)

% To do : estimates of the pipeline efficiency described here}

\subsubsection{Manual Inspection}

After the clustering, filtering, and cross-matching processes, we successfully eliminated most of the unwanted candidates, from a raw rate of $\sim5\times10^5$ candidates per day to $\sim30$ candidates per day after application of our heuristics. 
The candidates that are not pulses from unknown Galactic or extragalactic sources are typically caused by weak RFI, known sources aliased from well outside the field of view, or sidelobes from bright sources. We developed a web graphical user interface\footnote{\url{https://github.com/askap-craco/dash-craco-cand}} to manually check filterbank plots and synthesised images for all candidates in the final list to eliminate false positives. 

\subsection{Transient Localisation}\label{subsec:transient_loc}

Precise localisation is essential for studying transients, especially for their multi-wavelength follow-up observations. 
There are several approximations we make in the gridding and imaging steps to speed up the search. As a result, the pipeline reported positions become increasingly inaccurate (with errors of up to tens of arcsec) far from the centre of the field (see Section~\ref{subsubsec:astrometric_offset} for a detailed discussion). 
Additionally, not all antennas are used in the detection pipeline, meaning the opportunity exists to improve the S/N and hence localisation precision.
For both of those reasons, we developed post-processing scripts to enable arcsecond localisation for the transients we detected.
We first extracted $\sim$90~seconds of visibility data observed around the transient detection (hereafter termed ``field data'') with all available baselines, applied the calibration solution used for searching, and de-dispersed the visibility data with the detection DM. 
We also extracted the visibility data for the transient detection (hereafter termed ``event data'') out of the field data.
We imaged both the field and event data using the \texttt{tclean} procedure in \textsc{casa} \citep{2022PASP..134k4501C} with $w$-projection enabled, and used \textsc{PyBDSF} \citep{2015ascl.soft02007M} for source finding on both images.
Based on the raw transient position in the event data, we could produce a tied-array beam filterbank of the event with the highest possible SNR.
We cross-matched the detected sources in the field image against the RACS-low1 catalogue\footnote{We only selected isolated (the nearest source is more than 30\,arcsec away), unresolved (with an integrated to peak flux ratio less than 1.5) and bright (with a SNR larger than 10) sources.} to correct for systematic astrometric offsets. 
We also cross-matched the RACS-low1 catalogue against the ALLWISE catalogue \citep{2014yCat.2328....0C} to correct for the systematic offset in the RACS catalogue. A detailed discussion of this method, the final corrections for the whole RACS catalogue, and the estimation of the uncertainty in the residual RACS astrometric errors can be found in Jaini et al., (in prep.).
We then applied all these corrections to the event position. We typically achieve an uncertainty of several arcsec in the final localisation (compared to $\sim$30 arcsec uncertainty in the search pipeline). Unless otherwise specified, all coordinates reported in this work are those with corrections applied.

\subsection{Known Features or Issues}\label{subsec:known_issue}

% In CRACO-PS, we noticed several features or issues in our system. 

\subsubsection{Aliased false-positive candidates}\label{subsubsec:alias}

\begin{figure}[htb!]
    \centering
    \includegraphics[width=0.9\textwidth]{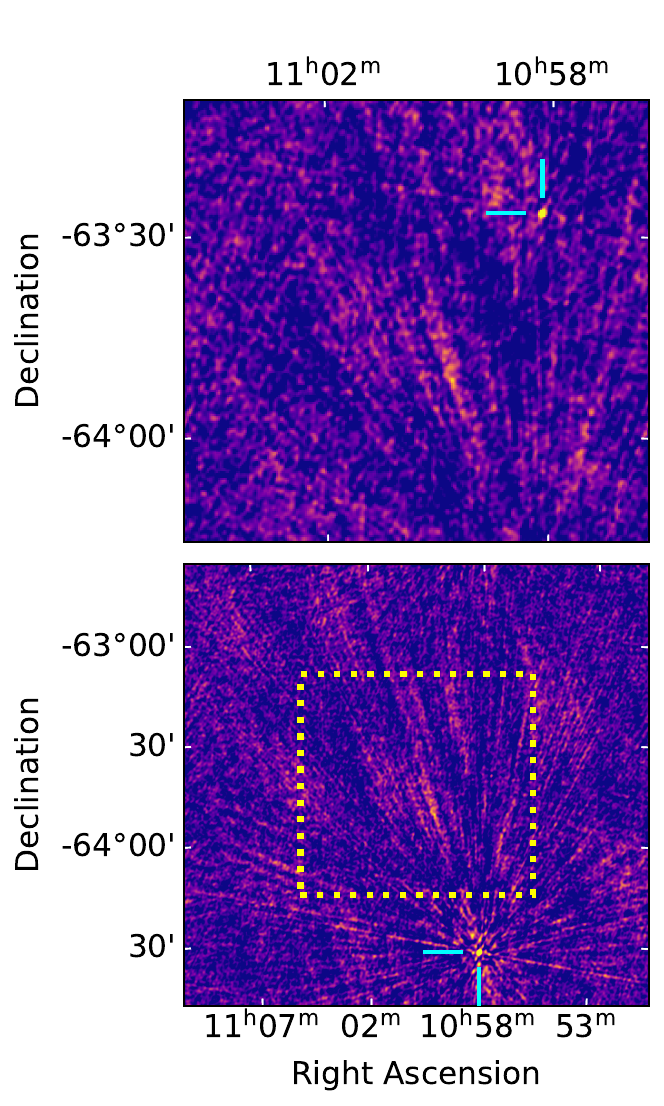}
    \caption{An example of a candidate detection due to aliasing. The top panel shows the detection image of the candidate produced by the pipeline, with a FoV of 1.1\,deg, and image dimensions of (256, 256) pixels. The bottom panel also shows the detection image, but with the image size doubled to (512, 512) pixels to yield a 2.2\,deg FoV. The detections are marked with cyan bars. The yellow dashed rectangle in the lower panel shows the region covered in the top panel. This illustrates how a candidate can arise from a source whose true sky position lies outside the imaged region in the upper panel, but appears in it due to aliasing. } 
    \label{fig:alias_example}
\end{figure}

In many observed fields, we see false positive candidates that are due to a phenomenon called ``aliasing''. These are an artefact of our imaging process and arise when the image is smaller than the FoV of the primary beam. Sources outside the imaging region may appear in the image (see Figure~\ref{fig:alias_example}). Aliasing phenomenon will be more severe especially when using pillbox gridding \citep[e.g.,][]{2017isra.book.....T}.
In CRACO-PS, we used a FoV of 1.1\,deg to make synthesized images, which is smaller than the primary beam (for an observation centred at 943\,MHz, the primary beam is $\sim$1.8\,deg in diameter).
Importantly, real transients outside the imaging region may be detected as alias candidates albeit with a lower SNR, which can increase our effective search FoV. We note that due to the visibility data being stored on disk, we are still able to recover the correct positions after the post-processing stage.
We manually checked alias sources in CRACO-PS, but have implemented an alias source filter to remove aliases from known sources automatically in the current CRACO.

\subsubsection{Nonuniform system response across the FoV}

\begin{figure}[htb!]
    \centering
    \includegraphics[width=0.95\textwidth]{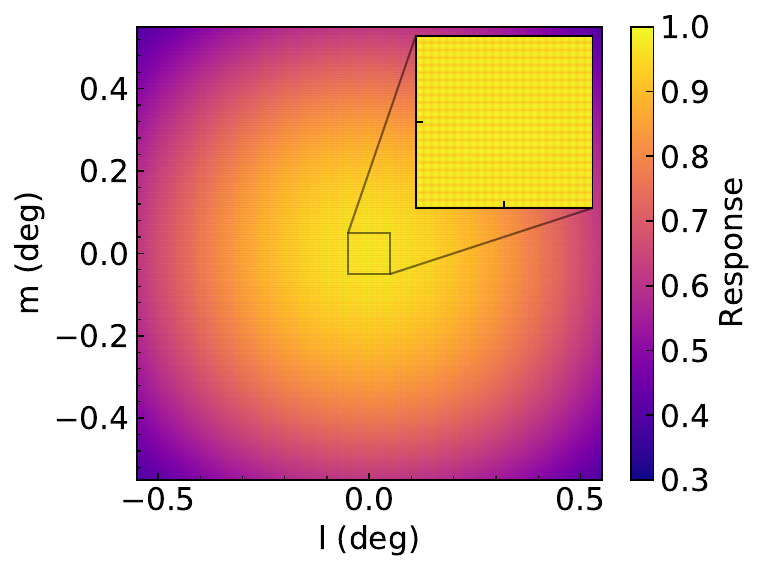}
    \caption{A map of the signal response across the FoV as a function of coordinate offset from the phase centre. The map was made using simulated visibility data with a point source at coordinate ($l$, $m$), and calculated response as the ratio of detection flux to the input flux. The inset shows a zoomed-in region around the phase centre.
    } 
    \label{fig:fov_response}
\end{figure}

Candidates far from the phase centre were detected with a lower SNR than the expected value. We made simulated visibility data of a point source at different coordinates and generated detection images to measure the system response. 
As is shown in Figure~\ref{fig:fov_response}, the SNR loss at the corners of the FoV reached $\sim$60\%. 

This is another limitation of using pillbox gridding. If the convolving function used for gridding is $C\left(u, v\right)$, the measured intensity of the synthesized image will follow $\widehat{C}\left(l, m\right)I'\left(l, m\right)$, where $\widehat{C}$ denotes the Fourier transform of $C$, and $I'$ denotes the real intensity map convolved with the synthesised beam.
For the pillbox gridding, the convolving function can be written as 
\begin{equation}
C\left(u, v\right) = \cfrac{1}{\Delta u\Delta v}\Pi\left(\cfrac{u}{\Delta u}\right)\Pi\left(\cfrac{v}{\Delta v}\right)
\end{equation}
where $\Delta u$ and $\Delta v$ are gridding spacing, and $\Pi\left(x\right)$ is the unit rectangle function. The Fourier transform of $C\left(u, v\right)$ is a sinc function,
\begin{equation}
\widehat{C}\left(l, m\right) = \cfrac{\sin\left(\pi\Delta ul\right)}{\pi\Delta ul} \cfrac{\sin\left(\pi\Delta vm\right)}{\pi\Delta vm}.
\end{equation}
At the edge of the synthesised image, $l = (2\Delta u)^{-1}$ and $m = (2\Delta v)^{-1}$, therefore $\widehat{C} = 4/\pi^2 \approx 0.4$.
The stripy pattern shown in Figure~\ref{fig:fov_response} (see inset) can be explained by insufficient oversampling across the primary beam, which can cause a maximum loss of another $\sim$4\%. 
The SNR reduction can be suppressed with a more complicated gridding function (e.g., Gaussian function, and Spheroidal functions) and larger oversampling factor (i.e., smaller pixel size). However, implementing these solutions requires FPGA code changing which is challenging and can shrink the CRACO effective FoV by eliminating aliases (see Section~\ref{subsubsec:alias}). 
Some faint events which should have been detected can be missed by the search pipeline due to the nonuniform response across the FoV.
For the events CRACO detects, we can recover their expected SNR in the post-processing step, e.g., by making synthesised images with \textsc{casa}, generating a tied-array beam filterbank towards the event position.

\subsubsection{Astrometric offset due to approximations}\label{subsubsec:astrometric_offset}

To allow faster searching, we made several approximations in making synthesised images. 
\corrs{There are three primary approximations, each of which can lead to an offset as large as $\sim$10\,arcsec:
\begin{itemize}
    \item We assumed the \texttt{CRACOPlan} would not change within a short time. In CRACO-PS, we updated \texttt{CRACOPlan} every 12 blocks (about 5 mins). This could lead to as much as a $\sim$10\,arcsec offset in astrometry because some data could be assigned to the wrong $uv$ cells. 
    \item We assumed that all baselines could be approximated as a plane. According to Equation~3.12 in \citet{2017isra.book.....T}, this approximation will not be held and we need to consider the effect of the $w$-term when the FoV is larger than 15\,arcmin for the array with a maximum baseline of 2\,km observed at 1\,GHz. The $w$-term can introduce an offset of $\sim$10\,arcsec for a source near the edges of the image.
    \item We approximated the source coordinate as the centre position of the brightest pixel in the image, which caused an error of $\sim$10\,arcsec caused by the sparse sampling.
\end{itemize}
}

\noindent
Although these astrometric offsets will not hugely impact the search pipeline performance, the coordinate information is critical in understanding the nature of the transient events. 
As we have discussed in Section~\ref{subsec:transient_loc}, we can achieve an uncertainty of several arcsec in the localisation with CRACO data in the post-processing steps.

\subsubsection{Other features or bugs}\label{subsubsec:other_bugs}

There are still several features or bugs in the pipeline \corrs{we used for CRACO-PS which are not fixed} or fully understood. We will discuss the workarounds to avoid them and how they affected our search capability in this section.
% We worked out strategies in avoiding them, but these workaround would affect the capability of the search pipeline. 

{\bf High-DM spurious candidates} are those high significance candidates with a high DM, but for which there was no convincing signal in either the synthesised images or filterbanks. 
This kind of candidate was usually detected with a \texttt{dm} (in samples, see Table~\ref{tab:search_coldescr}) larger than the number of channels of the data. Compared to a brute force de-dispersion, the FDMT algorithm takes intra-channel dispersion into account. During boxcar filtering, there could be overlaps between the adjacent samples. The higher the DM, the more overlaps there would be. The noise cannot be simply approximated as the standard deviation because of the correlation in the data. This is the same effect identified by \citet{2024MNRAS.528.1583H} in ``v2'' of the Fast Real-time Engine for Dedispersing Amplitudes search algorithm used in CRAFT/ICS observations, and it leads to an underestimation of the intrinsic noise fluctuations in the searches, and hence an overestimate of the SNR. This will limit the DM range the pipeline can search up to, and the impact will be more severe with a higher time resolution. For example, we can search up to $\sim$10\,000\,pc\,cm$^{-3}$ at a 110\,ms resolution, but only $\sim$1000\,pc\,cm$^{-3}$ at a 13.8\,ms resolution.
\corrs{This issue will be addressed in the future, ensuring it no longer affects our ability to search for high-DM signals.}

{\bf Phase centre fake candidates} are similar to high-DM fake candidates but appear exactly at the phase centre. 
This kind of candidate may arise from \texttt{CRACOPlan} update as most of them were detected in a plan update block, but we have not figured out the exact reason. 
In the candidate post-processing, we filtered out all candidates within 2 pixels of the phase centre. This will cause an effective FoV loss, but only $0.04\%$ in total area is masked out.

{\bf Buffer saturation} happens when the number of candidates in one block (256 samples) is larger than 8192, and we can only get the first 8192 candidates in the final output. The high number of candidates can happen due to various reasons. If the pulse is too bright, such as consecutive single pulses from the Vela pulsar, it will produce a large number of candidates at different DMs, positions (main lobe and side lobes), and boxcar widths, which also occurs when there is RFI in the data. Some issues we discussed above (e.g., high DM spurious candidates) can also result in a large number of unwanted candidates, which can saturate the buffer as well. To minimise the saturation occurrence, we flagged the known RFI frequencies before the data was run through the pipeline, and we set \texttt{ndm} close to the number of channels (see Table~\ref{tab:survey_param} for example) to suppress the high-DM spurious candidates. $\sim$10\% observing time at the mid-band (mainly caused by the satellite RFIs) and $\lesssim$0.1\% observing time at the low-band are affected by the saturation issue.
% Except for the fields containing the Vela pulsar, we only experience the severe saturation issue in the mid-band due to the satellite RFIs. 

\section{System Performance}\label{sec:system_perform}

\subsection{Sensitivity validation of CRACO/ASKAP}\label{subsec:sefd}

%description of the data acquired using CRACO and UWL
% check more details in this thread - https://craft-askap.slack.com/archives/C03GEV11D4M/p1693528200339929
We use methods described in \cite{2019PASA...36....9J} to validate the sensitivity of CRACO. To assess the sensitivity of CRACO, we performed an ASKAP observation (Schedule Block 52211) of the bright pulsar PSR~J1644$-$4559 contemporaneous with Parkes/Murriyang Ultra Wideband Low \cite[UWL,][]{2020PASA...37...12H} to verify the system equivalent flux density (SEFD). The ASKAP observation was performed with CRACO with a temporal and spectral resolution of 1.728 ms and 1.0 MHz, respectively, at a centre frequency of 944 MHz. Only 21 antennas were used for CRACO recording.  The Parkes/Murriyang UWL observation was performed in a pulsar search mode \citep{2020PASA...37...12H} with a temporal and spectral resolution of 64$\mu$s and 0.5 MHz, respectively. The search mode data were reduced to \code{PSRFITS} archive files using \code{dspsr} and further analysis was performed using the pulsar processing package \code{psrchive}.  

\begin{figure}[htb!]
    \centering
    \includegraphics[width=\textwidth]{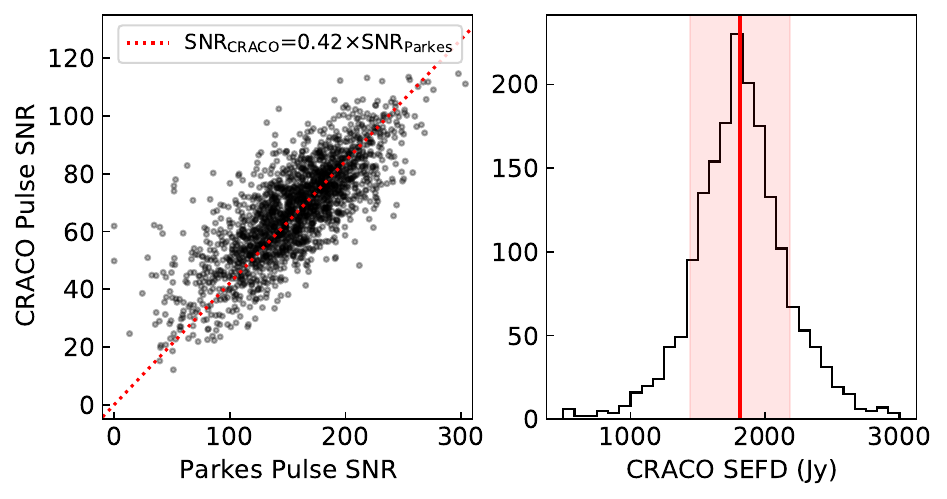}
    \caption{{\bf Left:} Comparison of pulse SNR detected from PSR~J1644$-$4559 using CRACO and Parkes UWL. The red dotted line represents the best linear fit without an intercept, with a slope of 0.42. {\bf Right:} Histogram of derived CRACO single-dish SEFD from single pulses. The red solid line indicates the median value of the derived CRACO SEFD (1810\,Jy), while the red shaded region represents the 1-$\sigma$ error range (370\,Jy).}
    \label{fig:PARKES_CRACO_single_pulses}
\end{figure}

The UWL single pulses were flux-calibrated using UWL calibrator observation of J0407$-$458. The UWL data are impacted by more hostile RFI environment compared to the ASKAP in the same band. We excise RFI using the native excision algorithm in \code{psrchive}. For comparison, we use the common frequency channels and average the data to the same time resolution, 26.8\,ms, for both UWL and CRACO. The SNR for the individual single pulses for CRACO and UWL are calculated from the frequency-averaged pulse profiles. The time offset between UWL and ASKAP observations was corrected using frequency-averaged profile cross-correlation between both data streams. 
The SNR derived from CRACO is on-average 42\% that of Parkes UWL (see Figure~\ref{fig:PARKES_CRACO_single_pulses}). Since the pulse duration for PSR~J1644-4559 ($\sim50\,ms$ at 900\,MHz) is greater than the temporal resolution of both systems, and we use identical frequencies for our analysis, we conclude that the primary effect on the SNR is the SEFD of each system, i.e., the ratio between CRACO SNR and Parkes SNR follows
\begin{equation}
    \cfrac{{\rm SNR}_{\rm CRACO}}{{\rm SNR}_{\rm Parkes}} = \cfrac{\sqrt{N_{\rm ant}\left(N_{\rm ant}-1\right)}{\rm SEFD}_{\rm Parkes}}{{\rm SEFD}_{\rm CRACO}}
    \label{eqn:snr_ratio}
\end{equation}
, where $N_{\rm ant}$ is the number of the antenna used for CRACO recording. 
The SEFD of Parkes UWL for the frequency channels we use for comparison is $\sim$38\,Jy\footnote{See \url{https://www.parkes.atnf.csiro.au/observing/Calibration_and_Data_Processing_Files.html}.}.
Therefore, we conclude that the SEFD of the CRACO system for a single antenna is $1820\pm370\,$Jy, which is consistent with what was reported in \citet{2021PASA...38....9H}.
% Thus we conclude that the SEFD of the CRACO system for these observations is (0.42^{-1}=2.38 times greater than that of Parkes, i.e., it is ???? Jy.
% The SNR of the individual pulses are measured using the extracted pulses from frequency-averaged single pulses: 

% %The S/N of the single pulses are correlated to calculate the time offset between UWL and CRACO observations which is further used to correct for the offset while comparing the single pulses. The SEFD of a telescope can be related to the S/N of the single pulses by  

% \begin{equation}
%     \mathrm{S/N_{\rm pulse}} = \frac{2S_{\rm avg}}{\rm SEFD}  \frac{P}{W} \sqrt{B\tau_s} , 
%     \label{eqn:SEFD_SNR}
% \end{equation}

% \noindent
% where $S_{\rm avg}$ is the average flux density of the pulsar, $P$ and $W$ are the period and width of the pulse, respectively, $B$ is the bandwidth of the recording instrument and $\tau_s$ is the averaging time. Using Equation \ref{eqn:SEFD_SNR} we estimate the SEFD of ASKAP-CRACO from the average SEFD measured for Parkes to be 
% %1867.17$\pm$380.21 Jy. 
% $1870 \pm 380$ Jy.

% Figure \ref{fig:PARKES_CRACO_single_pulses} shows the comparison of the SNRs of the single pulses from UWL and ASKAP-CRACO. 

\begin{figure*}[htb!]
    \centering
    \includegraphics[width=0.8\textwidth]{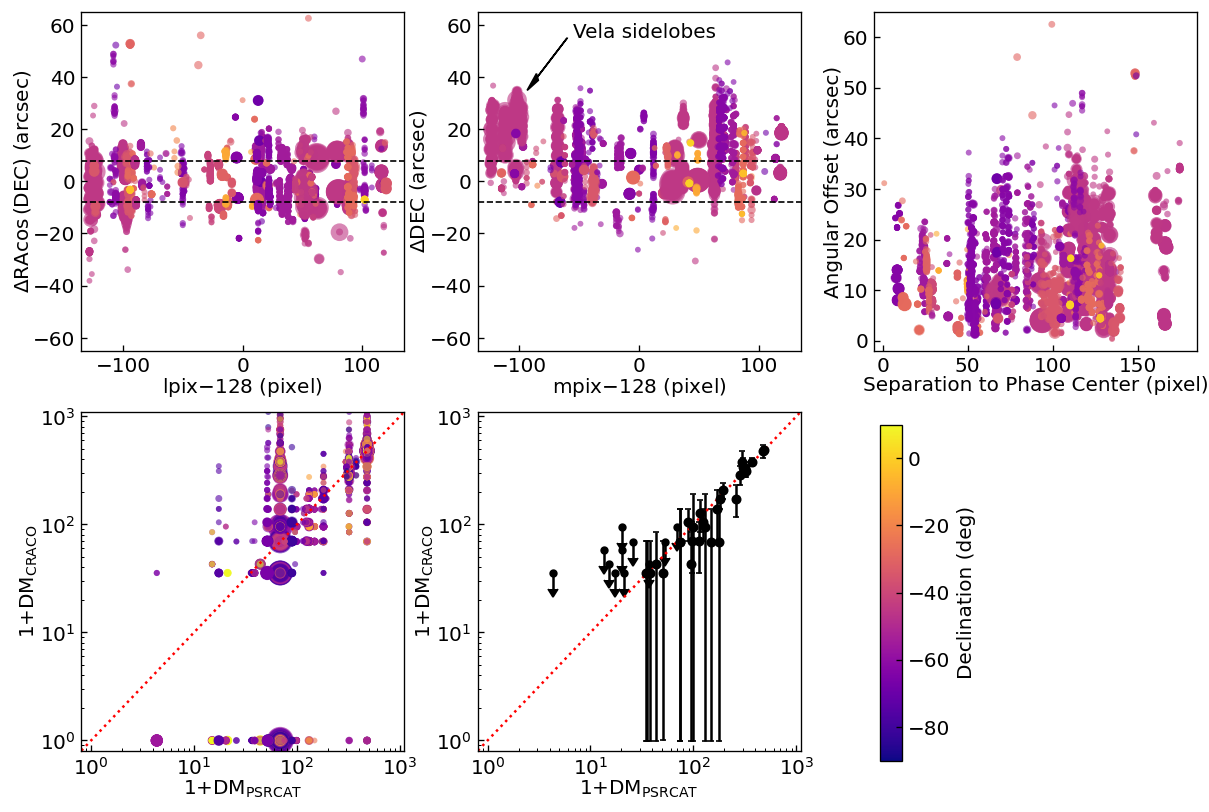}
    \caption{Coordinate and DM value comparison for known pulsars detected in the CRACO 110-ms pilot survey between the pipeline and PSRCAT. On the top panel, we show $\Delta$RA$\cos$DEC, $\Delta$DEC, and angular separation values for all CRACO candidates matched with a known pulsar against the separation to the phase center on $l$-axis, $m-$axis, and $lm-$space, respectively. Two horizontal dashed lines represent the size of one pixel ($\sim$15.5\,arcsec). On the bottom panel we show the comparison of the DM values for all CRACO candidates matched with a known pulsar between the one reported by CRACO and listed in PSRCAT on the left. We also show the same comparison but only for the candidates with the highest SNR for each pulsar on the right.}
    \label{fig:craco_coord_dm_comp}
\end{figure*}

\subsection{DM and Coordinate Measurement Accuracy}

We used known pulsars detected in the 110-ms survey (see Table~\ref{tab:cracopsr}) to estimate the accuracy of the pipeline measurement on the DM and coordinate.
We selected all bright (SNR > 8) candidates with a known pulsar matched in the candidate post-processing output (i.e., after clustering, and cross-matching), and compared their DM and coordinates against the values in the PSRCAT using the \texttt{psrqpy} package \citep{2018JOSS....3..538P}.

We measured an offset weighted by the SNR of $0.26\pm7.60\,$arcsec, $4.83\pm9.36\,$arcsec, and $11.36\pm6.30\,$arcsec for right ascension, declination, and angular offset, respectively (see Figure~\ref{fig:craco_coord_dm_comp} top panel). We observe an asymmetric distribution for declination, of which the underlying cause is still unclear.  We also noted that there were false matches due to sidelobes which would cause systematic offsets. However, these offsets are all within the expectation (compared to the typical pixel size of $\sim$15\,arcsec, and also see discussion in Section~\ref{subsubsec:astrometric_offset}). Even in the case of significant outliers, we are still able to measure a precise coordinate with the steps discussed in Section~\ref{subsec:transient_loc} for new discoveries.

We also compared the DM value measured by the pipeline against the value in the PSRCAT. Due to the low time resolution of the data (110\,ms), the detection DMs are highly quantised. We therefore only made qualitative comparisons. As is shown in the bottom panel of Figure~\ref{fig:craco_coord_dm_comp} the DM values reported by the pipeline largely agreed with the catalogued values. The scatter in the left plot is mainly caused by the false matches and imperfect clustering in our post-processing procedure.

\subsection{Candidate Pipeline Performance}

We performed regular manual checks to qualitatively understand the performance of the candidate pipeline. However, we did not perform comprehensive quantitative checks as the system was still in development.
We saw $\sim10^3$ candidates per day due to RFI, of which typically $\sim$10 per day were misclassified as potential real sources. 
The majority of the leaked RFI candidates are associated with either RFI with a slow variation (on a timescale of several seconds) in phase, or moving objects (i.e., from a satellite or aeroplane).
Other notable sources of false positives were aliases from known sources ($\sim$10 candidates per day) and sidelobes of bright sources ($\sim$1 candidate per day). Around 30 candidates per day occurred due to Galactic pulsating sources, of which $\sim$98\% were correctly classified and $\sim$2\% were missed mainly due to their poor localisation.
False negatives were mainly candidates from real pulsars classified as RFI, because either: (1) the candidate was detected at the same time as lots of real RFI; (2) due to the issues discussed in Section~\ref{subsubsec:other_bugs}; or (3) the pulsar signal was so bright (with a single pulse SNR $\gtrsim$50) that it produced several sidelobe candidates.
% \todo{also mention that we will do injection later, but not this survey}
We did not do systematic injections to test the performance of the pipeline robustly in our 110-ms survey. However, this will be included in future operations once the CRACO system becomes stable.

\section{Pilot Survey Description}\label{sec:survey}

We conducted the CRACO-PS observations as part of the system commissioning procedure between April and November of 2023. 
During the survey the CRACO system was piggybacked onto all available ASKAP surveys including the First Large Absorption line Survey in HI \citep[FLASH;][]{2022PASA...39...10A}, the Variables And Slow Transients \citep[VAST;][]{2013PASA...30....6M, 2021PASA...38...54M} survey, the Evolutionary Map of the Universe \citep[EMU;][]{2011PASA...28..215N, 2021PASA...38...46N} survey, and the Widefield ASKAP L-band Legacy All-sky Blind surveY \citep[WALLABY;][]{2020Ap&SS.365..118K}.
Parameters for these four surveys onto which we have piggybacked are listed in Table~\ref{tab:askap_obsvar}.

\begin{table}[hbt!]
    \centering
    \caption{Observation parameters for the ASKAP surveys for which CRACO operated in piggyback mode. The column marked Time shows the total number of hours we recorded with the CRACO backend.}
    \begin{threeparttable}
        \begin{tabular}{c c c c c}
        \hline\hline
        Survey & Frequency Range$^\ast$ & Pitch & Footprint & Time \\
         & (MHz) & (degree) & & (hrs) \\
        \hline
        FLASH & 711.5--999.5 &  1.05 & square 6$\times$6 & 52.4 \\
        VAST & 743.5--1031.5 &  1.05 & square 6$\times$6 & 144.5 \\
        EMU & 799.5--1087.5 & 0.9 & closepack 36 & 193.8 \\
        WALLABY & 1151.5--1439.5 & 0.9 & square 6$\times$6 & 8.0 \\
        \hline
        \end{tabular}
        \begin{tablenotes}
        \item[$^\ast$] This indicates the frequency range of the survey itself, and not necessarily the frequency band searched with CRACO
        \end{tablenotes}
    \end{threeparttable}
    \label{tab:askap_obsvar}
\end{table}

\begin{figure*}
    \centering
    \includegraphics[width=0.65\textwidth]{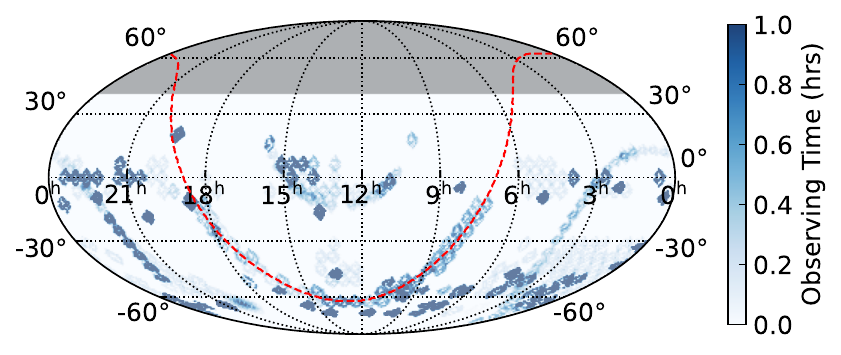}
    \caption{CRACO 110-ms Pilot Survey coverage map. The grey shaded region shows the sky that ASKAP cannot observe due to the elevation limit. The red dashed line indicates the position of the Galactic Plane.}
    \label{fig:coverage}
\end{figure*}

During the commissioning we developed the CRACO backend incrementally, such as by improving the bandwidth, optimising the search algorithm, and changing the search parameters, to meet the proposed capacity. 
Because of that, the survey consisted of two parts with different sets of search parameters listed in Table~\ref{tab:survey_param}.
The maximum DM in the physical unit we searched up to depends on the parameter \texttt{ndm} and the frequency range. As a reference, for the observations centred at 943.5\,MHz (i.e., EMU observations), the maximum DM we searched up to was $\sim$12\,000\,pc\,cm$^{-3}$ for both parts of the survey. 
We recorded data from the inner 30 antennas (with a maximum baseline of $\sim$2\,km), but we only searched data from the inner 23 antennas (with a maximum baseline of $\sim$1\,km). This decision was made to balance the sensitivity and FoV. 
We chose the minimum oversampling factor (with regard to the synthesised beam) to be 2.1 to minimise the SNR loss due to sparse sampling in the $uv$ plane, and used a FoV of 1.1\,deg to ensure the overlapping between adjacent beams for the search. 
Including longer baselines would decrease the synthesised beam, and therefore decrease the FoV if we held the oversampling factor fixed.
Not all 23 antennas were operational throughout the entire survey: there could be several (typically up to three) antennas non-operational due to maintenance and other various reasons.
We show the coverage map of the pilot survey in Figure~\ref{fig:coverage}. We could not search all 36 beams (the corresponding beams are coloured white in the figure) because of hardware issues in some FPGA cards at the time of the search.

\section{Pilot Survey Results}\label{sec:result}

The CRACO pipeline is able to detect transient events on timescales from a hundred milliseconds to a second. FRBs, pulsars/RRaTs, and ULPOs can all show burst-like emission on these timescales. 

\subsection{FRBs}

In CRACO-PS we detected one FRB, \cracofrb, through the CRACO pipeline, and also confirmed the detection of a CRAFT/ICS-detected FRB, FRB\,20230902A, in the CRACO data. We list their properties in Table~\ref{tab:craco_frb}.

\begin{table}[]
    \caption{
    Measured and derived parameters for two FRBs in CRACO-PS. All values are based on the post-processing refined analysis. 
    }
    \centering
    \begin{threeparttable}
        \begin{tabular}{l c c}
\hline\hline
                    &      FRB20230902A$^\ast$ &       \corrs{FRB20231027A}  \\
\hline
MJD$_{\rm TOPO}$    &      60189.033914 &       60244.277344 \\
UTC$_{\rm TOPO}$    & 2023-09-02 00:48:50.1& 2023-10-27 06:39:22.5\\
R.A. (J2000)        &    03h28m34.5(2)s &     17h00m18.10(7)s \\
Decl. (J2000)       &   $-$47d20m05(3)s & $-$07d16m44.2(12)s \\
Gl (deg)            &    256.991        &     12.691         \\
Gb (deg)            &   $-$53.336       &     20.716         \\
S/N                 &   \corrs{6.0}             &      15.2$\dagger$          \\
Width (ms)          &   <110.6          &   18.9$^\dag$      \\
DM (pc\,cm$^{-3}$)  &  426(85)          & 996(2)$^\dag$      \\
\hline
Fluence (Jy\,ms)    &    \corrs{22}$^{\mathsection\mathparagraph}$ &      27$^{\mathsection\mathparagraph}$   \\
DM$_{\rm MW, NE2001}$ (pc\,cm$^{-3}$)&   34.1           &      96.5          \\
DM$_{\rm MW, YMW16}$ (pc\,cm$^{-3}$)&    25.5           &      144.5         \\
\hline
$N_{\rm chan}$      &    120            &        240         \\
$N_{\rm ant}$       &     26            &         30         \\
\hline
\end{tabular}
    \begin{tablenotes}
        \item[$^\ast$] More accurate values for FRB20230902A are reported in \citet{ShannonICS}.
        \item[$^\dag$] Derived from CRAFT/ICS filterbank data.
        \item[$^\mathsection$] Both FRBs were detected with a width of one time sample in the CRACO system. We included the response factor $\eta\approx0.6$ reported in \citet{2023MNRAS.523.5109Q} in the calculation.
        \item[$^\mathparagraph$] Fluences were corrected for primary beam response.
    \end{tablenotes}
    \end{threeparttable}
    \label{tab:craco_frb}
\end{table}

\subsubsection{FRB\,20230902A}

\begin{figure}[htb!]
    \centering
    \begin{subfigure}[b]{0.49\linewidth}
    \centering
        \includegraphics[width=\textwidth]{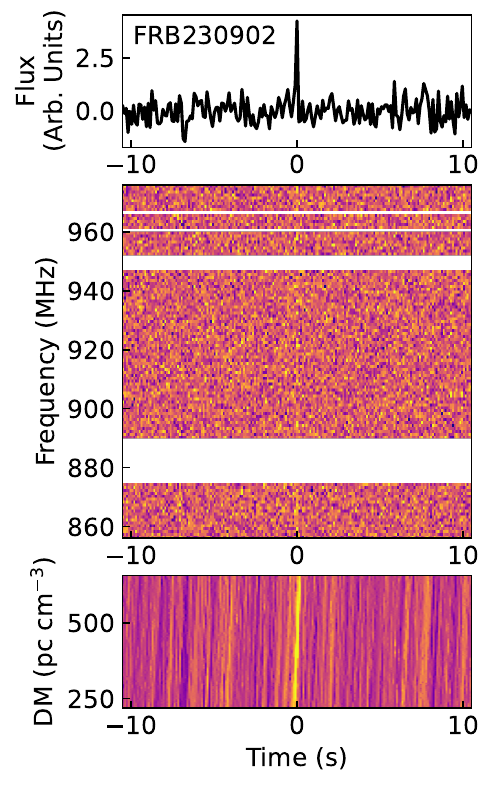}
        \caption{FRB20230902A}
        \label{fig:frb230902_filterbank}
    \end{subfigure}
    \hfill
    \begin{subfigure}[b]{0.49\linewidth}
    \centering
        \includegraphics[width=\textwidth]{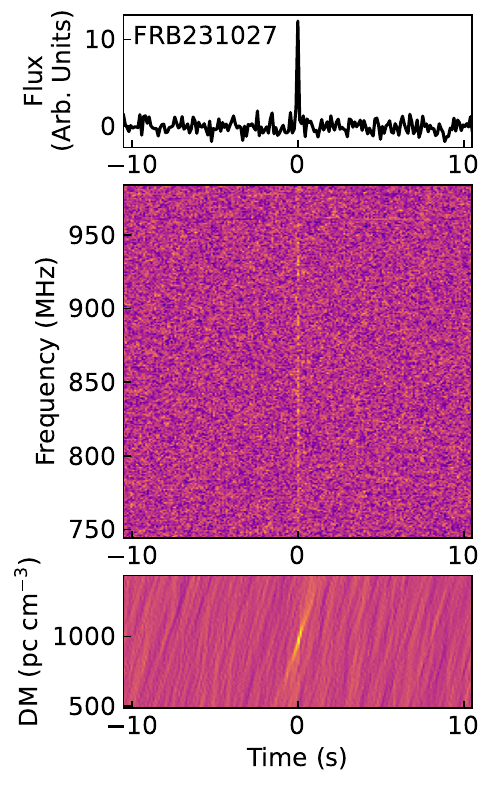}
        \caption{\cracofrb}
        \label{fig:frb231027_filterbank}
    \end{subfigure}
    \label{fig:frb_filterbank}
    \caption{
    Filterbank plots for FRBs detected in the CRACO pilot survey. For each burst, we show the dedispersed pulse profile for the burst (top), the dynamic spectrum for the dedispersed pulse (middle), and signal strength as a function of DM and time (bottom). 
    The horizontal stripes correspond to the parts of the band that have been removed due to the presence of RFI.
    }
\end{figure}

FRB\,20230902A\footnote{\url{https://www.wis-tns.org/object/20230902A}} was detected in real time by the CRAFT/ICS \citep[][]{2019Sci...365..565B} system (SB52520, beam 17) with a SNR of 10.1 (with 22 antennas available), a pulse width of $\sim8\,$ms, and a DM of 438.23\,pc\,cm$^{-3}$.
At the time of FRB\,20230902A, we were recording CRACO data with 26 available antennas, 110-ms time resolution, and a 120\,MHz bandwidth (from 856 to 976\,MHz).
% \sqrt{\cfrac{N_{\rm ant,CRACO}(N_{\rm ant,CRACO}-1)}{N_{\rm ant,ICS}}}
We expected an SNR of $\eta\sqrt{\cfrac{\Delta\nu_{\rm CRACO}}{\Delta\nu_{\rm ICS}}}\cfrac{N_{\rm ant,CRACO}}{\sqrt{N_{\rm ant,ICS}}}\sqrt{\cfrac{W_{\rm obs, ICS}}{W_{\rm obs, CRACO}}}{\rm SNR}_{\rm ICS} \approx 5.3$ detection for FRB\,20230902A in the CRACO data, where $\Delta\nu$ is the observing bandwidth, $N_{\rm ant}$ is the number of antennas, $W_{\rm obs}$ is the observed width, and $\eta\approx0.6$ is the response factor for a one time sample width pulse \citep[e.g.,][]{2023MNRAS.523.5109Q}.
We used \texttt{CASA} to image the CRACO data to confirm the detection. We identified a transient detected with an SNR of 5.4 at the FRB position (see Figure~\ref{fig:frb230902_interimg} left panel). In the normalised data (similar to what should be done in the search pipeline internally), the detection SNR increases to 6.0 (see Figure~\ref{fig:frb230902_interimg} right panel). There is also a clear burst in the dynamic spectrum (see Figure~\ref{fig:frb230902_filterbank}) from the CRACO data.
% The position of the FRB was 03h28m34.23s $-$47d20m02.07s (J2000) based on the \texttt{CASA} image.

Given its low SNR in the CRACO data, there was no detection of this FRB in the default search (with 21 out of the inner 23 antennas). We ran another search using all available antennas, and found a $6\sigma$ candidate at the FRB time and position.
The pipeline coordinate of the FRB was 03h28m34.30s, $-$47d20m00.87s (J2000), which is 1.4\,arcsec away from the \texttt{CASA} image coordinate and 8 arcsec away from the CELEBI pipeline position (with a phase calibration applied).

\subsubsection{\cracofrb}\label{subsubsec:frb231027}

\begin{figure*}[htb!]
    \centering
    \begin{subfigure}[b]{\linewidth}
        \centering
        \includegraphics[width=0.9\textwidth]{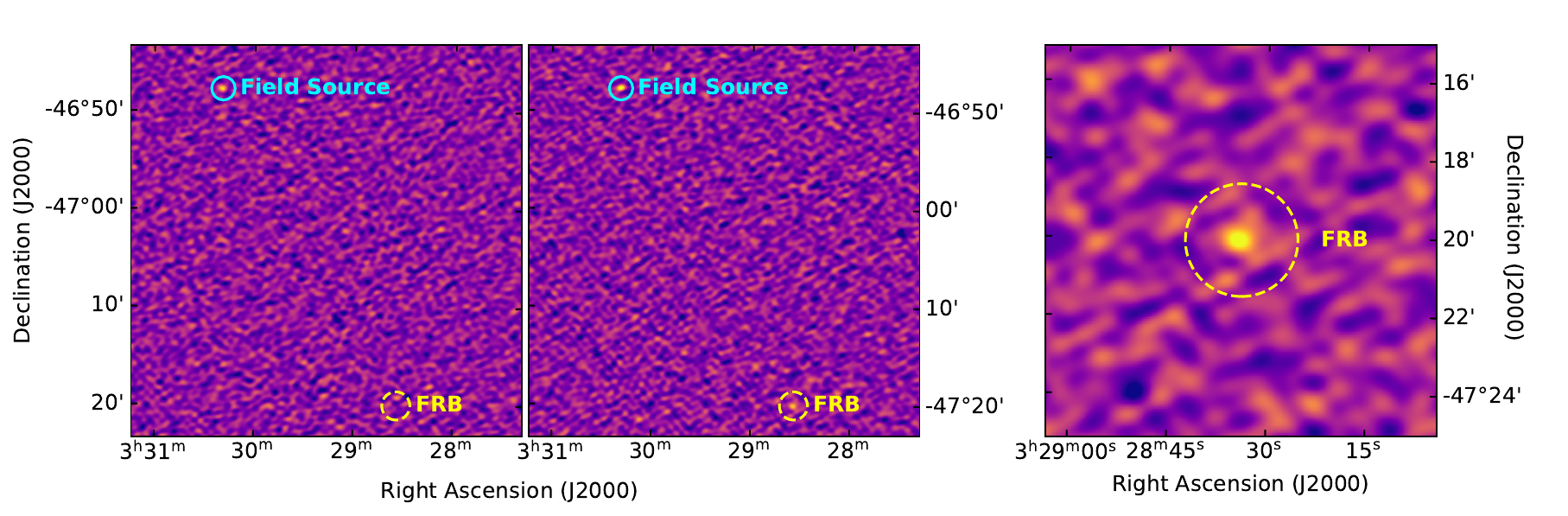}
        \caption{Interferometric Detection Images of FRB20230902A}
        \label{fig:frb230902_interimg}
    \end{subfigure}
    \vspace{1em}
    \begin{subfigure}[b]{\linewidth}
        \centering
        \includegraphics[width=0.9\textwidth]{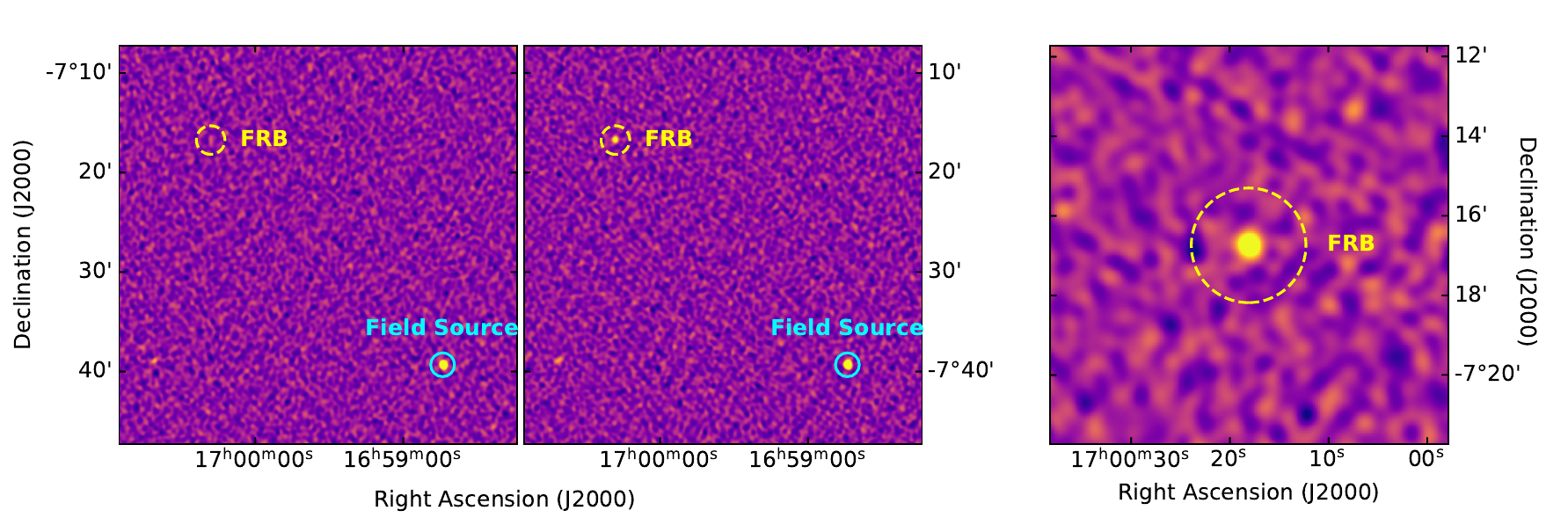}
        \caption{Interferometric Detection Images of \cracofrb}
        \label{fig:frb231027_interimg}
    \end{subfigure}
    \caption{Interferometric images of FRBs detected in the CRACO Pilot Survey. The left panel displays the unnormalised images before (left) and at the time of (right) the FRB detection. 
    The right panel presents the normalised (field source subtracted) images at the time of the FRB detection. 
    All images are imaged with a single sample of data with an integration time of 110 ms. Field sources are marked with cyan solid circles, and FRB locations are indicated by yellow dashed circles. All data have been de-dispersed using the optimal DM value available.
    }
    \label{fig:frb_interimg}
\end{figure*}

\cracofrb\ was the first FRB detected initially by the CRACO pipeline.  It was detected in a commensal observation with the VAST survey (SB54236 Beam 33) with an SNR of 9.7 (with 24 antennas available), a DM of 967\,pc\,cm$^{-3}$ at the coordinate of 17h00m18.2s $-$07d16m37s (J2000). We show the detection images of \cracofrb\ in Figure~\ref{fig:frb231027_interimg} and Figure~\ref{fig:frb231027_filterbank}.

The CRAFT/ICS system was also operating at the time of the detection. Although \cracofrb\ was detected in the ICS data, the event did not trigger a voltage dump due to the width being too great \footnote{We might miss SNR$\lesssim$15 events due to system effect in the ICS system.}. However, we were still able to obtain 1\,ms time resolution filterbank data from the ICS system, from which we measured a DM of 996$\pm$2\,pc\,cm$^{-3}$ and a pulse width of 18.9\,ms.

% \todo{make comparison and justify the SNR for CRACO and ICS filterbanks}

% We could achieve a better localisation based on the full CRACO data.
We recorded the data with all 30 available antennas, 110-ms time resolution, and a 240\,MHz bandwidth (from 744 to 984\,MHz).
% Instead of using the calibration solution for processing\footnote{The calibration solution used for processing was derived from SB54204, which was observed $\sim$9 hours before SB54236}, we applied the method described in \ref{subsec:calib} (but without any fitting) and derived a new calibration solution based on the 1-minute data observed around the FRB detection (hereinafter field observation data).
\cracofrb\ was detected in both beam 32\footnote{We did not see any candidate from beam 32 in the pipeline output because the (hardware) card processing the beam was non-functional.} and 33, with the FRB showing a higher SNR in beam 32 (12.8) compared to beam 33 (11.6). 
We therefore used the data from beam 32 for subsequent analysis.
% this may be changed depends on the value I can get from Adam
% We first imaged the field observation data using \texttt{CASA}, with w-projection enabled.
% We used \texttt{aegean} \citep{2012MNRAS.422.1812H, 2018PASA...35...11H} for source finding on the field image and then cross-matched the detected sources against the RACS-low1 catalogue to correct for any systematic astrometric offsets.
% We then imaged the burst observation data (i.e., the data taken during the FRB detection) using the same parameters as those for the field observation.
% After applying the offset correction, we detected \cracofrb\ at coordinate 17h00m18.09s$\pm$0.14s $-$07d16m44.5s$\pm$2.1s (J2000).
We used the post-processing scripts discussed in Section~\ref{sec:survey}, and measured coordinates for \cracofrb\ of 17h00m18.10s$\pm$0.07s $-$07d16m44.2s$\pm$1.2s (J2000).

\begin{figure}[hbt!]
    \centering
    \includegraphics[width=\linewidth]{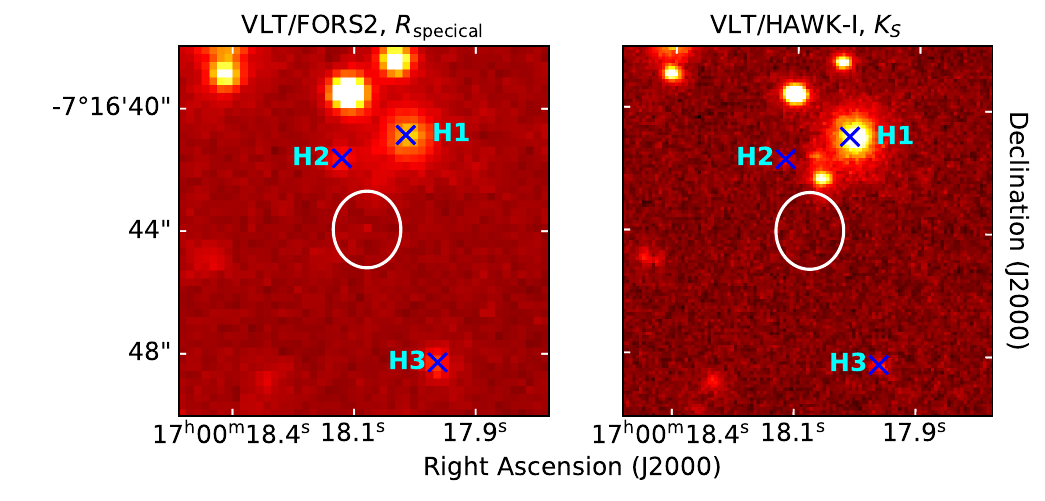}
    \caption{VLT FORS/$R$-band (left) and HAWK-$K_s$-band (right) imaging at the position of \cracofrb. Blue crosses and cyan labels denote PATH candidates with a posterior $>0.01$. The white ellipse in each image outlines the localisation region (1-$\sigma$ uncertainty) of the FRB.}
    \label{fig:frb_host}
\end{figure}

There was no host galaxy detection in the Dark Energy Camera Legacy Survey \citep[DR10;][]{2019AJ....157..168D} towards the FRB position. Galactic extinction along this sightline is high ($E(B-V)=0.44$, according to the IRSA Dust Tool\footnote{\url{https://irsa.ipac.caltech.edu/applications/DUST/}}).
We conducted deeper imaging observations in the $R_\mathrm{special}$ filter on the FOcal Reducer and low dispersion Spectrograph \citep[FORS2;][]{FORS} and $K_s$ on the High Acuity Wide field K-band Imager \citep[HAWK-I;][]{HAWK-I} instruments on the VLT \citep[][]{1992Msngr..67...18A}. 
% \textbf{Lachlan paragraph here}
Using the $R$-band VLT image, we apply the Probabilistic Association of Transients to their Hosts \citep{AggarwalPATH} code\footnote{\url{https://github.com/FRBs/astropath}} to attempt a host association. We apply standard priors and configuration as described in \citep{ShannonICS}, with the prior of the host being unseen set to 0.1. Three objects (see Figure~\ref{fig:frb_host}) have significant posterior probabilities of hosting the FRB: J170018.02-071640.86 (Candidate H1), J170018.16-071641.62 (Candidate H2), and J170017.95-071648.28 (Candidate H3). Their posteriors are given in Table~\ref{tab:path}. The posterior for an unseen host is 0.03. A spectroscopic observation was also performed using X-shooter on the VLT \citep[][]{2011A&A...536A.105V}, finding a redshift for the Candidate H3, from the [O\textsc{ii}] doublet and H$\alpha$ line. Additionally, five extended sources in the field were observed with the Low-Resolution Imaging Spectrometer \citep[LRIS;][]{LRIS} on Keck; as well as confirming the redshift of Candidate H3, redshifts were obtained for the objects Candidate H5 (H$\alpha$) and Candidate H4 ([O\textsc{ii}] and H$\beta$). These redshifts were obtained using Marz \citep{Marz}, and are tabulated in Table~\ref{tab:path}. No obvious features were found in the spectrum of Candidate H2, and due to difficulties with sky subtraction, a redshift has not yet been obtained for Candidate H1.
Therefore, while the preferred host is Candidate H1, it is not a highly secure association at a posterior of 0.67, and we do not currently have a redshift for it.

\corrs{The estimated redshift for FRB~20231027A using the Macquart relation is 0.86 (0.80), assuming contributions from the host galaxy of $100 (1+z)^{-1}$\,pc\,cm$^{-3}$, Milky Way (MW) halo of 50\,pc\,cm$^{-3}$, and the MW interstellar medium (ISM) of 96.5 (144.5)\,pc\,cm$^{-3}$. Given limiting image magnitudes of $m_R^{\rm lim}=25.8$ and $m_{K_s}^{\rm lim}=23.5$, the redshift--magnitude scaling for nine FRB host galaxies calculated by \citet{2023MNRAS.525..994M} suggests that six (five) of these galaxies would have been detected in the $R$-band ($K_s$-band) follow-up observations for the redshift range $0.80$--$0.86$. Therefore, the true host galaxy with a low luminosity may remain undetected at this redshift range.} % We also note that the inferred redshift from the Macquart relation is highly uncertain. It is also possible that the true host galaxy is undetected due to its high redshift.}

\begin{table*}
\caption{Host candidates of \cracofrb. Magnitudes are not corrected for Galactic extinction.}
\label{tab:path}
\begin{tabular}{ccccccc}
\hline\hline
ID & R.A. & Decl. & PATH & $z$ & $R$ & $K_s$ \\
 & (J2000) & (J2000) & posterior & & (AB mag) & (AB mag) \\
 \hline
H1 & 17$^{\rm h}$00$^{\rm m}$18$^{\rm s}$.02 & $-$07$^{\rm d}$16$^{\rm m}$40$^{\rm s}$.86 & 0.67 & -- & 22.5 & 23.3 \\
H2 & 17$^{\rm h}$00$^{\rm m}$18$^{\rm s}$.16 & $-$07$^{\rm d}$16$^{\rm m}$41$^{\rm s}$.62 & 0.17 & -- & 24.5 & -- \\
H3 & 17$^{\rm h}$00$^{\rm m}$17$^{\rm s}$.95 & $-$07$^{\rm d}$16$^{\rm m}$48$^{\rm s}$.28 & 0.06 & 0.99 & 24.5 & 22.8 \\
H4 & 17$^{\rm h}$00$^{\rm m}$17$^{\rm s}$.80 & $-$07$^{\rm d}$16$^{\rm m}$53$^{\rm s}$.11 & 0.0 & 0.74 & 24.6 & 23.1 \\
H5 & 17$^{\rm h}$00$^{\rm m}$17$^{\rm s}$.66 & $-$07$^{\rm d}$16$^{\rm m}$57$^{\rm s}$.75 & 0.0 & 0.38 & 24.1 & 22.3 \\
\hline
% Object & PATH & $z$ & $R$ & $K_s$ & Label\\
%   & posterior &   & (AB mag) & (AB mag) & \\ \hline
% \hline
% J170018.02$-$071640.86 & 0.67 & -- & 22.5 & 23.3 & H1\\
% J170018.16$-$071641.62 & 0.17 & -- & 24.5 & -- & H2\\
% J170017.95$-$071648.28 & 0.06 & 0.99 & 24.5 & 22.8 & H3\\
% J170017.80$-$071653.11 & 0.0 & 0.74 & 24.6 & 23.1 & H4\\
% J170017.66$-$071657.75 & 0.0 & 0.38 & 24.1 & 22.3 & H5\\
\hline
\end{tabular}
\end{table*}

% \todo{modify this part based on the later spectroscopic observations}
% A detailed analysis on the host association and its implication will be presented in Z. Wang et al., (in prep.)

% \begin{itemize}
%     \item $z=0.989$ host, PATH=???\%
%     \item derived luminosity of the burst
%     \item void discussion
% \end{itemize}

\subsection{Pulsars and RRATs}

\begin{figure*}
 \centering
 \begin{tabular}{ccc}
\includegraphics[width=2in]{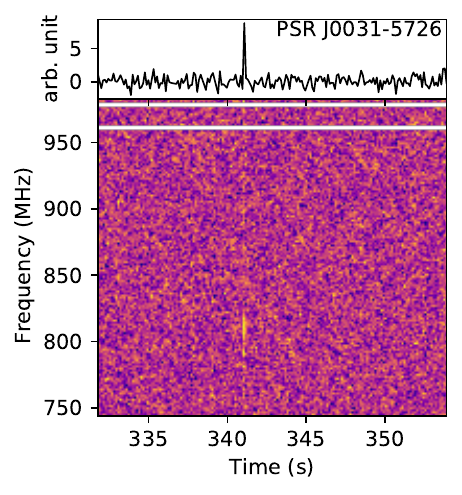} & 
\includegraphics[width=2in]{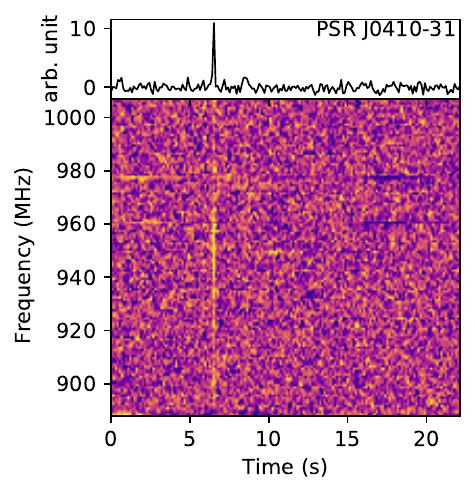} &
\includegraphics[width=2in]{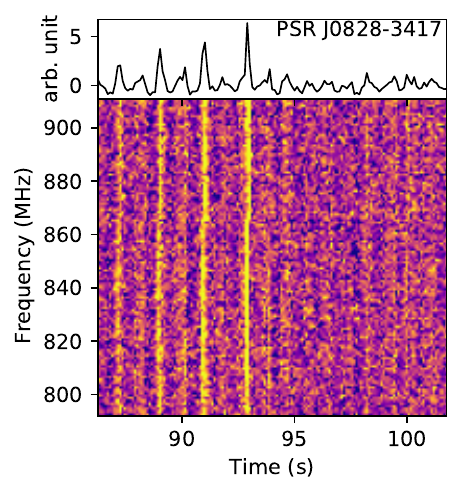} 
\\
\includegraphics[width=2in]{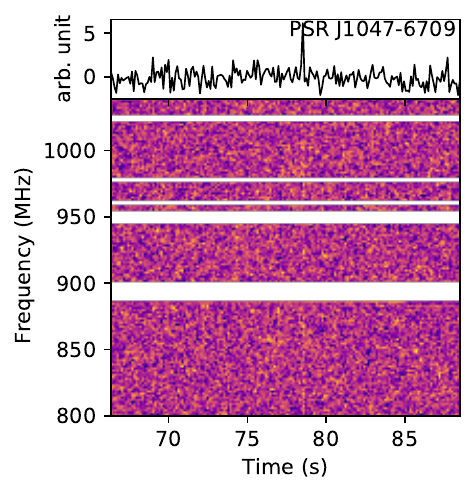} & 
\includegraphics[width=2in]{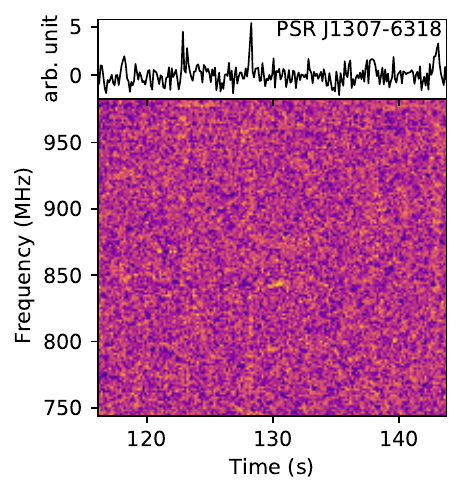} &
\includegraphics[width=2in]{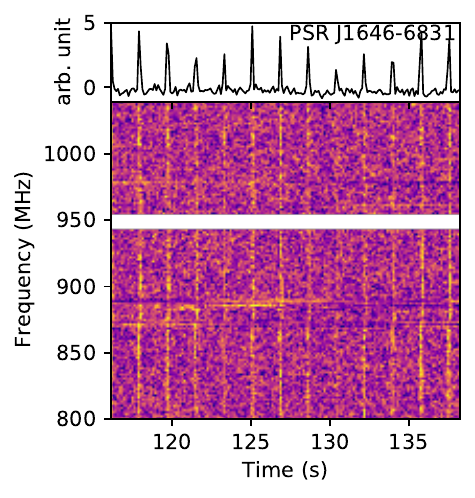} 
\\
\includegraphics[width=2in]{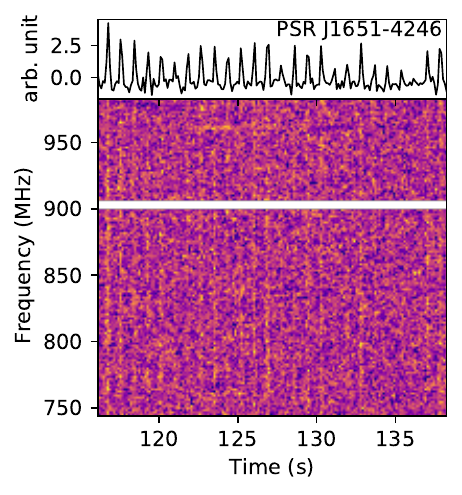} & 
\includegraphics[width=2in]{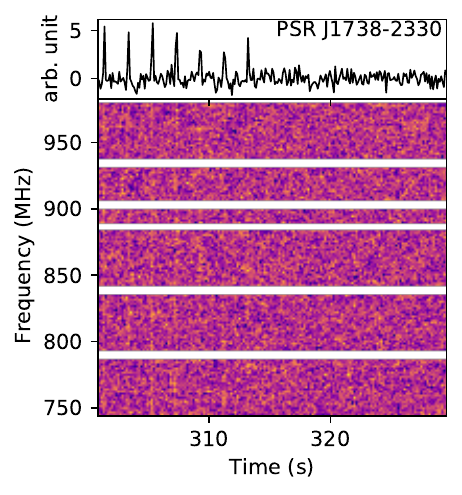} &
\includegraphics[width=2in]{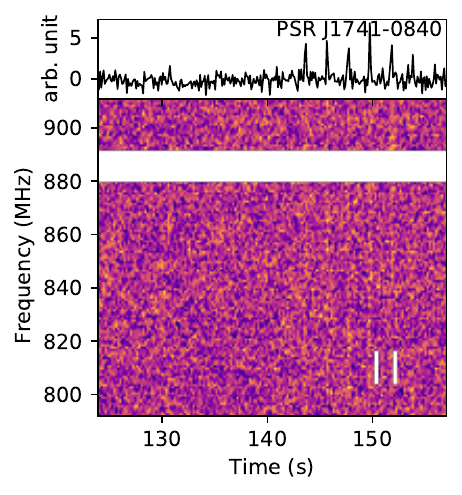} 
\\
\includegraphics[width=2in]{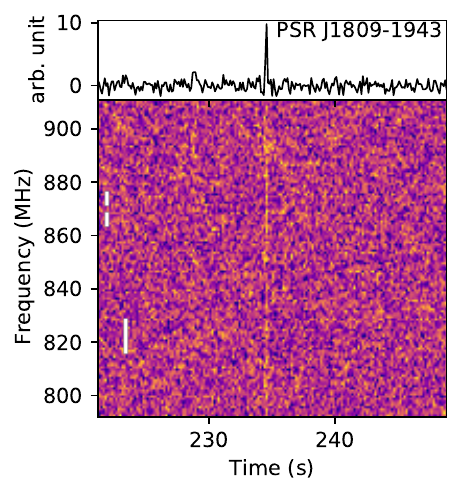} & 
\includegraphics[width=2in]{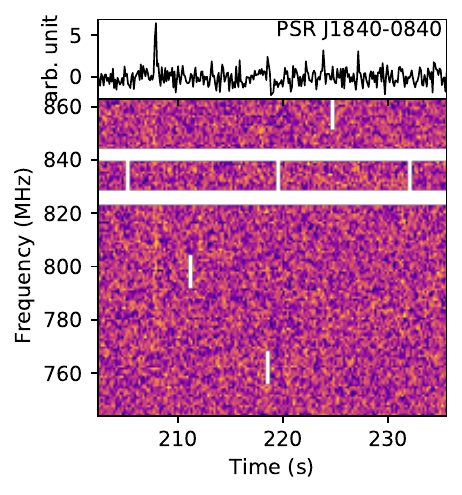} &
\includegraphics[width=2in]{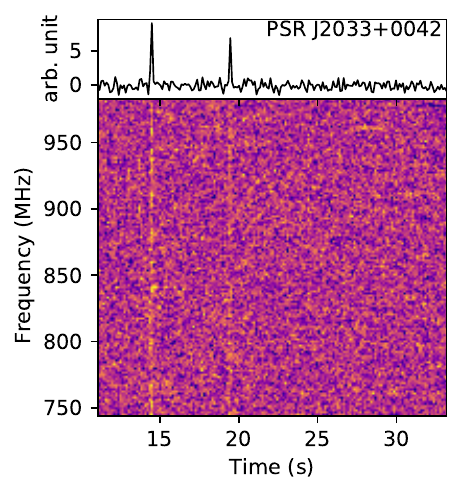} 
\\

\end{tabular}
\caption{Examples of dynamic spectra for single pulse detections with CRACO from known Milky Way pulsars. PSR~J0410--31 and PSR~J2033+0042 are RRATs; PSR~J1838--2330, PSR~J1741--0840, and PSR~J1840-0840 are nulling pulsars;  XTE~J1810--197 (PSR~J1809--1943) is a bright radio magnetar. White regions in the dynamic spectra have been flagged out due to the presence of RFI \corrs{or to missing packets during the recording}.}
\label{fig:psrfil}
\end{figure*}

Single pulses from pulsars and RRATs can be detected as candidates through the pipeline. 
In our pilot survey, a total of 42 known pulsars/RRATs were detected (see Figure~\ref{fig:psrfil} for single pulse detections for selected pulsars, and Appendix~\ref{sec:pulsar_cat} for the full list of pulsars). 

Pulsars with relatively constant pulse amplitudes and periods comparable to or less than our time resolution of 110\,ms do not appear as transient sources, and our pipeline is insensitive to them. All detected pulsars are with a relatively long spin period (longer than two time samples, i.e., $P_0 > 220$\,ms). All of the detected pulsars except PSR~J0835$-$4510 (the Vela pulsar) and PSR~J1047$-$6709 have a spin period larger than 220\,ms. 
The Vela pulsar has a spin period of 89.3\,ms. 
The peak flux density for Vela was measured to be $2.4\,$Jy\,beam$^{-1}$ at 888\,MHz \citep{2021PASA...38...58H}. Flux variation due to scintillation or intrinsic pulse-to-pulse flux changes can lead to candidate detections in our pipeline. 
% Detections from it can be explained by the fact that it is superthe pulse to pulse flux variation due to some intrinsic and/or external reasons (e.g., pulse nulling and scintillation). \\
PSR~J1047$-$6709 has a spin period of 198.4\,ms. This pulsar can have strong pulse to pulse flux variations, and is especially known to emit giant pulses \citep{2021MNRAS.501.3900S}. The sporadic single pulse detections in our data were consistent with those shown in \citet{2021MNRAS.501.3900S}. 

Among the 42 known pulsars/RRATs detected in CRACO-PS, we detected pulses from two RRATs (PSR~J0410$-$31, PSR~J2033+0042), several nulling pulsars (e.g., PSR~J1738$-$2330, PSR~J1741$-$0840, and PSR~J1840$-$0840), and one magnetar (XTE~J1810$-$197).
Most of the pulsars we detected do not appear to be remarkable, so we will not discuss them in detail. % However, these pulsars are useful tools to verify the performance of our new system.

% \todo{add RA, DEC, and DM comparison}

\subsubsection{Precise pulsar localisation}

Precise pulsar localisation can be obtained via long-term timing observations with a single-dish radio telescope, which is hard for pulsars with extreme nulling behaviours. In contrast, interferometric data can be used to localise pulsars upon detection.  With CRACO data, we can get better coordinates for previously poorly localised pulsars.  In our pilot survey, we improved the localisation for four known pulsars/RRATs: PSR~J0031$-$5726, PSR~J0410$-$31, PSR~J1738$-$2330, and PSR~J2228$-$65 (see Table~\ref{tab:psr_localisation}).
% To maximize the precision, we followed the same procedure mentioned in Section~\ref{sec:survey} to image the source.
We compared the DM and spin frequency derived from CRACO data to those in the literature to confirm the associations.
Though the DM uncertainty reported by CRACO is large (the typical uncertainty in DM is $\sim$35\,pc\,cm$^{-3}$), the DM measurements for all four pulsars are consistent with those in the literature (see Table~\ref{tab:cracopsr}).
We made filterbanks of the whole observation for them to measure their rotation frequency. As most pulses are not resolved at a 110-ms resolution, we therefore extracted pulse peak times as an approximation for times of arrival (ToAs).
We then fitted the ToAs in \textsc{pint} \citep{2021ApJ...911...45L}. We fit only for the rotation frequency $F_0$ and report the fitted value in Table~\ref{tab:psr_localisation}.

\begin{table*}[]
    \caption{Updated positions for the poorly localised pulsars detected in CRACO-PS. We also update the pulsar names based on their new positions, with previous names listed in parentheses. R.A. and Decl. columns are the coordinates derived from the post-processing images as discussed in Section~\ref{sec:survey}. $F_0$ is the frequency derived from the CRACO pulse detections, and N$_{\rm ToA}$ indicates the number of ToAs used in the derivation. For comparison, the catalogued frequency is listed as F$_{0, {\rm cat}}$ (values are from ATNF pulsar catalogue \citep[v 2.1.1,][]{2005AJ....129.1993M} unless otherwise specified).}
    \centering
    \begin{threeparttable}
    \begin{tabular}{l c c c c c l l }
\hline\hline
      Name &            R.A. &             Decl. &      Gl &      Gb &  $N_{\rm ToA}$ &   $F_0$ &      $F_{0, {\rm cat}}$ \\
& (J2000) & (J2000) & (deg) & (deg) & & (Hz) & (Hz) \\
\hline
J0031$-$5726                & 00h31m35.2(1)s & $-$57d26m36.7(13)s & 308.192 & $-$59.482 &            5 &  0.6368(2) &           0.63681$^\ast$ \\
J0409$-$3110 (J0410$-$31)   & 04h09m55.4(1)s &  $-$31d10m42.5(8)s & 230.633 & $-$46.832 &            5 &  0.5323(2) &        0.53234(6) \\
J1738$-$2331 (J1738$-$2330) & 17h38m08.8(1)s &    $-$23d31m22(4)s &   3.723 &   4.275   &            9 &  0.5054(3) & 0.505344668829(9) \\
J2227$-$6508 (J2228$-$65)   & 22h27m38.8(2)s &    $-$65d08m12(3)s & 323.659 & $-$45.779 &            11& 0.36415(6) &          0.364169 \\
\hline
\end{tabular}
    \begin{tablenotes}
    \item[$^\ast$] The value was retrieved from \url{https://mwatelescope.atlassian.net/wiki/spaces/MP/pages/24970773/SMART+survey+candidates} %J0031$-$5726 was discovered by MWA and has not been recorded in PSRCAT yet.
    \end{tablenotes}
    \end{threeparttable}
    \label{tab:psr_localisation}
\end{table*}

% see - https://docs.google.com/spreadsheets/d/1Iyp98Vakglz5R3WaJ9OE_U4oy1kBtiS3VQe2hnHoOmw/edit#gid=0

\subsubsection{Pulsar/RRATs discoveries}

We discovered two new pulsars/RRATs in CRACO-PS, with their measured and derived parameters listed in Table~\ref{tab:craco_newpsr}.
% Table~\ref{tab:craco_newpsr} contains the attributes of the discovery pulses, and Figure~\ref{fig:craco_newpsr} shows the filterbank plot of the discovery pulses.

\begin{table}[]
    \caption{Measured and derived parameters for two new pulsars discovered in CRACO-PS. We measured the coordinates for both pulsars with CRACO data after post-processing. All other information for J1319$-$4536 was derived from CRACO data, while that for J0755$-$7026 was derived from the Parkes/Murriyang observation.}
    \centering
    \begin{threeparttable}
        \begin{tabular}{l c c}
\hline\hline
                    &      J1319$-$4536 &       J0755$-$7026 \\
\hline
R.A. (J2000)        &    13h19m48.6(1)s &     07h55m42.9(1)s \\
Decl. (J2000)       & $-$45d36m04.4(8)s & $-$70d26m58.4(16)s \\
$F_0$ (Hz)          &       0.534500(2) &       0.3094631(4) \\
$\chi^2$/DOF        &           1.7/19  &             8.2/18 \\
DM (pc\,cm$^{-3}$)  &  < 68.45          &       54.7(9)      \\
RM (rad\,m$^{-2}$)  &                   &       $+$14.11(4)  \\
\hline
Gl (deg)            &    308.115        &     282.878        \\
Gb (deg)            &   16.985          &     $-$20.320      \\
$P_0$ (s)           &   1.870908(5)     &    3.231403(5)     \\
\hline
\end{tabular}
    \end{threeparttable}
    \label{tab:craco_newpsr}
\end{table}

\textit{PSR~J1319$-$4536/MTP0023} was first discovered in the MeerTRAP (\textit{More} TRAnsients and Pulsars, \citealt{2020SPIE11447E..0JR}) survey using the MeerKAT telescope, and independently discovered in CRACO-PS later.
It was rediscovered on 12 August 2023 during commensal observations with the EMU project (SB51948, Beam 28). The source was detected at R.A. 13h19m49.7s Decl. -45d36m13s with a SNR of 10.9 and a DM of 0\,pc\,cm$^{-3}$(i.e., $<68.45$\,pc\,cm$^{-3}$)\footnote{All values are reported by the pipeline.}.
We created the filterbank of PSR~J1319$-$4536 with the full 10-hour observation with all available antennas. We detected 20 single pulses with SNR larger than 8, and measured a frequency of 0.534500(2)\,Hz based on the ToAs using \textsc{pint}.
Its coordinates, DM, and period are all consistent with the MeerTRAP discovery (J. D. Turner, private communication). A detailed analysis of PSR~J1319$-$4536 will be given in J. D. Turner et al., (in prep.).

\textit{PSR~J0755$-$7026} was discovered on 15 September 2023 during commensal observations with the EMU project (SB53201, Beam 09). The source was detected at R.A. 07h55m42.3s Decl. $-$70d26m59.0s with a SNR of 12.8 at a DM of 41.9\,pc\,cm$^{-3}$ with a boxcar width of 1. 
% We imaged the visibility data with \texttt{CASA} and measured an updated position of R.A. 07h55m43.2s Decl. $-$70d27m01.0s. 
We got more precise coordinates for PSR~J0755$-$7026 using the post-processing script and report these in Table~\ref{tab:craco_newpsr}.  
Besides the discovery pulse, there were another four weak single pulses over the 90-minute observations.

We conducted a follow-up observation of PSR~J0755$-$7026 with the 64-m Parkes/Murriyang telescope, on 30 September 2023 (project code PX113) with the Ultra-Wideband Low receiver \citep{2020PASA...37...12H} with full Stokes parameters along with a noise diode observation.
We searched for single pulses using \textsc{heimdall-fetch} based multi-tiered sub-band search pipeline \citep{2012PhDT.......465B,2021MNRAS.500.2525K} with the DM searching up to 100\,pc\,cm$^{-3}$. We identified 20 single pulses in the 5-hour observation.
% par file - /fred/oz002/rshannon/craco/rrat/rrat_fit.par
We generated the ToAs based on the \textsc{heimdall-fetch} search pipeline output and fitted the rotation frequency $F_0$ using \textsc{pint} based on them.
PSR~J0755$-$7026 is a relatively slow spinning pulsar, with a spin period of 3.23\,s
% We derived the spin period of PSR~J0755$-$7026 to be 3.23\,s.

\begin{figure}[htb!] 
    \centering
    \begin{subfigure}[b]{0.49\linewidth}
    \centering
        \includegraphics[width=\textwidth]{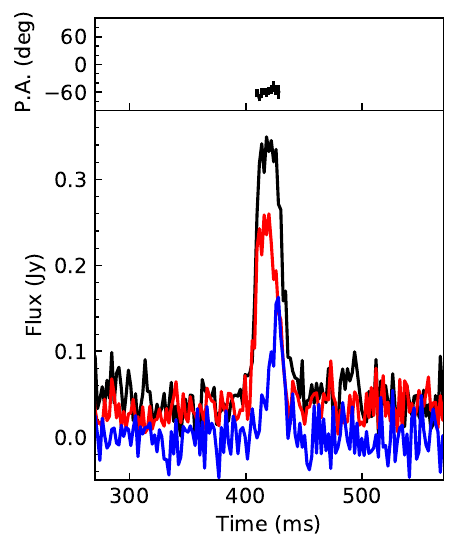}
    \end{subfigure}
    \hfill
    \begin{subfigure}[b]{0.49\linewidth}
    \centering
        \includegraphics[width=\textwidth]{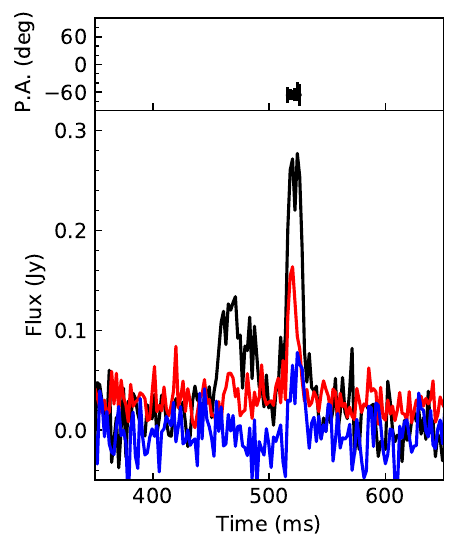}
    \end{subfigure}
    \caption{
    Examples of polarised pulse profiles of two types of single pulse from PSR~J0755$-$7026 at 1.45\,GHz after RM corrections. Left and right plots show the pulse profile for a single-peak pulse and a double-peak pulse respectively. We show the total intensity, linear polarisation and circular polarisation in black, red, and blue lines, respectively. The top panels show the polarisation position angle (P.A.) variation for each pulse. The uncertainties of P.A. are plotted in 3$\sigma$.
    }
    \label{fig:J0755_7026_single_pulse}
\end{figure}

Based on the brightest single pulse detected at MJD 60217.741325, we measured a DM of 54.7\,pc\,cm$^{-3}$ using \texttt{pdmp}, which is consistent with the CRACO detection.
We performed polarisation calibration using the 2-min observation of a linearly polarised noise diode obtained at the start of the observation. We used PKS~B1934$-$638 as a flux density reference for the flux calibration. We used \textsc{psrchive}'s \texttt{pac} pulsar archive calibration program to perform the calibration. We then used the \texttt{rmfit} tool of \textsc{psrchive} to fit for the RM. We found a strong detection with a RM of +14.11\,rad\,m$^{-2}$. We found that the pulse was highly polarised (see Figure~\ref{fig:J0755_7026_single_pulse} left plot), with a linear polarisation fraction of $\sim$65\% and a circular polarisation fraction of $\sim$25\%.
%%% get polarisation fraction here
We noted that there was a known pulsar, PSR~J0750$-$6846, with a similar DM (54.6\,pc\,cm$^{-3}$), $\sim$1.7\,deg away from the detection position \citep{2005AJ....129.1993M, 2020MNRAS.496.4836S}. However, neither the RM nor the spin period of PSR~J0755$-$7026 is the same as that of PSR~J0750$-$6846. % We concluded that PSR~J0755$-$7026 was a new discovery.

Similar to some other RRaTs \citep[e.g.,][]{2023arXiv230602855Z}, single pulse properties of PSR~J0755$-$7026 varied from pulse to pulse. 
While the emission for most pulses was confined to the lower frequency range ($\lesssim$2\,GHz), several pulses have emission extending up to $\sim$3\,GHz.
The pulse profiles also changed over time. We could classify all pulses into two categories based on their morphology: single-peak pulses and double-peak pulses (see Figure~\ref{fig:J0755_7026_single_pulse}). The separations between two peaks are all around 50\,ms for those double-peak pulses. 
A detailed analysis of PSR~J0755$-$7026 will be presented in future work.

\subsection{Ultra-long Period Objects}

\begin{figure}[htb!]
    \centering
    \includegraphics[width=0.55\textwidth]{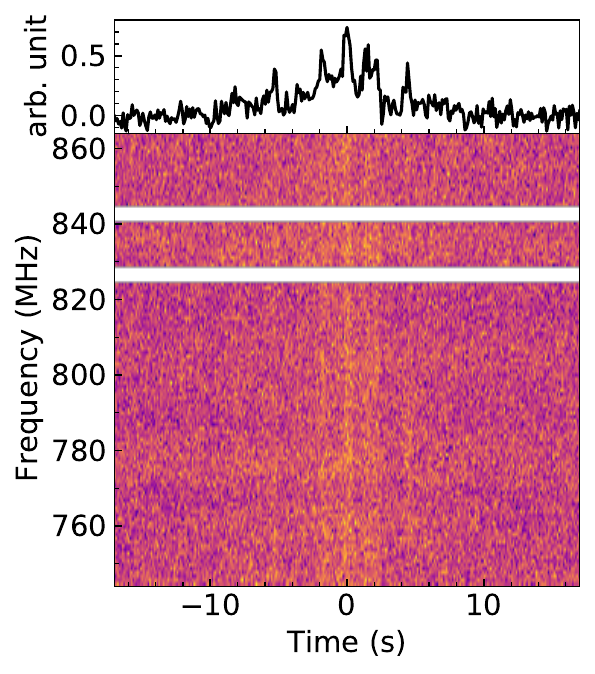}
    \caption{CRACO filterbank plot for GPM~J1839$-$10. We show the pulse profile on the top and the dynamic spectrum for the pulse on the bottom. All data were dedispersed at DM=273.5\,pc\,cm$^{-3}$.}
    \label{fig:GPMJ1839-10_fil}
\end{figure}

\citet{2022Natur.601..526H, 2023Natur.619..487H, 2024NatAs.tmp..107C, 2024arXiv240707480D} recently discovered four ultra-long period objects (ULPOs), whose nature has not been fully understood.
Although their main pulses (with timescales of minutes) are too wide to be detected in the CRACO pipeline, the pipeline is still able to detect them through their millisecond-timescale sub-pulse structures.
On 2023 April 28 (UT), we conducted a three-hour observation (SB49744) targeting one ULPO, GPM~J1839$-$10. Pulses from GPM~J1839$-$10 were detected \corrs{in an untargeted search} by the pipeline\footnote{There was a bug in the pipeline causing it to report the wrong coordinates. We will not discuss the quality of the astrometry from the pipeline for this source.} (see Figure~\ref{fig:GPMJ1839-10_fil}). 
The pipeline reported a DM of 172$\pm$57\,pc\,cm$^{-3}$, which is a bit off compared to its known DM of 273.5$\pm$2.5\,pc\,cm$^{-3}$ reported in \citep{2023Natur.619..487H}. This can be explained by the wide pulse profile, the low time resolution of the data, and a relatively narrow bandwidth (120\,MHz). 

\subsection{Scintillating Continuum Sources}

\begin{figure}[htb!] 
    \centering
    \begin{subfigure}[b]{0.49\linewidth}
    \centering
        \includegraphics[width=\textwidth]{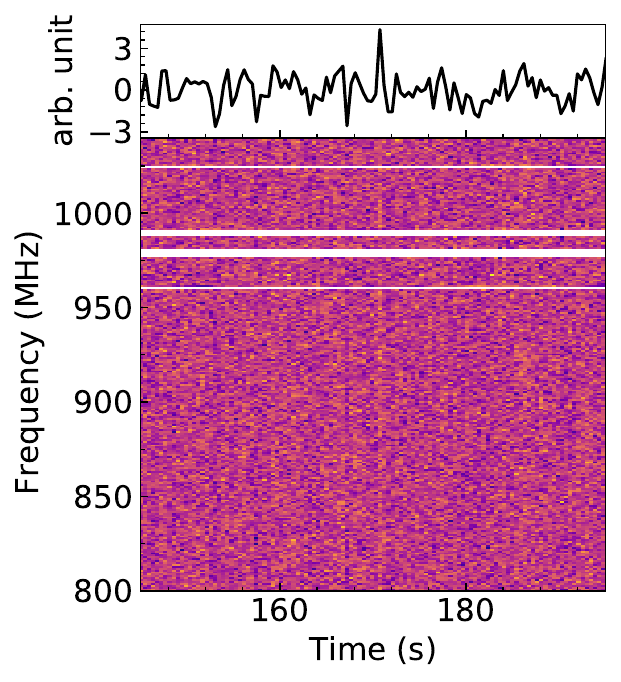}
    \end{subfigure}
    \hfill
    \begin{subfigure}[b]{0.49\linewidth}
    \centering
        \includegraphics[width=\textwidth]{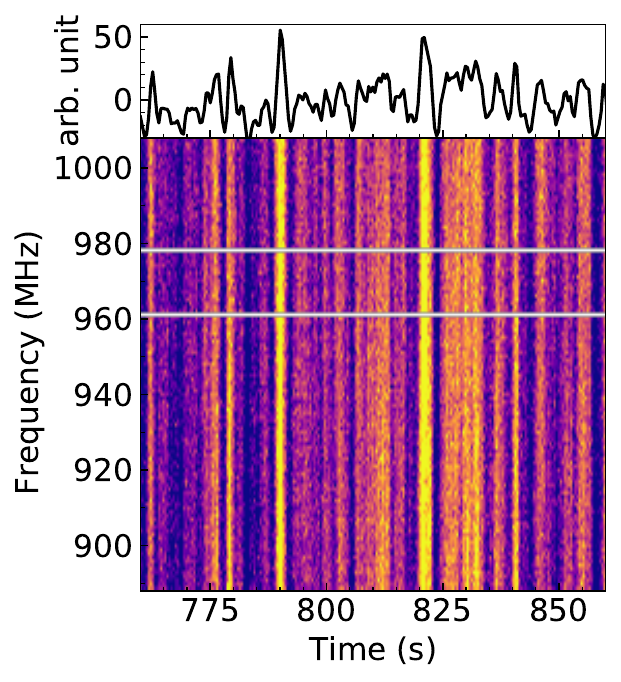}
    \end{subfigure}
    \caption{Filterbanks of example CRACO candidates caused by scintillation effects. We show the candidate caused by normal interstellar scintillation on the left and by interplanetary scintillation on the right.}
    \label{fig:scint_filt}
\end{figure}

At radio frequencies, any compact object (with an angular size $\lesssim 1\,$mas at GHz frequencies) can exhibit flux variations due to the effect of interstellar scintillation (ISS) caused by the inhomogeneities in the ionised interstellar medium (ISM) of the Milky Way.
The typical variation timescales range from minutes to months \citep[e.g.,][]{1972ApL....12..193H, 1998MNRAS.294..307W}.
For compact bright sources (e.g., radio-loud active galactic nuclei), the slow flux fluctuation caused by the ISS can be detected as candidates by the CRACO pipeline. 
Scintillation candidates usually show vertical stripes in their filterbanks.
We show this for one of the scintillation candidates we detected as an example in Figure~\ref{fig:scint_filt}. 

Besides the inhomogeneities in the ISM, solar activity (e.g., solar wind and coronal mass ejection) can also cause electron density irregularities in the interplanetary medium (IPM). 
Similar to the effect of the ISM, compact objects can show amplitude scintillation at radio frequencies due to the interplanetary scintillation (IPS). 
Compared to ISS, IPS is expected to have a flux variation on a shorter timescale (i.e., $\sim$ seconds) and a larger modulation index \citep[e.g.,][]{2018MNRAS.473.2965M}. 
We show the filterbank from an IPS candidate in Figure~\ref{fig:scint_filt} in the right panel, where a more dramatic flux variation can be seen compared to the normal ISS shown in the left panel. 

In the pilot survey, we detected one scintillating continuum source in every five fields on average.
Searching for scintillators is not the primary science goal for CRACO real time search, so as discussed already in Section~\ref{sec:system}, we classified all candidates that are spatially associated with a known RACS source brighter than 0.3\,Jy as known sources.
This criterion works for most observations except those close to the Sun (i.e., within $\sim30\,$deg).
In these observations, relatively faint continuum sources ($\sim$50\,mJy) could also show huge flux variations due to the IPS and lead to candidate detections in the CRACO pipeline. 
These candidates can be easily identified in the human inspection step.

\section{Discussion}\label{sec:dis}

\subsection{Inferred FRB all-sky rates}

We first estimated the flux densities of events in our survey based on a modified version of the interferometry sensitivity equation \citep{2017isra.book.....T} 
\begin{equation}
    \overline{S}_{\rm peak}\left({\rm S/N}, N, W_{\rm obs}\right) = \cfrac{{\rm S/N}}{\sqrt{N(N-1)}} \cfrac{{\rm SEFD}}{\sqrt{N_{\rm pol}\Delta\nu W_{\rm obs}}}
\end{equation}
where $S_{\rm peak}$ is the peak flux density, $N$ is the number of the antennas, $W_{\rm obs}$ is the observed width, $N_{\rm pol}$ is the number of polarisations summed, and $\Delta\nu$ is the observing bandwidth. 
The observed width is given by $W_{\rm obs} = \sqrt{W^2 + W_{\rm disp}^2 + W_{\rm int}^2}$, where $W$ is the intrinsic pulse width, $W_{\rm disp}$ is the smearing time, and $W_{\rm int}$ is the time resolution.
% $a_{\rm beam}$ is the attenuation factor of the primary beam, 
In Table~\ref{tab:craco_frb}, we show the inferred fluence limits $\mathcal{F} = S_{\rm peak}W_{\rm obs} / \eta$ for the FRBs we detected, where we used the SEFD estimated in Section~\ref{subsec:sefd}, and post-processing refined measured values S/N, $W_{\rm obs}$ to derive the fluence limits.
The response factor $\eta$ is a function of pulse width due to scalloping. The response factor for unresolved, or barely resolved pulses (i.e., those detected with only one or two time sample width) can be $\sim60-70\%$ \citep[e.g.,][]{2023MNRAS.523.5109Q}

\begin{figure}[htb!]
    \centering
    \begin{subfigure}[b]{\linewidth}
        \centering
        \includegraphics[width=0.95\textwidth]{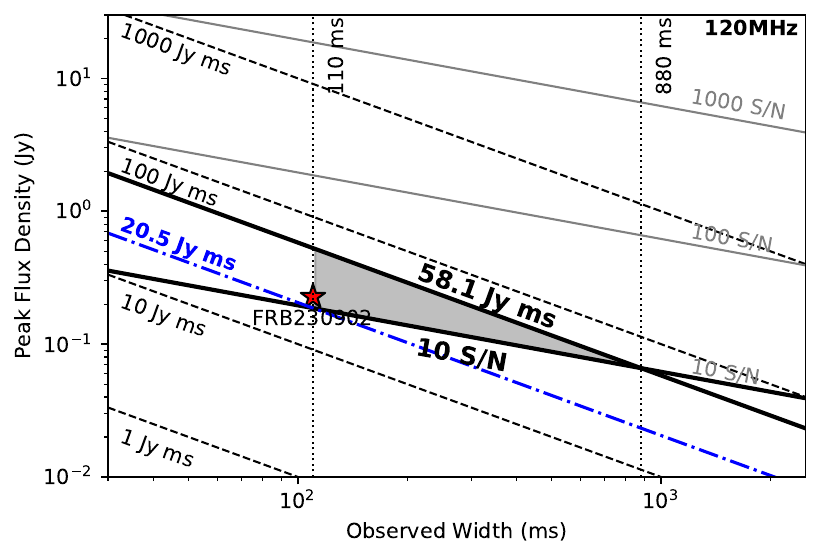}
        % \caption{Interferometric Detection Images of FRB20230902A}
        % \label{fig:frb230902_interimg}
    \end{subfigure}
    \vspace{1em}
    \begin{subfigure}[b]{\linewidth}
        \centering
        \includegraphics[width=0.95\textwidth]{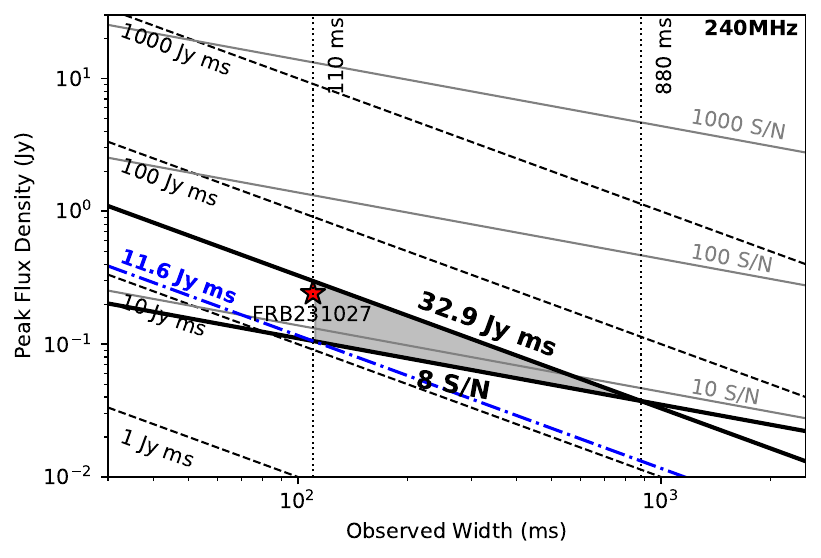}
        % \caption{Interferometric Detection Images of \cracofrb}
        % \label{fig:frb231027_interimg}
    \end{subfigure}
    \caption{Peak flux density versus observed width for two phases of CRACO-PS. Solid gray lines represent lines of constant S/N and dashed black lines represent lines of constant fluence. The range of widths searched is enclosed by the dashed vertical lines. The lowest fluence limits of the survey are shown with blue dash-dotted lines. We highlight the search S/N threshold and the fluence completeness threshold curves with thick lines. The shaded region is the fluence incomplete region. We show the parameters of the two FRBs presented in this work as red stars.}
    \label{fig:survey_perform}
\end{figure}

Following \citet{2015MNRAS.447.2852K}, we estimated the CRACO-PS survey fluence completeness threshold $\mathcal{F}_c$ (see Fig~\ref{fig:survey_perform}). We derived the fluence completeness threshold for two phases with the parameters listed in Table~\ref{tab:survey_param} respectively. We assumed that pulse widths would be less than 880\,ms, the widest pulse we can detect. 
During the search, we used data from the inner 23 antennas that were available. To be conservative, we used $N=20$ to calculate the fluence limits to account for the bad antennas during the observation.
This corresponds to fluence completeness for bursts less than 880ms duration of 58.1\,Jy\,ms and 32.9\,Jy\,ms for the survey phase 1 and phase 2, respectively.
For bursts at 110\,ms duration and below, sensitivity is uniform, with a completeness threshold of 20.5\,Jy\,ms and 11.6\,Jy\,ms for the phase 1 and phase 2, resepctively.

\begin{figure}[htb!]
    \centering
    \includegraphics[width=\textwidth]{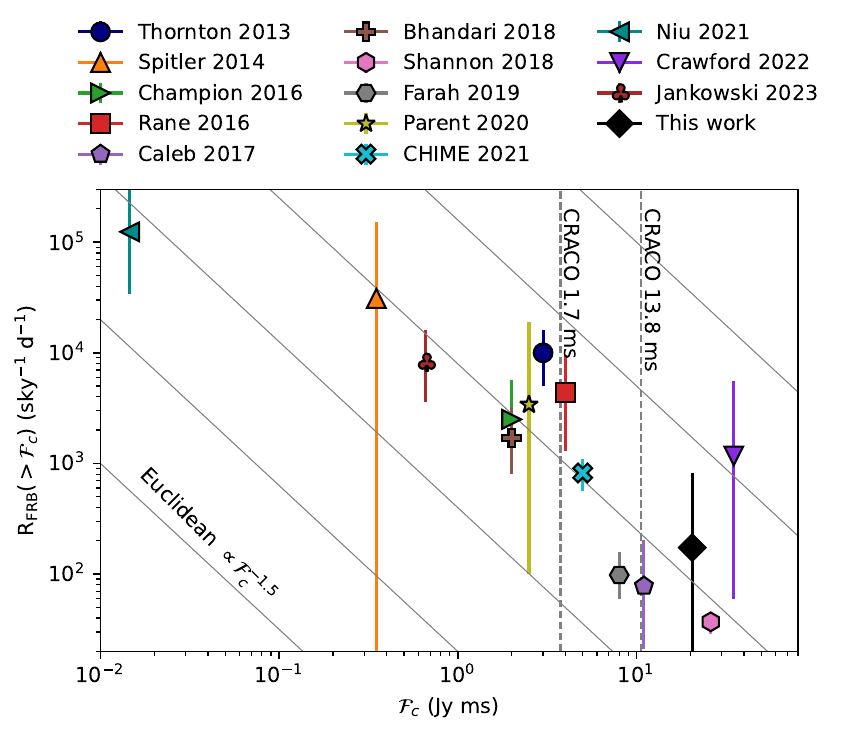}
    \caption{FRB all-sky rates as a function of fluence inferred from CRACO-PS and literature surveys (see references below). The black diamond denotes the upper limit of the rate inferred from CRACO-PS. Gray solid lines show the relation between rate and fluence threshold under the Euclidean distribution. We also show the expected fluence completeness threshold for further CRACO (data recorded at a time resolution of 13.8\,ms and 1.7\,ms) in gray dashed lines. Note that the inferred all-sky FRB rate for Crawford 2022 was re-estimated (see the discussion in the main text). 
    \textbf{References:} Thornton 2013 \citep{2013Sci...341...53T}, Spitler 2014 \citep{2014ApJ...790..101S}, Champion 2016 \citep{2016MNRAS.460L..30C}, Rane 2016 \citep{2016MNRAS.455.2207R}, Caleb 2017 \citep{2017MNRAS.468.3746C}, Bhandari 2018 \citep{2018MNRAS.475.1427B}, Shannon 2018 \citep{2018Natur.562..386S}, Farah 2019 \citep{2019MNRAS.488.2989F}, Parent 2020 \citep{2020ApJ...904...92P}, CHIME 2021 \citep{2021ApJS..257...59C}, Niu 2021 \citep{2021ApJ...909L...8N}, Crawford 2020 \citep{2022MNRAS.515.3698C}, and Jankowski 2023 \citep{2023MNRAS.524.4275J}}
    \label{fig:frb_rate}
\end{figure}

We calculated the inferred FRB all-sky rates ($\mathcal{R}_{\rm FRB}$) for phase 1 and phase 2 separately. 
During phase 1, the search volume is 4688\,deg$^2$\,hr and we detected no FRB. This corresponds to a rate upper limit of
$$
\mathcal{R}_{\rm FRB}\left(\mathcal{F} > 58.1\,{\rm Jy}\,{\rm ms}\right) < 6.32\times10^2\,{\rm events}\,{\rm sky}^{-1}\,{\rm d}^{-1}
$$
where we used a Poisson upper limit at the 95 per cent confidence level given a non-detection \citep{1986ApJ...303..336G}.
During phase 2, we detected one FRB with a search volume of 2452\,deg$^2$\,hr. Based on this, we estimate a detectable event rate of
$$
\mathcal{R}_{\rm FRB}\left(\mathcal{F} > 11.6\,{\rm Jy}\,{\rm ms}\right) = 4.03_{-3.82}^{+19.15}\times10^2\,{\rm events}\,{\rm sky}^{-1}\,{\rm d}^{-1}
$$
at a 95 per cent confidence level. 
% This rate is given as a lower limit as our phase 2 survey is complete for a fluence above 32.9\,Jy\,ms in the fluence-width plane. 
We detected no FRB above the fluence completeness threshold, which gives us a rate upper limit of
$$
\mathcal{R}_{\rm FRB}\left(\mathcal{F} > 32.9\,{\rm Jy}\,{\rm ms}\right) < 1.21\times10^3\,{\rm events}\,{\rm sky}^{-1}\,{\rm d}^{-1}.
$$
Other FRB detections are mostly sensitive to bursts at less than 110\,ms duration. Hence, we calculate our rate estimate for the whole CRACO-PS based on our completeness thresholds of 20.5\,Jy\,ms to bursts with a width of 110 ms and less. We use an assumed Euclidean distribution to scale down the effective survey area for phase 2 by a factor of $\left(\mathcal{F}_2/\mathcal{F}_1\right)^{3/2} \approx 0.42$. This gives as a limit on the rate of FRBs of duration $\leq$110\,ms of 
\begin{equation}
    \label{eq:frb_rate}
    \mathcal{R}_{\rm FRB}\left(\mathcal{F} > 20.5\,{\rm Jy}\,{\rm ms}\right) = 1.72_{-1.64}^{+6.47}\times10^2\,{\rm events}\,{\rm sky}^{-1}\,{\rm d}^{-1}
\end{equation}
, which is largely consistent with FRB survey rates in the literature (see Figure~\ref{fig:frb_rate}).

Only a small fraction ($\lesssim2\%$) of FRBs have been observed with a pulse width $\gtrsim$100\,ms \citep[e.g.,][]{2022A&ARv..30....2P}.
% The lack of long FRB detections is intriguing. 
Hunting for these long FRBs can help us map the baryons in the Universe and better characterise the FRB population.
Wide pulse widths can be caused by the propagation effects such as scattering \citep[e.g.,][]{2021ApJ...911L...3P}, or they could be intrinsic to the emission mechanism \citep[e.g.,][]{2022Natur.607..256C}. Detections of highly scattered FRBs can help in measuring the physical properties of the scattering material and constraining the DM budget, which can be used to estimate the FRB redshifts \citep[e.g.,][]{2022ApJ...931...87O}.
The lack of long FRB detections is intriguing. It can be explained by the assumption that the intrinsically long FRBs only represent a small fraction of the entire population, but biases in previous surveys can also explain it.
Most surveys were conducted at a time resolution of order $\sim 1\,$ms (e.g., 0.983\,ms for CHIME \citep{2018ApJ...863...48C}, 0.864\,ms for ASKAP-CRAFT \citep{2019Sci...365..565B}, and 0.306\,ms for MeerKAT-MeerTRAP \citep{2022MNRAS.514.1961R}), and only searched for pulses with a width of a few dozen samples (i.e., $\lesssim50\,$ms), and therefore are insensitive to wide pulses.
\citet{2021ApJS..257...59C} also conceded that a huge number of wide FRBs (especially those with a width beyond 100\,ms, either intrinsic or highly scattered) could be missed by the current CHIME FRB detection pipeline \corrs{based on sensitivity testing with injections}.

There have been a few attempts to search for long FRBs. For example, \citet{2019ARep...63...39F, 2019ARep...63..877F} found 12 FRB candidates, all of which had a pulse width $\gtrsim300\,$ms, at 111\,MHz with the Large Phased Antenna of the Lebedev Physical Institute with the data recorded at 100\,ms time resolution. However, it is difficult to confirm the astrophysical origin of these candidates due to the narrow bandwidth (2.5\,MHz over 6 channels) and low SNR reported for the events.
\citet{2022MNRAS.515.3698C} found four FRB candidates in a search of the Parkes 70-cm pulsar survey archival data at a time resolution of 0.3\,ms. They searched for pulsed signals with a boxcar width ranging from 1 to 512 samples (i.e., $\sim150\,$ms). All four candidates were detected with a pulse width exceeding $\sim50\,$ms. However, we consider three candidates to be marginal because of either RFI-like patterns or low SNR estimations. 
Based on this, we re-estimated the inferred all-sky FRB rate from this survey to be $\mathcal{R}_{\rm FRB}\left(\mathcal{F} > 35\,{\rm Jy}\,{\rm ms}\right) = 1.2^{+4.4}_{-1.1}\times10^3\,{\rm events}\,{\rm sky}^{-1}\,{\rm d}^{-1}$.
The most promising one, FRB~19910730A, was detected with a pulse width of 113.4\,ms. \citet{2022MNRAS.515.3698C} argued that the wide pulse from FRB~910730 was unlikely to be caused by scattering, and could be intrinsic.
In Figure~\ref{fig:frb_rate}, we notice that the inferred FRB rate based on \citet{2022MNRAS.515.3698C} is about an order of magnitude higher compared to other surveys (assuming a Euclidean distribution), which indicates that long FRBs may have a different origin to the short FRB population. 

It is important to realise however that FRB thresholds quoted at different FRB durations $t_{\rm FRB}$ must be treated with caution. For an experiment searching for FRBs, the fluence threshold up to a minimum time resolution $t_{\rm res}$ is constant with $t_{\rm FRB}$; it then decreases as $t_{\rm FRB}^{-0.5}$ up to a maximum search time window $t_{\rm max}$; and then decreases as $t_{\rm FRB}^{-1}$ above that. This makes the rate dependent on the FRB width distribution. This has been modelled by both \citet{James2022Meth}, who analyse the total (intrinsic plus scattered) width distribution of FRBs observed by ASKAP and Parkes; and by \citet{2021ApJS..257...59C}, who separately model the intrinsic and scattered widths. Both use log-normal distributions, finding the mean and standard deviation ($\mu_w$,$\sigma_w$) of $\log w\,[{\rm ms}]$ to be (1.7, 0.9) and (0.7, 1.72) respectively.  Assuming the rate dependence of equation~(\ref{eq:frb_rate}), we find that a CRACO-PS-like experiment searching for FRBs between 110\,ms and 880\,ms, with threshold $\mathcal{F}_0$ at $110$\,ms, should detect 2.5 (1.5) times as many FRBs assuming $\mu_w$, $\sigma_w$=1.7, 0.9 (0.7, 1.72) as a CRAFT/ICS-like experiment with the same threshold $\mathcal{F}_0$ at $t_{\rm res}=1.182$\,ms, searching up to $12 \times 1.182 = 14.184$\,ms. This estimate ignores DM-smearing, which increases the observed width, and would further increase the rate of experiments with long integration times, relative to those short integration times.
While our measured rate is significantly more than a factor of 1.5--2.5 above the trend from experiments search at ms scales evident from Figure~\ref{fig:frb_rate}, the large errors are still consistent with that rate.
%\textcolor{red}{Make conclusive statement once detection point is placed on above figure}

To better understand the long FRBs and model the full FRB population, FRB surveys that are sensitive to pulse widths from a hundred milliseconds to seconds are needed.
% say something about CRACO-PS
CRACO-PS can search for pulses lasting $\sim$seconds. Compared to \citet{2019ARep...63...39F} and \citet{ 2019ARep...63..877F, 2022MNRAS.515.3698C}, CRACO-PS has a much wider bandwidth ($>100\,$MHz) and the capacity of recording visibility data, which enables us to rule out most non-astrophysical signals and localise the events.
Although the FRB we detected in CRACO-PS was a short one (16\,ms), we can still put a limit on the long FRB event rates (see discussions above). The rate is consistent with that inferred from \citet{2022MNRAS.515.3698C} but not very constraining.
Further similar surveys can be used in estimating a more robust all-sky long FRB event rate and therefore can be beneficial in modelling the full FRB population and understanding its progenitors.
%%%% further SKA - from microseconds to second

% Due to the relatively low time resolution compared to , CRACO-PS is more 
% Most FRBs were detected with a pulse width of $\lesssim 10\,$ms \citep[e.g.,][]{2021ApJS..257...59C}.
% However, this may caused by the limited parameter space previous FRB surveys searched. Typical FRB surveys were conducted at a time resolution of an order of $\lesssim 1\,$ms (\todo{add some examples here}), and conservatively were sensitive to the pulse width $\lesssim 50\,$ms (\todo{add some citations}).
% The expected detection rate with a 110-ms time resolution system is low (but with a large uncertainty).
% We estimated a detection rate of \todo{XXX} according to the four large-width FRBs detected in \citet{2022MNRAS.515.3698C}.

\subsection{Long period pulsars and RRATs Discoveries}

Most pulsars have been discovered through periodicity searches on time series data.
However, these searches are not sensitive to pulsars with exotic and extreme properties, such as high intermittency, and long periodicity.
Many of the recently discovered RRATs and long period pulsars (i.e., with a spin period larger than 10\,s) are first identified in single pulse searches \citep[e.g.,][]{2010MNRAS.401.1057K, 2015ApJ...809...67K, 2016ApJ...821...10D, 2018ApJ...866...54T, 2022NatAs...6..828C}.
Though CRACO performed searches on synthesised images, it works equivalently as a single pulse searching machine on $256\times256$ coherent tied-array beams.

CRACO discovered two new RRATs during $\sim$250 hrs of CRACO-PS observation.
There are $\sim$400 RRATs discovered so far (see \texttt{RRATalog}\footnote{See \url{https://rratalog.github.io/rratalog/} and \url{https://github.com/David-McKenna/RRATCat}.}) with $\sim$20 years of searching for RRATs since their discovery in 2005 \citep{2006Natur.439..817M}.
Other single pulse searches with wide field-of-view instruments have resulted in a large number of discoveries as well. For example, \citet{2021ApJ...922...43G} and \citet{2023MNRAS.524.5132D} detected 28 new pulsars (including 18 RRATs) with CHIME, and \citet{2022MNRAS.512.1483B} detected 12 new Galactic transients (include 7 confirmed pulsars) with MeerKAT.

Compared to the normal pulsar surveys with single-dish telescopes, CRACO has a much wider instantaneous field of view (e.g., a factor of 50 compared to the Parkes MultiBeam Pulsar Survey), which is useful in finding previously missed pulsars.
Most pulsar surveys only target regions that are believed to have more pulsars, for example, the low Galactic latitude region (i.e., $|b|\lesssim 5\,$deg), globular clusters, and the Magellanic Clouds.
There have been two all-sky pulsar surveys using single dish telescopes: the High Time Resolution Universe North \citep[HTRU-N;][]{2013MNRAS.435.2234B} survey with the Effelsberg 100-m Radio Telescope, and the High Time Resolution Universe South \citep[HTRU-S;][]{2010MNRAS.409..619K} survey with the Parkes/Murriyang 64-m telescope.
These surveys were limited in the amount of time per pointing because of the huge number of pointings to cover the sky. 
There are 218\,371 and 42\,641 pointings for HTRU-N and HTRU-S, respectively.
As a comparison, only 1\,493 tiles are needed to cover the whole visible sky with ASKAP using the \texttt{closepack36} footprint \citep[e.g.,][]{2023PASA...40...34D}. 
For single-dish pulsar surveys, those with limited coverage can easily miss pulsars at high galactic latitude, and those with full sky coverage can miss pulsars with a significant nulling fraction (e.g., intermittent pulsars), and faint pulsars with only a few bright pulses.
Single pulse searches with wide field-of-view instruments (e.g., CHIME, ASKAP, LOFAR, and MeerKAT) offer us opportunities to find previously missed pulsars, which can help us better understand the \corrs{underlying pulsar} population.

We quantified the potential new detections following the same method described in \citet{2021ApJ...922...43G}. We first used the pulsar population synthesis package \textsc{PsrPopPy} \citep{2014MNRAS.439.2893B} to simulate the pulsar population using the default parameters with 1038 pulsar detections in the Parkes Multi-Beam Pulsar Survey \citep[PKMBS;][]{2001MNRAS.328...17M}. For a specific pulsar, we drew a series of single pulse luminosities from a lognormal distribution \citep[e.g.,][]{2018MNRAS.479.5413M}, 
\begin{equation}
    p\left(L\right) = \cfrac{1}{L\sigma_L\sqrt{2\pi}}\exp\left(-\cfrac{1}{2}\cfrac{\left(\ln L - \mu_L\right)^2}{\sigma_L^2}\right)
\end{equation}
where the mean luminosity $\mu_L$ is the value from the \textsc{PsrPopPy} simulation, and the scale parameter $\sigma_L$ is drawn from a normal distribution \citep[e.g.,][]{2018MNRAS.479.5413M},
\begin{equation}
    f\left(\sigma_L\right) = \cfrac{1}{\sigma_{\sigma_L}\sqrt{2\pi}}\exp\left(-\cfrac{1}{2}\cfrac{\left(\sigma_L - \mu_{\sigma_L}\right)^2}{\sigma_{\sigma_L}^2}\right).
\end{equation}
We calibrated $\mu_{\sigma_L}$ and $\sigma_{\sigma_L}$ with the PKMBS single-pulse detections from 278 sources \citep{2018MNRAS.479.5413M}, and the result was consistent with \citet{2021ApJ...922...43G}.
The number of single pulses in the series was calculated as $N = T_{\rm obs}/P$, where $T_{\rm obs}$ is the observation duration for each pointing, and $P$ is the pulsar period. We used $T_{\rm obs}=30\,$mins for PKMBS to calibrate, and $T_{\rm obs}=400\,$mins for CRACO to estimate the expected detections in two years\footnote{We assumed that ASKAP would observe the whole observable sky equally with a 50\% efficiency in two years.}.
The sky temperature was retrieved from the global diffuse sky model described in \citet{2008MNRAS.388..247D} using the \textsc{PyGDSM} package\footnote{\url{https://github.com/telegraphic/pygdsm}} \citep{2016ascl.soft03013P}.
In the simulation, we estimated that we could detect single pulses from $\sim$3000--4000 objects (including on the order of 2500--3000 known pulsars) in a two-year time span.  
We note that this number is a very rough estimation because of the limited knowledge of the single-pulse statistics and the underlying pulsar population, and the approximation on the ASKAP observing time, but it does show a huge potential of discovering more new pulsars with final CRACO in the next few years.

\subsection{ULPO Discoveries}

Ultra-long period objects (ULPOs) are radio transients detected with a period significantly longer than that of the known pulsar population. There have only been six ULPOs discovered so far: GLEAMX~J162759.5--523504.3 with a $\sim$18~minute period \citep{2022Natur.601..526H}; GPM~J1839$-$10 with period of $\sim$21 minutes \citep{2023Natur.619..487H}; ASKAP~J1935$+$2148 with a $\sim$54~minute period \citep{2024NatAs.tmp..107C}; CHIME~0630+25 with a $\sim$7-minute period \citep{2024arXiv240707480D}; \corrs{ILT~J1101+5521 with a $\sim$126-minute period \citep{2024arXiv240811536D}; and GLEAM-X~J0704$-$37 with a $\sim$2.9-hour period \citep{2024arXiv240815757H}}. 
All of them were discovered through general image domain transient searches. 
Both GLEAMX~J162759.5--523504.3 and GPM~J1839$-$10 show millisecond to second timescale sub-pulse structures \citep{2022Natur.601..526H, 2023Natur.619..487H} which could be detectable by single-pulse searches, and indeed
%machine like CRACO. 
the blind detection of sub-pulse structures within GPM~J1839$-$10 showed the capability of finding ULPOs with CRACO.

The nature of ULPOs is still not clear, and several possible models have been suggested, including magnetars \citep{2023MNRAS.520.1872B, 2024arXiv240604135C}, binary neutron star systems \citep{2005ApJ...628L..49T}, and white dwarf pulsars \citep{2005ApJ...631L.143Z, 2022Ap&SS.367..108K, 2024arXiv240905978Q}. The ULPOs may in fact span multiple object classes. 
Discoveries of more of these systems are vital for understanding their nature and emission mechanism. 
While CRACO can detect ULPOs through their sub-pulse structures,
the ASKAP image-domain transient search project, VAST, can also detect ULPOs but mainly via their second- to minute-timescale pulse structures and polarised emission \citep{2021PASA...38...54M, 2023MNRAS.523.5661W}.

A combination of CRACO and VAST will be an excellent machine for discovering and initially characterising new ULPOs.
CRACO can detect radio emission from ULPOs in (near) real-time with an accurate measurement of dispersion measure.
VAST can provide polarisation information of the source, and we can use multiple detections to measure the periodicity of the source if there is any.
All this information is helpful in planning and scheduling multi-wavelength follow-up observations in a timely manner as the radio emission from ULPOs may turn off abruptly; for example, GLEAMX~J162759.5--523504.3 was only active for 3 months \citep{2022Natur.601..526H}.

\subsection{Further CRACO developments}

The final goal for the CRACO system is to search for dispersed signals from the data at 1.7\,ms resolution in real time. 
The CRACO-PS survey used our pilot computing cluster \texttt{seren} and search pipeline version 1.0\footnote{\url{https://github.com/askap-craco/craco-python/releases/tag/v1.0}}, which were about three times slower (i.e., it took three times the duration of the observation to search the data) at 110\,ms resolution.
A new cluster \texttt{skadi}, dedicated to CRACO operation, has been installed on site. \texttt{skadi} is equipped with 72 FPGA cards, about 3 times more than that of the pilot cluster. The FPGA card on \texttt{skadi} is more powerful than the card on \texttt{seren} (Alveo U55C card compared to Alveo U280 card).
With the new cluster \texttt{skadi} and a new version of the search pipeline, we are currently able to run the pipeline at 13.8\,ms resolution in real time and search the number of DM trails of 240 (DM value up to 1000\,pc\,cm$^{-3}$; as of Jun 2024). 
We are now working on improvements to the search pipeline to make it fast enough for higher time resolutions, and resolving the known issues described in  Section~\ref{subsubsec:other_bugs}.

We are also planning for the next version of CRACO, CRACO 2.0, aiming at improving the search capability of the system as follows: 
\begin{itemize}
    \item {\bf Gridding function} - As discussed in Section~\ref{subsec:known_issue}, the simple pillbox gridding function can cause SNR loss of 60\% near the corners of the FoV. This can be improved by using a more complicated gridding function. Although implementing a new gridding function will eliminate aliased detections, using extra cards to cover the sky near the FoV edges will solve this issue.
    \item {\bf Field of view} - The current system can cover the half power beam width for all inner beams (1.7\,deg at 1\,GHz), but there are $\sim$7\,deg$^2$ of sky observable by ASKAP but not searched by the CRACO pipeline because of the 1.1\,deg$^2$ FoV we set in the search pipeline. In \texttt{skadi}, we have 36 extra FPGA cards available, which can be useful in searching the region near the edges of the FoV. 
    \item {\bf Number of antennas} - To balance the oversampling factor and FoV, we cannot use the data from the all inner 30 antennas CRACO records during the search. We can optimise the number of antennas being used in the search by simulating $uv$-coverage for sources at different hour angle and declination. 
    \item {\bf Maximum boxcar width} - The current system only supports searching for signals with a maximum boxcar width of 8, which can easily miss signals with a (potential) wide width such as intrinsically wide FRBs, super scattered FRBs, slow rotating pulsars, and sub-pulse structure from ULPOs. At the target 1.7\,ms time resolution, CRACO will only be sensitive to signals with a width of $\lesssim$20\,ms, but it will increase to $\sim$2\,s if we can increase the maximum boxcar width to 1024. For transients with a timescale larger than $\sim$2\,s, image domain searches (without de-dispersion or with very coarse de-dispersion) will be sufficient \citep[e.g.,][]{2021PASA...38...54M, 2023MNRAS.523.5661W}.
\end{itemize}

\section{Conclusions}\label{sec:conclusion}

CRACO, a coherent upgrade of the current ASKAP ICS system, searches for dispersed signals in the visibility data at millisecond time resolution.
We conducted a 110\,ms pilot survey (CRACO-PS) as part of the commissioning. During the 400\,hr survey, we detected one ICS-discovered FRB (FRB\,20230902A), and discovered one FRB (\cracofrb). The detection of \cracofrb\ gives us a detectable event rate of $\mathcal{R}_{\rm FRB}\left(\mathcal{F} > 20.5\,{\rm Jy}\,{\rm ms}\right) \gtrsim 1.72_{-1.64}^{+6.47}\times10^2\,{\rm events}\,{\rm sky}^{-1}\,{\rm d}^{-1}$, which is broadly consistent with rates estimated in the previous literature. 
We corrected the localisation for four previous poorly localised pulsars, and discovered two new RRATs by searching for single pulses. We also demonstrated the CRACO capability of detecting new ULPOs. 

A 13.8-ms search mode has already been achieved and is currently in operation (December 2023), after the 110-ms transient survey concluded.
At the target 1.7\,ms resolution, CRACO will be $\sim$5 times more sensitive than the previous ICS system, and therefore we expect to detect $\sim$100 localised FRBs within a year.
These discoveries will be crucial for our understanding of the FRB underlying population, the surrounding environment of the FRB progenitor, and probing the universe by constraining cosmology parameters such as the Hubble constant. 
CRACO is a powerful machine for detecting FRBs with arcsecond-level uncertainty in their localisations, extreme neutron stars (e.g., pulsars with a high nulling fraction, slow spin-period pulsars, and RRATs), and sources of an unknown nature such as ULPOs.

\begin{acknowledgement}

We are grateful to the ASKAP engineering and operations team for their assistance in developing fast radio burst instrumentation for the telescope and supporting the survey. 
This work uses data obtained from Inyarrimanha Ilgari Bundara / the CSIRO Murchison Radio-astronomy Observatory. We acknowledge the Wajarri Yamaji People as the Traditional Owners and native title holders of the Observatory site. CSIRO’s ASKAP radio telescope is part of the Australia Telescope National Facility (https://ror.org/05qajvd42). Operation of ASKAP is funded by the Australian Government with support from the National Collaborative Research Infrastructure Strategy. ASKAP uses the resources of the Pawsey Supercomputing Research Centre. Establishment of ASKAP, Inyarrimanha Ilgari Bundara, the CSIRO Murchison Radio-astronomy Observatory and the Pawsey Supercomputing Research Centre are initiatives of the Australian Government, with support from the Government of Western Australia and the Science and Industry Endowment Fund.
CRACO was funded through Australian Research Council Linkage Infrastructure Equipment, and Facilities grant LE210100107.

Murriyang, the Parkes radio telescope, is part of the Australia Telescope National Facility (https://ror.org/05qajvd42) which is funded by the Australian Government for operation as a National Facility managed by CSIRO.
We acknowledge the Wiradjuri people as the Traditional Owners of the Parkes Observatory site.
% The Australia Telescope Compact Array is part of the Australia Telescope National Facility (https://ror.org/05qajvd42) which is funded by the Australian Government for operation as a National Facility managed by CSIRO.
% We acknowledge the Gomeroi people as the Traditional Owners of the Paul Wild Observatory site.
Part of this work was based on observations collected at the European Southern Observatory under ESO programme 108.21ZF.
Some of the data presented herein were obtained at Keck Observatory, which is operated as a scientific partnership among the California Institute of Technology, the University of California, and the National Aeronautics and Space Administration. The Observatory was made possible by the generous financial support of the W. M. Keck Foundation. The authors wish to recognize and acknowledge the very significant cultural role and reverence that the summit of Maunakea has always had within the Native Hawaiian community. We are most fortunate to have the opportunity to conduct observations from this mountain. 

Part of this work was performed on the OzSTAR national facility at Swinburne University of Technology. The OzSTAR program receives funding in part from the Astronomy National Collaborative Research Infrastructure Strategy (NCRIS) allocation provided by the Australian Government, and from the Victorian Higher Education State Investment Fund (VHESIF) provided by the Victorian Government.

RMS, and YW acknowledge support through Australian Research Council Future Fellowship FT\,190100155.  RMS, ATD, AJ, YWJL, and TM acknowledge support through Australian Research Council Discovery Project DP\,220102305. MG and CWJ acknowledge support from the Australian Government through the Australian Research Council Discovery Project DP210102103.  
MG is also supported through UK STFC Grant ST/Y001117/1. For the purpose of open access, the author has applied a Creative Commons Attribution (CC BY) licence to any Author Accepted Manuscript version arising from this submission.
Parts of this research were conducted by the Australian Research Council Centre of Excellence for Gravitational Wave Discovery (OzGrav), through project numbers CE170100004 and CE230100016.
ACG and the Fong Group at Northwestern acknowledges support by the National Science Foundation under grant Nos. AST-1909358, AST-2308182 and CAREER grant No. AST-2047919.
ACG acknowledges support from NSF grants AST-1911140, AST-1910471 and AST-2206490 as a member of the Fast and Fortunate for FRB Follow-up team.
MC acknowledges support of an Australian Research Council Discovery Early Career Research Award (project number DE220100819) funded by the Australian Government.

This publication makes use of data products from the Wide-field Infrared Survey Explorer, which is a joint project of the University of California, Los Angeles, and the Jet Propulsion Laboratory/California Institute of Technology, funded by the National Aeronautics and Space Administration.

% ESO ackowledgement, if we use VLT data
% on FRB 20231027

\end{acknowledgement}

% \section{New sources}

% \begin{itemize}
%     \item MTP source
%     \item New RRAT candidate
%     \item Several previously undiscovered 8-sigma candidate
% \end{itemize}

% \section{System Performance}

% \subsection{Validation}
% injections
% known pulsars
% 1ms comparison

\bibliography{references}

\begin{appendix}

\section{Data Verification}

\begin{figure*}[h!]
    \centering
    \includegraphics[width=0.95\textwidth]{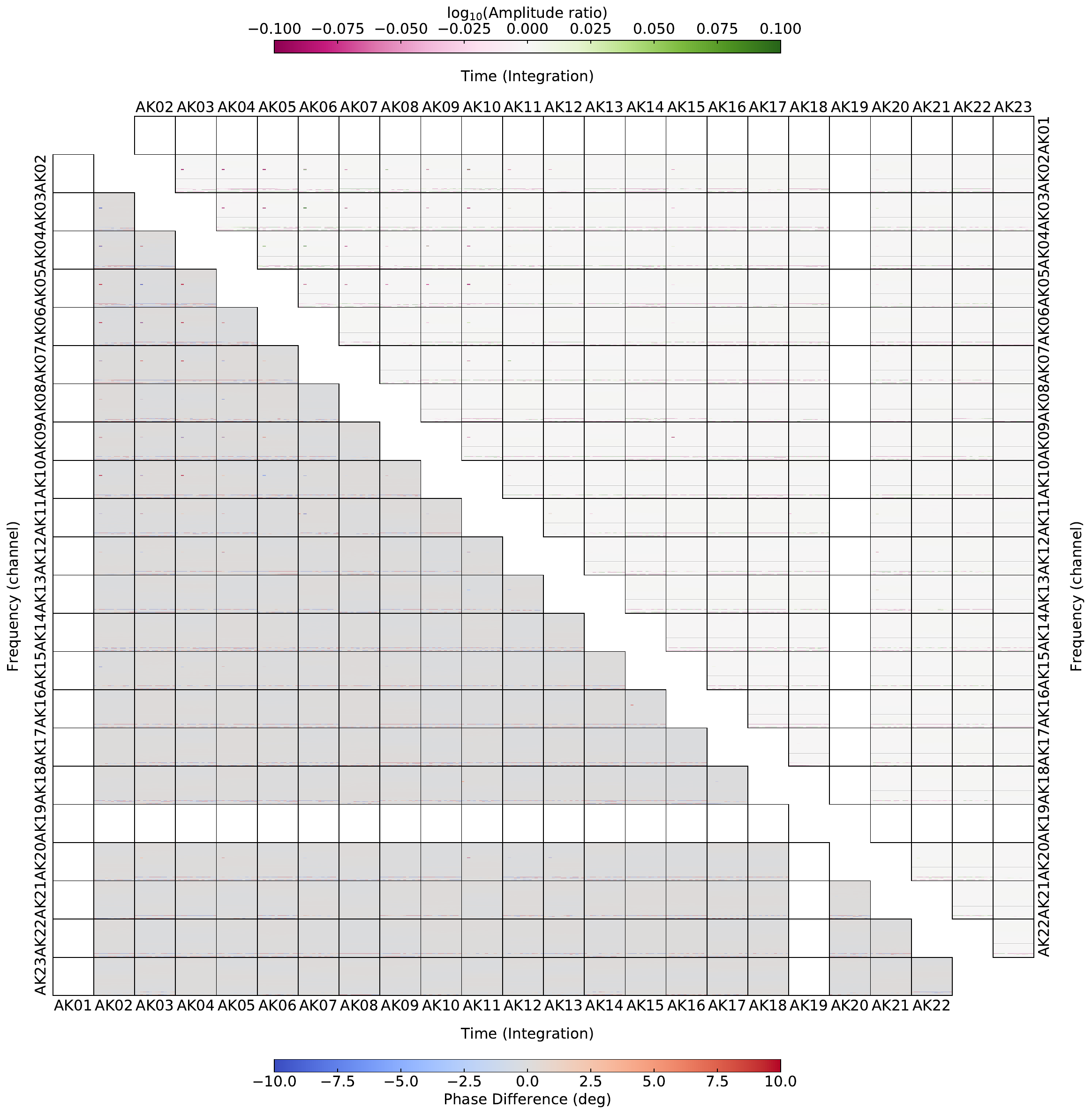}
    \caption{Amplitude and phase comparison between CRACO and ASKAP hardware visibility data. We averaged CRACO data by a factor of 90 to match the hardware time resolution. We show the phase differences and amplitude ratio between the two datasets in the lower left and upper right, respectively. For each small plot, the x-axis shows the time and the y-axis shows the frequency. Antenna AK01 and AK19 were flagged during the observation so therefore there are no data from them in either CRACO or ASKAP hardware data.} 
    \label{fig:craco_hw_comp}
\end{figure*}

The visibility data for both CRACO and general ASKAP hardware are averaged based on the same correlator output. 
We verified the CRACO data quality by averaging CRACO data to 9.95\,s time resolution and comparing it to the ASKAP hardware data.
As is shown in Figure~\ref{fig:craco_hw_comp}, the CRACO data was consistent with the hardware data. The reason for minor differences is the random packet loss during the data transposing in the CRACO recording system.

\section{Detected known pulsars}\label{sec:pulsar_cat}

We list the 42 known pulsars/RRaTs detected in CRACO-PS in Table~\ref{tab:cracopsr} with their measured properties reported by the CRACO pipeline and in the literature.
% The detection DMs are highly quantised due to the coarseness of the DM steps and the low time resolution of the survey (110 ms).

\begin{table*}[hbt!]
    \centering
    \caption{Known pulsars detected in the CRACO 110-ms pilot survey. The SBID and Beam columns indicate the observation information of the brightest single pulse detected with the CRACO pipeline. DM$_{\rm CRACO}$, R.A.$_{\rm CRACO}$, and Decl.$_{\rm CRACO}$ are the DM and coordinates reported by the pipeline. DM$_{\rm cat}$, R.A.$_{\rm cat}$, and Decl.$_{\rm cat}$ are the information from the previous work, which are retrieved from the ATNF pulsar catalogue unless specified.
    }
    \begin{threeparttable}
    \begin{tabular}{c c c c c c c c c}
\hline\hline
JName & SBID & Beam & DM$_{\rm CRACO}$$^\ast$ & DM$_{\rm cat}$ & R.A.$_{\rm CRACO}$$^\mathsection$ & Decl.$_{\rm CRACO}$$^\mathsection$ & R.A.$_{\rm cat}$ & Decl.$_{\rm cat}$ \\
 & & & (pc\,cm$^{-3}$) & (pc\,cm$^{-3}$) & (J2000) & (J2000) & (J2000) & (J2000) \\
\hline
J0031$-$5726$^\dagger$ & 54393 & 10 & $<34.53$ & 6.83$^\dagger$ & $00^{\rm h}31^{\rm m}35.3^{\rm s}$ & $-57^{\rm d}26^{\rm m}16.2^{\rm s}$ & ... & ... \\
J0343$-$3000 & 54281 & 33 & $<34.53$ & $20.2\pm0.3$ & $03^{\rm h}43^{\rm m}32.0^{\rm s}$ & $-30^{\rm d}00^{\rm m}24.3^{\rm s}$ & $03^{\rm h}43^{\rm m}27.9^{\rm s}\pm0.4^{\rm s}$ & $-30^{\rm d}00^{\rm m}27.5^{\rm s}\pm0.4^{\rm s}$ \\
%%%%%%%%%%%%%%%%%%%%%%%%%%%%%%%%%%%%%%%%%%%%%%%%%%
J0410$-$31 & 52470 & 26 & $<68.45$ & $9.2\pm3$ & $04^{\rm h}10^{\rm m}02.5^{\rm s}$ & $-31^{\rm d}10^{\rm m}38.0^{\rm s}$ & $04^{\rm h}10^{\rm m}39^{\rm s}\pm33^{\rm s}$ & $-31^{\rm d}07^{\rm m} \pm 7^{\rm m}$ \\
%%%%%%%%%%%%%%%%%%%%%%%%%%%%%%%%%%%%%%%%%%%%%%%%%%
J0820$-$4114 & 53273 & 18 & $69.06\pm34.53$ & $113.4\pm0.2$ & $08^{\rm h}20^{\rm m}15.9^{\rm s}$ & $-41^{\rm d}14^{\rm m}39.9^{\rm s}$ & $08^{\rm h}20^{\rm m}15.5^{\rm s}\pm0.2^{\rm s}$ & $-41^{\rm d}14^{\rm m}35.2^{\rm s}\pm0.1^{\rm s}$ \\
J0828$-$3417 & 51548 & 28 & $<68.45$ & $51.6\pm0.5$ & $08^{\rm h}28^{\rm m}16.4^{\rm s}$ & $-34^{\rm d}17^{\rm m}11.2^{\rm s}$ & $08^{\rm h}28^{\rm m}16.6^{\rm s}\pm1.2^{\rm s}$ & $-34^{\rm d}17^{\rm m}7.1^{\rm s}\pm3.5^{\rm s}$ \\
J0835$-$4510 & 52189 & 25 & $<94.49$ & $67.8\pm0.0$ & $08^{\rm h}35^{\rm m}21.5^{\rm s}$ & $-45^{\rm d}10^{\rm m}35.9^{\rm s}$ & $08^{\rm h}35^{\rm m}20.6^{\rm s}\pm0.0^{\rm s}$ & $-45^{\rm d}10^{\rm m}34.9^{\rm s}\pm0.0^{\rm s}$ \\
J0837$-$4135 & 51571 & 22 & $68.45\pm68.45$ & $147.2\pm0.0$ & $08^{\rm h}37^{\rm m}21.3^{\rm s}$ & $-41^{\rm d}35^{\rm m}5.3^{\rm s}$ & $08^{\rm h}37^{\rm m}21.2^{\rm s}\pm0.0^{\rm s}$ & $-41^{\rm d}35^{\rm m}14.6^{\rm s}\pm0.0^{\rm s}$ \\
J0856$-$6137 & 54101 & 09 & $41.89\pm41.89$ & $95.0\pm3.0$ & $08^{\rm h}56^{\rm m}58.9^{\rm s}$ & $-61^{\rm d}37^{\rm m}49.2^{\rm s}$ & $08^{\rm h}56^{\rm m}59.3^{\rm s}\pm0.2^{\rm s}$ & $-61^{\rm d}37^{\rm m}52.7^{\rm s}\pm0.1^{\rm s}$ \\
J0942$-$5552 & 53277 & 08 & $172.66\pm34.53$ & $180.2\pm0.0$ & $09^{\rm h}42^{\rm m}15.4^{\rm s}$ & $-55^{\rm d}52^{\rm m}56.1^{\rm s}$ & $09^{\rm h}42^{\rm m}14.9^{\rm s}\pm0.9^{\rm s}$ & $-55^{\rm d}52^{\rm m}55.1^{\rm s}\pm0.5^{\rm s}$ \\
J1001$-$5507 & 51912 & 04 & $94.49\pm94.49$ & $130.3\pm0.2$ & $10^{\rm h}01^{\rm m}38.4^{\rm s}$ & $-55^{\rm d}07^{\rm m}14.5^{\rm s}$ & $10^{\rm h}01^{\rm m}37.9^{\rm s}\pm0.7^{\rm s}$ & $-55^{\rm d}07^{\rm m}7.8^{\rm s}\pm0.5^{\rm s}$ \\
J1047$-$6709 & 54771 & 03 & $125.67\pm41.89$ & $116.2\pm0.0$ & $10^{\rm h}47^{\rm m}28.2^{\rm s}$ & $-67^{\rm d}09^{\rm m}46.8^{\rm s}$ & $10^{\rm h}47^{\rm m}28.3^{\rm s}\pm0.0^{\rm s}$ & $-67^{\rm d}09^{\rm m}51.1^{\rm s}\pm0.0^{\rm s}$ \\
J1056$-$6258 & 54771 & 33 & $335.11\pm41.89$ & $320.6\pm0.0$ & $10^{\rm h}56^{\rm m}26.9^{\rm s}$ & $-62^{\rm d}58^{\rm m}52.5^{\rm s}$ & $10^{\rm h}56^{\rm m}25.6^{\rm s}\pm0.0^{\rm s}$ & $-62^{\rm d}58^{\rm m}47.7^{\rm s}\pm0.0^{\rm s}$ \\
J1157$-$6224 & 53282 & 24 & $310.79\pm34.53$ & $325.2\pm0.5$ & $11^{\rm h}57^{\rm m}13.7^{\rm s}$ & $-62^{\rm d}24^{\rm m}36.6^{\rm s}$ & $11^{\rm h}57^{\rm m}15.2^{\rm s}\pm0.1^{\rm s}$ & $-62^{\rm d}24^{\rm m}50.9^{\rm s}\pm0.0^{\rm s}$ \\
J1243$-$6423 & 52169 & 10 & $377.96\pm94.50$ & $297.1\pm0.0$ & $12^{\rm h}43^{\rm m}15.6^{\rm s}$ & $-64^{\rm d}23^{\rm m}6.5^{\rm s}$ & $12^{\rm h}43^{\rm m}17.1^{\rm s}\pm0.0^{\rm s}$ & $-64^{\rm d}23^{\rm m}23.7^{\rm s}\pm0.0^{\rm s}$ \\
J1307$-$6318 & 53284 & 02 & $379.85\pm34.53$ & $374.0\pm8.0$ & $13^{\rm h}07^{\rm m}54.4^{\rm s}$ & $-63^{\rm d}18^{\rm m}34.1^{\rm s}$ & $13^{\rm h}07^{\rm m}54.7^{\rm s}\pm9.0^{\rm s}$ & $-63^{\rm d}18^{\rm m}35.0^{\rm s}\pm4.0^{\rm s}$ \\
J1320$-$3512 & 54305 & 19 & $<34.53$ & $16.4\pm0.1$ & $13^{\rm h}20^{\rm m}12.6^{\rm s}$ & $-35^{\rm d}12^{\rm m}21.6^{\rm s}$ & $13^{\rm h}20^{\rm m}12.7^{\rm s}\pm0.6^{\rm s}$ & $-35^{\rm d}12^{\rm m}26.0^{\rm s}\pm0.8^{\rm s}$ \\
J1327$-$6222 & 53284 & 08 & $310.79\pm34.53$ & $318.5\pm0.0$ & $13^{\rm h}27^{\rm m}17.6^{\rm s}$ & $-62^{\rm d}22^{\rm m}27.1^{\rm s}$ & $13^{\rm h}27^{\rm m}17.2^{\rm s}\pm0.1^{\rm s}$ & $-62^{\rm d}22^{\rm m}45.5^{\rm s}\pm0.1^{\rm s}$ \\
J1401$-$6357 & 52177 & 12 & $94.49\pm94.49$ & $98.0\pm0.5$ & $14^{\rm h}01^{\rm m}52.4^{\rm s}$ & $-63^{\rm d}57^{\rm m}49.1^{\rm s}$ & $14^{\rm h}01^{\rm m}52.5^{\rm s}\pm0.2^{\rm s}$ & $-63^{\rm d}57^{\rm m}42.0^{\rm s}\pm0.1^{\rm s}$ \\
J1534$-$5334 & 50690 & 16 & $<68.45$ & $24.8\pm0.0$ & $15^{\rm h}34^{\rm m}8.2^{\rm s}$ & $-53^{\rm d}34^{\rm m}14.8^{\rm s}$ & $15^{\rm h}34^{\rm m}8.3^{\rm s}\pm0.1^{\rm s}$ & $-53^{\rm d}34^{\rm m}19.7^{\rm s}\pm0.1^{\rm s}$ \\
J1602$-$5100 & 50692 & 03 & $136.90\pm68.46$ & $170.8\pm0.0$ & $16^{\rm h}02^{\rm m}18.6^{\rm s}$ & $-50^{\rm d}59^{\rm m}58.2^{\rm s}$ & $16^{\rm h}02^{\rm m}18.7^{\rm s}\pm0.2^{\rm s}$ & $-51^{\rm d}00^{\rm m}6.1^{\rm s}\pm0.2^{\rm s}$ \\
J1605$-$5257 & 53804 & 28 & $34.53\pm34.53$ & $34.9\pm0.1$ & $16^{\rm h}05^{\rm m}17.4^{\rm s}$ & $-52^{\rm d}57^{\rm m}28.2^{\rm s}$ & $16^{\rm h}05^{\rm m}16.3^{\rm s}\pm0.0^{\rm s}$ & $-52^{\rm d}57^{\rm m}34.8^{\rm s}\pm0.0^{\rm s}$ \\
J1644$-$4559 & 50694 & 28 & $479.16\pm68.46$ & $478.7\pm0.0$ & $16^{\rm h}44^{\rm m}48.8^{\rm s}$ & $-45^{\rm d}59^{\rm m}9.9^{\rm s}$ & $16^{\rm h}44^{\rm m}49.3^{\rm s}\pm0.0^{\rm s}$ & $-45^{\rm d}59^{\rm m}9.7^{\rm s}\pm0.0^{\rm s}$ \\
J1645$-$0317 & 54234 & 18 & $34.53\pm34.53$ & $35.8\pm0.0$ & $16^{\rm h}45^{\rm m}1.8^{\rm s}$ & $-03^{\rm d}17^{\rm m}54.8^{\rm s}$ & $16^{\rm h}45^{\rm m}2.0^{\rm s}\pm0.0^{\rm s}$ & $-03^{\rm d}17^{\rm m}57.8^{\rm s}\pm0.0^{\rm s}$ \\
J1646$-$6831 & 54099 & 25 & $41.89\pm41.89$ & $43.0\pm2.0$ & $16^{\rm h}46^{\rm m}54.9^{\rm s}$ & $-68^{\rm d}31^{\rm m}44.1^{\rm s}$ & $16^{\rm h}46^{\rm m}54.9^{\rm s}\pm0.4^{\rm s}$ & $-68^{\rm d}31^{\rm m}51.7^{\rm s}\pm0.1^{\rm s}$ \\
J1651$-$4246 & 53490 & 07 & $483.45\pm34.47$ & $482.0\pm3.0$ & $16^{\rm h}51^{\rm m}48.9^{\rm s}$ & $-42^{\rm d}45^{\rm m}57.5^{\rm s}$ & $16^{\rm h}51^{\rm m}48.8^{\rm s}\pm0.1^{\rm s}$ & $-42^{\rm d}46^{\rm m}10.0^{\rm s}\pm0.0^{\rm s}$ \\
J1701$-$3726 & 50696 & 08 & $342.26\pm68.45$ & $303.4\pm0.5$ & $17^{\rm h}01^{\rm m}18.3^{\rm s}$ & $-37^{\rm d}26^{\rm m}22.9^{\rm s}$ & $17^{\rm h}01^{\rm m}18.5^{\rm s}\pm0.2^{\rm s}$ & $-37^{\rm d}26^{\rm m}27.2^{\rm s}\pm0.5^{\rm s}$ \\
J1731$-$4744 & 53521 & 16 & $103.60\pm34.52$ & $123.1\pm0.0$ & $17^{\rm h}31^{\rm m}42.9^{\rm s}$ & $-47^{\rm d}44^{\rm m}39.8^{\rm s}$ & $17^{\rm h}31^{\rm m}42.2^{\rm s}\pm0.1^{\rm s}$ & $-47^{\rm d}44^{\rm m}36.3^{\rm s}\pm0.1^{\rm s}$ \\
J1738$-$2330 & 53319 & 05 & $69.06\pm34.53$ & $96.6\pm1.3$ & $17^{\rm h}38^{\rm m}8.3^{\rm s}$ & $-23^{\rm d}31^{\rm m}24.0^{\rm s}$ & $17^{\rm h}38^{\rm m}8.8^{\rm s}\pm4.5^{\rm s}$ & $-23^{\rm d}30.8^{\rm m}\pm1.7^{\rm m}$ \\
J1741$-$0840 & 50703 & 27 & $68.45\pm68.45$ & $74.9\pm0.1$ & $17^{\rm h}41^{\rm m}21.9^{\rm s}$ & $-08^{\rm d}40^{\rm m}21.8^{\rm s}$ & $17^{\rm h}41^{\rm m}22.6^{\rm s}\pm0.0^{\rm s}$ & $-08^{\rm d}40^{\rm m}31.7^{\rm s}\pm0.0^{\rm s}$ \\
J1741$-$2019 & 50701 & 32 & $68.45\pm68.45$ & $74.9\pm0.4$ & $17^{\rm h}41^{\rm m}6.0^{\rm s}$ & $-20^{\rm d}19^{\rm m}11.2^{\rm s}$ & $17^{\rm h}41^{\rm m}6.9^{\rm s}\pm0.4^{\rm s}$ & $-20^{\rm d}19^{\rm m}24.0^{\rm s}\pm5.0^{\rm s}$ \\
J1743$-$3150 & 53802 & 14 & $207.19\pm34.53$ & $193.1\pm0.1$ & $17^{\rm h}43^{\rm m}37.2^{\rm s}$ & $-31^{\rm d}50^{\rm m}16.7^{\rm s}$ & $17^{\rm h}43^{\rm m}36.7^{\rm s}\pm0.1^{\rm s}$ & $-31^{\rm d}50^{\rm m}22.7^{\rm s}\pm0.9^{\rm s}$ \\
J1745$-$3040 & 53802 & 15 & $103.60\pm34.53$ & $88.4\pm0.0$ & $17^{\rm h}45^{\rm m}56.5^{\rm s}$ & $-30^{\rm d}40^{\rm m}23.2^{\rm s}$ & $17^{\rm h}45^{\rm m}56.3^{\rm s}\pm0.2^{\rm s}$ & $-30^{\rm d}40^{\rm m}22.9^{\rm s}\pm1.1^{\rm s}$ \\
J1752$-$2806 & 53522 & 29 & $34.53\pm34.52$ & $50.4\pm0.0$ & $17^{\rm h}52^{\rm m}58.2^{\rm s}$ & $-28^{\rm d}06^{\rm m}35.9^{\rm s}$ & $17^{\rm h}52^{\rm m}58.7^{\rm s}\pm0.0^{\rm s}$ & $-28^{\rm d}06^{\rm m}37.3^{\rm s}\pm0.3^{\rm s}$ \\
J1809$-$1943 & 50706 & 33 & $68.45\pm68.45$ & $178.0\pm5.0$ & $18^{\rm h}09^{\rm m}51.3^{\rm s}$ & $-19^{\rm d}43^{\rm m}56.6^{\rm s}$ & $18^{\rm h}09^{\rm m}51.1^{\rm s}\pm0.0^{\rm s}$ & $-19^{\rm d}43^{\rm m}51.9^{\rm s}\pm0.0^{\rm s}$ \\
J1825$-$0935 & 52159 & 17 & $<94.49$ & $19.4\pm0.0$ & $18^{\rm h}25^{\rm m}31.4^{\rm s}$ & $-09^{\rm d}35^{\rm m}23.6^{\rm s}$ & $18^{\rm h}25^{\rm m}30.6^{\rm s}\pm0.0^{\rm s}$ & $-09^{\rm d}35^{\rm m}21.2^{\rm s}\pm0.2^{\rm s}$ \\
J1830$-$1135 & 49744 & 13 & $172.34\pm57.45$ & $257.0\pm6.0$ & $18^{\rm h}30^{\rm m}1.5^{\rm s}$ & $-11^{\rm d}35^{\rm m}32.2^{\rm s}$ & $18^{\rm h}30^{\rm m}1.7^{\rm s}\pm0.9^{\rm s}$ & $-11^{\rm d}35^{\rm m}32.0^{\rm s}\pm6.0^{\rm s}$ \\
J1840$-$0840 & 49744 & 04 & $287.23\pm57.45$ & $285.2\pm1.4$ & $18^{\rm h}40^{\rm m}48.4^{\rm s}$ & $-08^{\rm d}40^{\rm m}54.6^{\rm s}$ & $18^{\rm h}40^{\rm m}51.9^{\rm s}\pm6.0^{\rm s}$ & $-08^{\rm d}40^{\rm m}29.0^{\rm s}\pm15.0^{\rm s}$ \\
% J1840$-$1419 & 49744 & 26 & $<57.45$ & $19.4\pm1.4$ & $18^{\rm h}40^{\rm m}27.1^{\rm s}$ & $-14^{\rm d}19^{\rm m}21.0^{\rm s}$ & $18^{\rm h}40^{\rm m}33.0^{\rm s}\pm0.2^{\rm s}$ & $-14^{\rm d}19^{\rm m}6.5^{\rm s}\pm0.9^{\rm s}$ \\
J1921$+$2153 & 49745 & 32 & $<57.45$ & $12.4\pm0.0$ & $19^{\rm h}21^{\rm m}45.3^{\rm s}$ & $+21^{\rm d}53^{\rm m}5.9^{\rm s}$ & $19^{\rm h}21^{\rm m}44.8^{\rm s}\pm0.0^{\rm s}$ & $+21^{\rm d}53^{\rm m}2.2^{\rm s}\pm0.0^{\rm s}$ \\
%J1922$+$2110 & 49722 & 01 & $<57.45$ & $217.1\pm0.0$ & $19^{\rm h}21^{\rm m}55.2^{\rm s}$ & $+21^{\rm d}51^{\rm m}59.5^{\rm s}$ & $19^{\rm h}22^{\rm m}53.5^{\rm s}\pm0.1^{\rm s}$ & $+21^{\rm d}10^{\rm m}42.0^{\rm s}\pm0.1^{\rm s}$ \\
J2033$+$0042 & 54248 & 08 & $34.53\pm34.53$ & $37.8\pm0.1$ & $20^{\rm h}33^{\rm m}30.7^{\rm s}$ & $+00^{\rm d}42^{\rm m}23.3^{\rm s}$ & $20^{\rm h}33^{\rm m}31.1^{\rm s}\pm0.3^{\rm s}$ & $+00^{\rm d}42^{\rm m}24.1^{\rm s}\pm0.9^{\rm s}$ \\
J2144$-$3933 & 54317 & 29 & $<34.53$ & $3.4\pm0.0$ & $21^{\rm h}44^{\rm m}11.3^{\rm s}$ & $-39^{\rm d}34^{\rm m}1.4^{\rm s}$ & $21^{\rm h}44^{\rm m}12.1^{\rm s}\pm0.0^{\rm s}$ & $-39^{\rm d}33^{\rm m}56.9^{\rm s}\pm0.0^{\rm s}$ \\
J2228$-$65 & 53513 & 15 & $<41.88$ & $36.0$ & $22^{\rm h}27^{\rm m}39.5^{\rm s}$ & $-65^{\rm d}08^{\rm m}12.7^{\rm s}$ & $22^{\rm h}28^{\rm m}18.0^{\rm s}$ & $-65^{\rm d}11^{\rm m}47.0^{\rm s}$ \\
J2324$-$6054 & 53264 & 16 & $<41.89$ & $14.0\pm0.6$ & $23^{\rm h}24^{\rm m}26.8^{\rm s}$ & $-60^{\rm d}53^{\rm m}52.7^{\rm s}$ & $23^{\rm h}24^{\rm m}27.1^{\rm s}\pm0.2^{\rm s}$ & $-60^{\rm d}54^{\rm m}5.8^{\rm s}\pm0.0^{\rm s}$ \\
\hline\hline
\end{tabular}
    \begin{tablenotes}
    % \begin{footnotesize}
    \item[$^\ast$] The uncertainty in DM$_{\rm CRACO}$ corresponds to the DM of one sample. An upper limit is reported if the DM measured with the pipeline is 0. 
    \item[$^\mathsection$] The error on the coordinate depends on the pixel size of the images, which varies for different observations. The typical pixel size of CRACO images is $\sim$30-60\,arcsec.
    \item[$^\dagger$] J0031$-$5726 was discovered by MWA (see \url{https://mwatelescope.atlassian.net/wiki/spaces/MP/pages/24970773/SMART+survey+candidates}), and not recorded in ATNF Pulsar catalogue yet.
    % \end{footnotesize}
    \end{tablenotes}
    \end{threeparttable}
    \label{tab:cracopsr}
\end{table*}

\end{appendix}

\end{document}